\documentclass[nofootinbib,prd,aps]{revtex4}
\usepackage{graphicx}
\usepackage{dcolumn}
\usepackage{bm}
\usepackage{amsfonts}

\def\P{{\bf P}}

\def\P{\rm P}

\def\Hbar{\overline{H}}
\def\cc{$c\bar c$ }
\def\be{\begin{equation}}
\def\ee{\end{equation}}

\newcommand{\slj}[3]{\mbox{$^{#1}${\ifcase#2\or S\or 
         P\or D\or F\or G\fi}$_{#3}$}}

\begin{document}
\voffset = 0.3 true in
\topmargin = -1.0 true in 

\title{The New Heavy Mesons: A Status Report}

\author{Eric S. Swanson\footnote{{\tt swansone@pitt.edu}}}

\affiliation{Department of Physics and Astronomy, University of Pittsburgh\\
Pittsburgh, PA 15260, USA}

\begin{abstract}
A survey of the experimental, phenomenological, and theoretical status of the new heavy mesons
is presented. States discussed are the $B_c$, $h_c$, $\eta_c'$, $D_s(2317)$, $D_s(2460)$, 
$X(3872)$, $X(3940)$,
$Y(3940)$, $Z(3930)$, and $Y(4260)$. Quark models for spectra, strong decays, and hadronic interactions
are reviewed and used to interpret the new states. New results for strong decay models, bound
state decays, mesonic molecules, properties of the X(3872), and the chiral doublet model  are 
also presented.

\end{abstract}

\maketitle

\tableofcontents

\section{Introduction}   

More than 30 years after the November Revolution, charmonium spectroscopy continues to 
surprise and challenge. A new era began in April of 2003 when BaBar announced the
discovery of the enigmatic $D_s(2317)$.  Since then CLEO, Belle, Fermilab, and BES have 
joined the scrum, and the number of new states has risen to the double digits.


The formerly comfortable world of heavy quark spectroscopy is being severely
tested by new experiment\cite{quiggRev}. This report summarises the experimental status of eleven new open and 
hidden charm mesons\footnote{States not discussed here include the X(1835), putative glueballs, hybrids $\pi_1(1400)$, $\pi_1(1600)$,
and $\pi(1800)$, and baryons.}, the challenges these make to our understanding of strong QCD, and
the theoretical attempts to meet these challenges.

This report begins with a summary of the new eleven and a review of what is expected in charmonium
spectroscopy. For the latter, several models (new and old) are compared to the data, and three
models of strong decays are examined. This serves to illustrate what is known (and not known)
before examining the  new charmonia states. A similar introduction to heavy-light systems
is made in Sections IV.A and IV.B.

\subsection{Summary}

Table \ref{summaryTab} summarises the contents of this report for the impatient. A selection of 
interpretations and theoretical ideas are listed in the `comments' column. Theory
and more experimental detail can be found in the main text.

\begin{table}[h]
\caption{New Heavy Mesons}
\begin{tabular}{c|ccclc} 
\hline
state & mass (MeV) & width (MeV) & production/decay mode & comments & ref \\
\hline
\hline
$h_c$ &     $3524.4 \pm 0.6 \pm 0.4$  & --   & $\psi(2S) \to \pi^0 h_c \to (\gamma\gamma)(\gamma\eta_c)$  &  $\approx$ CQM / tests spin dependence   & CLEO\cite{cleoB}    \\
$\eta_c'$ & $3654\pm 6 \pm 8$ & $< 55$  & $B\to K\eta_c' \to KK_SK\pi$  & $\approx$ CQM / tests hyperfine splitting  & Belle\cite{BelleA} \\
     & $3642.9 \pm 3.1 \pm 1.5$ & $6.3 \pm 12.4 \pm 4.0$  & $e^+e^- \to \eta_c' J/\psi$   &            & CLEO\cite{cleoC} \\
     & $3630.8 \pm 3.4 \pm 1.0$  &  $17.0 \pm 8.0 \pm 2.5$ &  $\gamma\gamma \to \eta_c' \to K_SK\pi$  &                               & BaBar\cite{Aubert:2003pt} \\
$X(3872)$ & $3872.0 \pm 0.6 \pm 0.5$    &  $< 2.3$ 95\% C.L.  & $B \to KX \to K\pi\pi J/\psi$    &   molecule, cusp, tetraquark  & Belle\cite{belleX} \\
          & $3873.4 \pm 1.4$    &   --   & $B \to KX \to K\pi\pi J/\psi$    &           & BaBar\cite{BabarX} \\
          & --                  &   --   & $B \to X \to \pi\pi\pi J/\psi$    &         & Belle\cite{Abe:2005ix} \\
          & --                  &   --   & $B \to X \to \gamma J/\psi$    &         & Belle\cite{Abe:2005ix} \\
          & $3871.3 \pm 0.7 \pm 0.4$  &  --  &  $p\bar p \to X \to \pi\pi J/\psi$ &   & CDF\cite{cdfX} \\
          & $3871.8 \pm 3.1 \pm 3.0$    & -- &  $p\bar p \to X \to \pi\pi J/\psi$ &     &  D{\O}\cite{D0X} \\
          & avg = $3871.9 \pm 0.5$             &     &     &  \\

$X(3940)$ & $3943 \pm 6 \pm 6$   & $< 52$   & $e^+e^- \to J/\psi X \to J/\psi D\bar D^*$  & $\chi_{c1}'$, $\eta_c''$ / needs confirmation                  &  Belle\cite{belleC}   \\
$Y(3940)$ & $3943 \pm 11 \pm 13$    &  $87 \pm 22 \pm 26$  & $B \to KY \to K\pi\pi\pi J/\psi$   &  needs confirmation    & Belle\cite{belleB}  \\
$Z(3930)$ & $3931 \pm 4 \pm 2$   &  $20 \pm 8 \pm 3$  & $\gamma\gamma \to Z \to D\bar D$  &  $\chi_{c2}'$ / $\approx$ CQM    & Belle\cite{belleZ} \\
$Y(4260)$ & $4259 \pm 8 \pm 4$   & $88 \pm 23 \pm 5$  & $e^+e^- \to \gamma_{ISR}Y \to \gamma_{ISR}J/\psi \pi\pi$    & hybrid?/ needs confirmation     & BaBar\cite{babarB} \\
\hline
$D_s(2317)$ & $2317.3 \pm 0.4 \pm 0.8$   & $< 10$  & $e^+e^- \to D_s(2317) \to D_s \pi^0$  & molecule, tetraquark, shifted $c\bar s$  & BaBar\cite{babarDS} \\
            & $2319.8 \pm 2.1 \pm 2.0$   & $\approx 0$ & $B \to \bar D D_s(2317) \to \bar D D_s\pi^0$   &                               & Belle\cite{belleDS} \\
            & $2318.5 \pm 1.2 \pm 1.1$   &     & $D_s(2317) \to D_s\pi^0$   &                               & CLEO\cite{cleoDS} \\
$D_s(2460)$ & $2463.6 \pm 1.7 \pm 1.2$ & $< 7$ 90\% C.L. & $D_s(2460) \to D_s^*\pi^0$  & molecule, tetraquark, shifted $c\bar s$ & CLEO\cite{cleoDS} \\
            & $2458.0 \pm 1.0 \pm 1.0$    & resolution  & $D_s(2460) \to D_s\pi^0\gamma$  &                               & BaBar\cite{babarDS2460} \\
            & $2459.2 \pm 1.6 \pm 2.0$    & $\approx 0$ &  $B \to \bar D D_s(2460) \to \bar D D_s^*\pi^0$, $\bar D D_s \gamma$ &   & Belle\cite{belleDS} \\
$D_s(2630)$ & $2632.6 \pm 1.6$    & $< 17$ 90\% C.L.    & $D_s \to D^0K^+$ and $D_s\eta$   &  artefact                     & SELEX\cite{selex}  \\
$B_c$ & $6285.7 \pm 5.3 \pm 1.2$ & $0.474 \pm 0.07 \pm 0.33$ ps & $p\bar p \to B_c \to J/\psi \pi^\pm$ & $\approx$ CQM  & CDF\cite{Acosta:2005us} \\
\hline
\hline
\end{tabular}
\label{summaryTab}
\end{table}

\subsection{Charmonia}

This section describes the current status of charmonium and strong decay (mostly
constituent quark and lattice gauge theory). This is in an effort to set the 
stage for the new hadrons to be described in subsequent sections and to
establish the degree of reliability which we may attach to those expectations.

The utility of the nonrelativistic limit in describing charmonia has been
recognised since the days of the November Revolution\cite{oldcc,cornell}. 
The subsequent advent of the operator product\cite{Peskin}, nonrelativistic\cite{bbl},
 and potential nonrelativistic QCD\cite{hq} formalisms
has considerably strengthened the connection of heavy quark phenomenology with
QCD\footnote{It also reveals that the charm quark is uncomfortably light.}. 
For example, 
the hierarchy of scales $m_c >> m_c \alpha_s(m_c) >> m_c \alpha_s^2(m_c)$
permits one to choose a cutoff which removes transverse gluons from the effective description
of a meson. Thus an efficient potential description and Fock space expansion of charmonium
is feasible\cite{lepage}.

\subsubsection{Spin-dependent Interactions}
\label{spindepSec}

Typical nonrelativistic quark models contain the following ingredients: nonrelativistic quark
kinetic energy, a central confining potential, and a variety of spin-dependent $(v/c)^2$ corrections. 
Thus one has

\be
H_{CQM} = K(q) + K(\bar q) + V_{conf} + V_{SD}
\ee
where $K$ is a quark kinetic energy, $V_{conf}$ is the central confining potential, and $V_{SD}$ is the
spin-dependent potential.

The form of the spin-dependent interaction has been computed by Eichten and Feinberg\cite{EF} at the tree level using the Wilson loop methodology. This result was extended by
Pantaleone, Tye, and Ng\cite{PTN} in a one-loop computation. The net result is:

\begin{eqnarray}
V_{SD}(r) &=& \left( {\bm{\sigma}_q \over 4 m_q^2} +
{\bm{\sigma}_{\bar q} \over 4 m_{\bar q}^2} \right)\cdot {\bf L} \left( {1\over r}
{d \epsilon \over d r} + {2 \over r} {d V_1 \over d r} \right) +
\left( {\bm{\sigma}_{\bar q} +
        \bm{\sigma}_q \over 2 m_q m_{\bar q}} \right)\cdot {\bf L}
        \left( {1 \over r} {d V_2 \over d r} \right) \nonumber \\
&&+ {1 \over 12 m_q m_{\bar q}}\Big( 3 \bm{\sigma}_q \cdot \hat {\bf r} \,
 \bm{\sigma}_{\bar q}\cdot \hat {\bf r} -   \bm{\sigma}_q\cdot
 \bm{\sigma}_{\bar q} \Big) V_3
+ {1 \over 12 m_q m_{\bar q}} \bm{\sigma}_q \cdot \bm{\sigma}_{\bar q}
V_4 \nonumber \\
&&+ {1\over 2}\left[ \left( {\bm{\sigma}_q \over m_q^2} - {\bm{\sigma}_{\bar q}\over m_{\bar q}^2}\right)\cdot {\bf L} +
\left({\bm{\sigma}_q - \bm{\sigma}_{\bar q}\over m_q m_{\bar q}}\right)\cdot {\bf L} \right] V_5.
\label{VSD}
\end{eqnarray}
Here $\epsilon=\epsilon(r)$ is the static potential, ${\bf L} = {\bf L}_q = - {\bf L_{\bar q}}$, 
$r=|{\bf r}|= |{\bf r}_q - {\bf r}_{\bar q}|$ is the ${\bar Q Q}$ separation
and the $V_i=V_i(m_q,m_{\bar q}; r)$ are determined by
electric and magnetic
field insertions on quark lines in the Wilson loop matrix element.
$V_5$ vanishes in the equal quark mass case.
As shown by Gromes \cite{G} covariance under Lorentz transformation
leads to a  constraint
between the SD potentials,

\begin{equation}
\epsilon'(r) = V_2'(r) - V_1'(r).
\label{grom}
\end{equation}

In a model approach, ${Q\bar Q}$ interactions are typically derived from
a nonrelativistic reduction of a relativistic current-current interaction

\be
H_{int} = \frac{1}{2}\int d^3x d^3y \, J_\Gamma(x) K(x-y) J_\Gamma(y)
\label{JKJ}
\ee
where $J_\Gamma = \bar\psi\Gamma \psi$ specifies a quark current.

As far as long range potentials are concerned only time-like vector or
scalar currents are relevant since other currents do not yield static
central potentials\cite{Grev}. 
Vector confinement interactions are motivated as natural extensions of one-gluon-exchange.
More fundamentally, the instantaneous nonperturbative interaction of QCD in Coulomb gauge 
has the Lorentz structure $V^0 \otimes V^0$. Scalar interactions are also often considered
when model building for reasons to be discussed shortly.

Nonrelativistic reductions of vector or scalar interactions yield:

\begin{eqnarray}
V_1 &=& 0,\qquad V_2 = \epsilon, \qquad V_3 = \epsilon'/r - \epsilon'', \qquad V_4 =
2 \nabla^2\epsilon\qquad\qquad{\rm vector} \nonumber \\
V_1 &=& -\epsilon,\qquad  V_2= V_3 = V_4 = 0\qquad\qquad{\rm scalar}.
\label{scal}
\end{eqnarray}

Determining the Dirac structure of the long range potential yields important information
on the nonperturbative gluodynamics which gives rise to confinement. This can be accomplished
by examining spin-dependent mass splittings in quarkonia, in particular the ratio of P-wave 
splittings defined by 

\be
\rho = {2^{++} - 1^{++}\over 1^{++}-0^{++}} 
\ee
is a useful diagnostic\cite{schnitzer}.

Perturbative P-wave quarkonium splittings are given by

\begin{eqnarray}
2^{++} &=& E_0 - {2\over 5} T + L + {1\over 4} S \\
1^{++} &=& E_0 + 2T  - L + {1\over 4} S \\
0^{++} &=& E_0 - 4T - 2L + {1\over 4} S \\
1^{+-} &=& E_0 - {3\over 4} S
\end{eqnarray}
where 

\begin{eqnarray}
T &=& {1\over 12 m^2} \langle V_3 \rangle \\
L &=& {1 \over m^2}  \langle {\epsilon'\over 2r} + {V_1'\over r} + {V_2'\over r} \rangle \\
S &=& {1\over 3 m^2} \langle  V_4\rangle.
\end{eqnarray}

Using the results given in Eq. \ref{scal} gives vector and scalar spin splittings ratios of

\be 
\rho(V) = {\langle {16\over 5} {\alpha_s\over r^3} + {14\over 5} {b\over r} \rangle \over
           \langle 4 {\alpha_s \over r^3} + 2 {b \over r}\rangle }
\ee

and

\be 
\rho(S) = {\langle {16\over 5} {\alpha_s\over r^3} -  {b\over r} \rangle \over
           \langle 4 {\alpha_s \over r^3} - {b \over 2r}\rangle }.
\ee

The measured values for this parameter are $\rho(\chi_c) = 0.48(1)$ and 
$\rho(\chi_b) = 0.66(2)$.  Notice that the results given here imply that
$\rho(Q\bar Q) \to 4/5$ in the heavy quark limit. Thus, although the motion is
in the right direction, even the $\chi_b$ system is not yet heavy by this metric.

\begin{figure}[h]
  \includegraphics[width=7 true cm, angle=270]{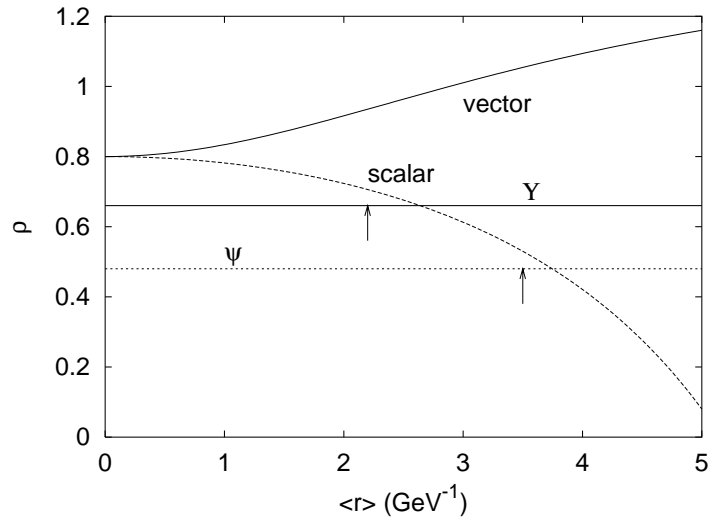}
  \caption{Mass Splitting Ratio vs. $r_{RMS}$. Experimental values are given as
horizontal lines.}
\label{rhoNewFig}
\end{figure}

Fig. \ref{rhoNewFig} displays the splitting ratio versus the RMS meson radius along with
the experimentally measured ratios for the $\psi$ and $\Upsilon$ systems. It is clear
that the vector confinement hypothesis cannot explain the data. Alternatively, the
scalar hypothesis decreases with increasing RMS radius as required. Arrows in the figure
show RMS radii for the $\chi_c$ and $\chi_b$ systems computed in a simple nonrelativistic
model, indicating that scalar confinement
is also quantitatively reliable.

The hyperfine interaction $V_4$ may be probed by comparing the mass splitting between the P-wave system
and the axial vector quarkonium state.  Specifically,
perturbative tensor and spin-orbit splittings are not present in the `centre of gravity'
of the $P$-wave system:

\be
M_{cog} = {1\over 9}\left( M(0^{++}) + 3  M(1^{++}) + 5 M(2^{++}) \right) = E_0 + {1\over 4} S.
\ee
Thus

\be
M_{cog} - M(1^{+-}) = S.
\ee
For one gluon exchange $S \propto \langle\delta(\vec r)\rangle$ and since this expectation
value is zero in P-waves, very little axial-cog mass splitting is expected (little because
the delta function is smeared by zitterbewegung effects). Alternatively, for a vector
confining potential $S = {4 b \over 3 m^2}\langle {1\over r}\rangle$ and a nonzero
splitting can be expected. The recent discovery of the $h_c$ permits this splitting to
be measured. The results are discussed in Section \ref{hcSec}.

Simple spectroscopy teaches us about the $v^2/c^2$ structure of the
confinement potential, and therefore something about the nonperturbative gluodynamics
associated with flux tubes. This is nontrivial and worthy of pursuit!

\subsubsection{Constituent Quark Model and Lattice Spectra}

The predictions of four representative constituent quark models are given in Table \ref{spectrumTab}.
These are a simple nonrelativistic model\cite{BGS}, an updated version of the
`relativised' Godfrey-Isgur model\cite{BGS}, a `relativistic' quark model\cite{EFG}, and
the original Cornell model\cite{cornell}.

The simple constituent quark model (labelled BGS in Table \ref{spectrumTab}) employs a nonrelativistic
quark kinetic energy, $K(q) = m_q - \nabla^2/2m_q$, a `Coulomb+linear' central potential, $V_{conf} = -4 
\alpha_s/(3r) + br$ and a spin-dependent interaction that corresponds to the vector one-gluon-exchange + scalar
confinement Ansatz in the equal quark mass case:

\be
V_1 = -br, \qquad V_2 = -{4\over 3}{\alpha_s\over r}, \qquad V_3 = 4 {\alpha_s\over r^3}, \qquad V_4 = 
\frac{32\pi\alpha_s}{3}\,\tilde \delta_{\sigma}(r), \qquad V_5 = 0
\ee
where
$\tilde \delta_{\sigma}(r) = (\sigma/\sqrt{\pi})^3\, e^{-\sigma^2 r^2}$ represents a `smeared' delta function.
Fitting the charmonium spectrum yields:
$\alpha_s= 0.5462$, $b = 0.1425$ GeV$^2$, $m_c = 1.4794$ GeV, and  $\sigma  = 1.0946$ GeV.

The Godfrey-Isgur model is a ``relativised'' extension of typical nonrelativistic quark models.
In particular, a relativistic dispersion relation for the quark kinetic energy $K(q) = \sqrt{m_q^2 + \nabla^2}$, 
a QCD-motivated running coupling $\alpha_s(r)$, a flavor-dependent
potential smearing parameter $\sigma \to \sigma(m_q,m_{\bar q})$, and additional factors of 
$m_q/K(q)$ are all used\cite{BGS,GI}.
Just as in the nonrelativistic model, the quark-antiquark potential
V$_{q\bar{q}} (\vec{p},\vec{r})$ incorporates the
Lorentz vector one gluon exchange interaction at
short distances and a Lorentz scalar linear confining interaction.
Note that the
string tension and quark mass ($b =  0.18$ GeV$^2$ and
$m_c=1.628$ GeV) are significantly larger than the values used in the nonrelativistic model.
Predictions of this model are in the column labelled GI in Table \ref{spectrumTab}.

Ebert, Faustov, and Galkin\cite{EFG} attempt a more formal (`quasipotential') approach to 
relativising the quark model. A nonrelativistic dispersion relation is used but with reduced
and total masses determined relativistically (for example, the reduced mass is given by
$\mu = E_qE_{\bar q}/(E_q + E_{\bar q})$ with $E_q = (M^2-m_{\bar q}^2 + m_q^2)/(2M)$ and $M$
is the eigenenergy). The interaction employs the complete Coulomb gauge gluon propagator
(so that the temporal and spatial vector interactions are included), a scalar confinement
potential, and a vector confinement potential with an anomalous magnetic moment $\kappa$ for the quark.
In the nonrelativistic limit, the confinement potential is $V_{conf}(r) = br + B$ where
the linear potential receives contributions from the vector and scalar confinement
terms of $(1-\delta)br +B$ and $\delta b r$ respectively where $\delta$ is a mixing coefficient.
The coupling is fixed for each quark flavour and is $\alpha_s = 0.314$ for charmonia.
Additional parameters are determined to be $m_c = 1.55$ GeV,  $b = 0.18$ GeV$^2$, $B = -0.16$ GeV, $\kappa = -1$. Finally, at ${\cal O}(v^2/c^2)$  the spin-dependent interaction  reduces to precisely that used
in the BGS model discussed above (when $\kappa = -1$). For example, the long range contribution to $V_4$
is

\be 
V_4(r) = 4(1+\kappa)^2 (1-\delta) {b\over r}
\ee
which is disfavoured by the data so
that that this choice of the 
anomalous magnetic moment eliminates unwanted long range chromomagnetic interactions\cite{Buch}.
The predictions of this model are labelled EFG in Table \ref{spectrumTab}.

The final model is that of the Cornell group\cite{cornell}, whose original predictions are 
reproduced in the sixth column of Table \ref{spectrumTab}. The starting point is an assumed
nonrelativistic kinetic energy and a quark-quark interaction mediated by 

\be
H_{int} = {1\over 2} \int d^3x\, d^3y \, \rho^a(x) K(x-y) \rho^a(y)
\label{Vcornell}
\ee
with the kernel given by $K(r) = \alpha_s/r - {3 \over 4} br$ and the current given by the colour
charge density, $\rho^a = \psi^\dagger T^a \psi$. Of course this is a vector confinement model
so that the nonrelativistic reduction of Eq. \ref{Vcornell} corresponds to 

\be
\epsilon(r) = -{4\over 3}K(r), \qquad V_1 = 0, \qquad V_2 = -{4\over 3}K(r), \qquad V_3 = 4{\alpha_s\over r^3} + {b\over r}, \qquad V_4 = -8\pi \alpha_s \delta(\vec r)+  4{b\over r}.
\ee

Fitted parameters are $m_c = 1.84$ GeV, $b = 0.183$ GeV$^2$, and $\alpha_s = 0.39$. Spin-dependent
effects are ignored in the original Cornell papers. However, the Cornell group does attempt to 
include the effects of open charm virtual states on the spectrum (this will be discussed in greater
detail in Sections \ref{decaySec} and \ref{spectrumCritiqueSec}). The resulting masses are shown in square 
brackets in Table \ref{spectrumTab}. We note that the coupled channel approach shifts most 
of the predicted masses a modest amount, except the $\psi(3S)$ and $\psi(4S)$ states, for which the
naive predictions have been considerably worsened.

The advent of QCD motivated quark models coincided with the invention of lattice gauge theory
and the lattice has provided an important counterpoint to models ever since. Recently, lattice
computations have become sufficiently robust that attempts have been made to compute extensive
spectra. Two such computations are presented in the last columns of Table \ref{spectrumTab}.
These computations have been made in the `quenched' approximation where the effects of 
virtual quarks are neglected on anisotropic lattices with a lattice volume fixed at (1.6 fm)$^3$ in
Ref. \cite{cppacs} or over lattice volumes ranging from (1.5 fm)$^3$ to (3.3 fm)$^3$ in Ref. \cite{col}.
Continuum extrapolated values are quoted in the table.


\begin{table*}[h]
\caption{Experimental and theoretical spectrum of \cc states.
States assigned in this review are denoted $^a$.}
\vskip 0.5cm
\begin{tabular}{l|c|cccccc}
\hline
State & PDG\cite{pdg} & BGS\cite{BGS} & GI\cite{BGS} & EFG\cite{EFG} & Cornell\cite{cornell}& CP-PACS\cite{cppacs} & Chen\cite{col} \\
\hline
\hline
 $J/\psi(1^3{\rm S}_1) $ & $ 3096.87 \pm 0.04 $ & 3090 & 3098 & 3096 & 3095 [3095] & $3085 \pm 1$ & $3084\pm 4$  \\
 $\eta_c(1^1{\rm S}_0) $ & $ 2979.2 \pm 1.3   $ &  2982 & 2975 & 2979 & 3095  & $3013 \pm 1$  & $3014\pm 4$ \\
\hline
 $\psi'(2^3{\rm S}_1) $ & $ 3685.96 \pm 0.09 $   & 3672 & 3676 & 3686 & 3684 [3684]& $3777 \pm 40$  & $3780\pm 43$  \\
 $\eta_c'(2^1{\rm S}_0) $ &  $ 3637.7 \pm 4.4  $ & 3630 & 3623 & 3588 & 3684  & $3739 \pm 46$  & $3707\pm 20$ \\
\hline
 $\psi(3^3{\rm S}_1) $ & $ 4040 \pm 10    $ &  4072 & 4100 & 4088 & 4110 [4225] & -- &  --\\
 $\eta_c(3^1{\rm S}_0) $ &                  &  4043 & 4064 & 3991 & 4110 & -- & -- \\
\hline
 $\psi(4^3{\rm S}_1) $ & $ 4415 \pm 6       $ &  4406 & 4450 & -- & 4460 [4625] & -- & --  \\
 $\eta_c(4^1{\rm S}_0) $ &                    & 4384 & 4425 & -- &  4460 & -- & -- \\
\hline
 $\chi_2(1^3{\rm P}_2) $ & $ 3556.18 \pm 0.13 $  & 3556 & 3550 & 3556 & 3522 [3523]& $3503\pm 24$ & $3488\pm 11$ \\
 $\chi_1(1^3{\rm P}_1 $) &  $ 3510.51 \pm 0.12 $  & 3505 & 3510 & 3510 & 3522 [3517]& $3472\pm 9$  & $3462\pm15$ \\
 $\chi_0(1^3{\rm P}_0) $ &  $ 3415.3 \pm 0.4   $  & 3424 & 3445 & 3424 & 3522 & $3408$ & $3413\pm 10$ \\
 $h_c(1^1{\rm P}_1) $ &  $3524 \pm 1 {}^a$           & 3516 & 3517 & 3526 & 3522 [3519]& $3474 \pm 40$ & $3474\pm 20$ \\
\hline
 $\chi_2(2^3{\rm P}_2) $ & $3931 \pm 5 {}^a$        & 3972 & 3979 & 3972 & --& $4030\pm 180$ & -- \\
 $\chi_1(2^3{\rm P}_1) $ &                        & 3925 & 3953 & 3929 & --& $4067\pm 105$ & $4010\pm 70$ \\
 $\chi_0(2^3{\rm P}_0) $ &                        & 3852 & 3916 & 3854 & --& $4008\pm 122$ & $4080\pm 75$ \\
 $h_c(2^1{\rm P}_1) $ &                           & 3934 & 3956 & 3945 & --& $4053 \pm 95$ & $3886\pm 92$ \\
\hline
 $\chi_2(3^3{\rm P}_2) $ &                        & 4317 & 4337 & -- & -- & -- \\
 $\chi_1(3^3{\rm P}_1) $ &                        & 4271 & 4317 & -- & -- & -- \\
 $\chi_0(3^3{\rm P}_0) $ &                        & 4202 & 4292 & -- & -- & -- \\
 $h_c(3^1{\rm P}_1) $ &                           & 4279 & 4318 & -- & -- & -- \\
\hline
 $\psi_3(1^3{\rm D}_3) $ &                        & 3806 & 3849 & 3815 & 3810 & -- & $3822\pm 25$  \\
 $\psi_2(1^3{\rm D}_2) $ &                        & 3800 & 3838 & 3811 & 3810 & -- & $3704 \pm 33$ \\
 $\psi(1^3{\rm D}_1) $ & $ 3769.9 \pm 2.5  $      & 3785 & 3819 & 3798 & 3810 [3755]& -- & -- \\
 $\eta_{c2}(1^1{\rm D}_2) $ &                     & 3799 & 3837 & 3811 & 3810 & -- & $3763\pm 22$ \\
\hline
 $\psi_3(2^3{\rm D}_3) $ &                        & 4167 & 4217 & -- & 4190 & -- & --\\
 $\psi_2(2^3{\rm D}_2) $ &                        & 4158 & 4208 & -- & 4190 & -- & --\\
 $\psi(2^3{\rm D}_1) $ & \quad  $ 4159 \pm 20  $  & 4142 & 4194 & -- & 4190 [4230]& -- & -- \\
 $\eta_{c2}(2^1{\rm D}_2) $ &                     & 4158 & 4208 & -- & 4190 & -- & -- \\
\hline
 $\chi_4(1^3{\rm F}_4) $ &                        & 4021 & 4095 & -- & -- & -- & -- \\
 $\chi_3(1^3{\rm F}_3) $ &                        & 4029 & 4097 & -- & -- & -- &  $4222\pm 140$\\
 $\chi_2(1^3{\rm F}_2) $ &                        & 4029 & 4092 & -- & -- & -- &  --\\
 $h_{c3}(1^1{\rm F}_3) $ &                        & 4026 & 4094 & -- & -- & -- &  $4224\pm 74$\\
\hline
 $\chi_4(2^3{\rm F}_4) $ &                        & 4348 & 4425 & -- & -- & -- & -- \\
 $\chi_3(2^3{\rm F}_3) $ &                        & 4352 & 4426 & -- & -- & -- & -- \\
 $\chi_2(2^3{\rm F}_2) $ &                        & 4351 & 4422 & -- & -- & -- & -- \\
 $h_{c3}(2^1{\rm F}_3) $ &                        & 4350 & 4424 & -- & -- & -- & -- \\
\hline
 $\psi_5(1^3{\rm G}_5) $ &                        & 4214 & 4312 & -- & --& -- & -- \\
 $\psi_4(1^3{\rm G}_4) $ &                        & 4228 & 4320 & -- & --& -- & -- \\
 $\psi_3(1^3{\rm G}_3) $ &                        & 4237 & 4323 & -- & --& -- & -- \\
 $\eta_{c4}(1^1{\rm G}_4) $ &                     & 4225 & 4317 & -- & --& -- & -- \\
\hline
\hline
\end{tabular}
\label{spectrumTab}
\end{table*}

Overall the agreement with the data is quite impressive. Average errors for the models are listed
in Table \ref{errorTab}. Surprisingly, the simple nonrelativistic BGS model is most accurate in 
describing the spectrum. It is somewhat disappointing that lattice has not caught up to models
yet, but this may change soon.

\begin{table}[h]
\caption{Average Model Errors.}
\begin{tabular}{cccccc}
\hline
BGS & GI & EFG & Cornell & CP-PACS & Chen \\
\hline
\hline
0.3\% & 0.6\% & 0.4\% & 1.4\% & 0.9\% & 1.3\% \\
\hline
\hline
\end{tabular}
\label{errorTab}
\end{table}

Spectra are not particularly precise determinants of model efficacy. Other observables such as
strong and leptonic widths and radiative transition rates provide sensitive probes to dynamics
and it is important to consider the full set of model predictions when making comparisons.

An example of these considerations concerns the strong and leptonic widths of the $\psi(3770)$.
The $\psi(3770)$ is generally assumed to be the $^3$D$_1$ \cc state; however, mixing with the 
$2^3$S$_1$ Fock component  is possible and would modify the predicted strong and leptonic widths.
Allowing this mixing brings a strong decay calculation into agreement with the measured width 
of the $\psi(3770)$ for a mixing angle of approximately $-17^o$. Assuming that leptonic widths
are dominated by S-wave wavefunction components then yields the prediction\cite{BGS}

\be
\frac{\Gamma_{e^+e^-}(\psi(3770))}
{\Gamma_{e^+e^-}(\psi(3686))}\bigg|_{thy.}
= 0.10 \pm 0.03,
\ee
which is in good agreement with the experimental value of $0.12 \pm 0.02$.

\subsubsection{Strong Decay Models and Predictions}
\label{decaySec}

Strong decays of mesons are driven by nonperturbative gluodynamics and are therefore (i)
difficult to compute (ii) good probes of strong QCD.  Unfortunately, lattice computations
of strong decays are very difficult (preliminary attempts are in Ref. \cite{lattDec}) and
one is forced to rely on models. The most popular are the `$^3P_0$' model\cite{3p0}
with a two-body transition operator given by

\be
\hat H = 2 m \gamma \int d^3x \bar \psi \psi|_{\rm nonrel.}
\label{3p0Eq}
\ee
and the Cornell model which takes Eq. \ref{Vcornell} seriously. The Cornell model has the advantage of
unifying the description of the spectrum and decays and completely specifies the  strength
of the decay.

Although, extensive comparisons of models have not been made, Ref. \cite{ABS} finds that
$^3P_0$ amplitudes tend to be much larger than those due to perturbative gluon pair production,
while Ref. \cite{GS}  finds that the $^3P_0$ model gives better predictions for amplitude ratios
than does a model which creates quark pairs with $^3S_1$ quantum numbers. Table \ref{decaysTab} is
an attempt
to partially rectify this situation by presenting the predictions of the $^3P_0$ model, the Cornell
model, and computations of strong widths made in a coupled
channel approach to the Cornell model\cite{cornell,quigg}. The first two computations are made in
first order perturbation theory, while the latter is nonperturbative. I shall refer to perturbative
Cornell model as the $\rho K \rho$ model to highlight this difference.

The $^3P_0$ and $\rho K\rho$ calculations
assume SHO wavefunctions with a universal scale parameter set to $\beta = 0.5$ GeV; meson masses 
are taken as shown. The $^3P_0$ model results are those of Ref. \cite{BGS} with the
exception that the decay strength has been taken to be $\gamma = 0.35$, giving somewhat
better agreement with PDG widths. The $\rho K\rho$ computation sets $K(r) = -3/4 br$ and uses
a string tension of $b = 0.18$ GeV$^2$. The last column displays an attempt to extract widths
from the coupled channel Cornell model (state masses and wavefunctions only approximate those
used in the other models). Table \ref{decaysTab} also lists PDG measured widths and the results
of a recent reanalysis of $R$ by Seth\cite{seth}.

\begin{table*}[h]
\caption{Open-flavor Strong Decays of Selected Charmonia. Widths assigned in this review are denoted ${}^a$.}
\vskip 0.3cm
\begin{tabular}{llccccc}
\hline
Meson & Mode & PDG & Seth & $^3P_0$ & $\rho K\rho$ & Cornell \\
\hline
\hline
$\chi_2'(3972)\ [2^3P_2]$ & $D\bar D$ &  &  & 32 & 8.4 & $\approx 9$ \\
                        & $D D^*$ &     & & 28 & 6.5 & $\approx 5$ \\
                        & $D_s D_s$ &  &  & 0.5 & 0.1 &  --  \\
\hline
$\chi_2'(3930)\ [2^3P_2]$ & $D\bar D$ &  &  & 26 &  &  \\
                        & $D D^*$ &     & & 9 &  &  \\
                        & $D_s D_s$ &  &  & -- &  &   \\
             & total    & $20 \pm 10 {}^a$ &  & 35 &  &  \\
\hline
$\chi_1'(3925)\  [2^3P_1]$ & $D\bar D^*$ & &   & 127 & 16 &  $\approx 130$ \\
\hline
$\chi_0'(3852)\  [2^3P_0]$ & $D\bar D$  & &   &  23 &  4  & $\approx 60$ \\
\hline
$h_c(3934)\  [2^1P_1]$ & $D\bar D^*$ &   & & 67 & 9 & $\approx 65$ \\
\hline
$\psi_3(3806)\  [1^3D_3]$ & $D\bar D$ &  &  & 0.3 & 0.04 & $\approx 0.2$ \\
\hline
$\psi(3770)\  [1^3D_1]$ & $D\bar D$ & $24\pm 3$&  & 33 & 11 & 20 \\
\hline
$\psi(4159)\  [2^3D_1]$ &  $D \bar D$ & & &   12 & &  \\
                      &  $D \bar D^*$ &  & &  0.3  &   &   \\
                      &  $D^*\bar D^*$ &  & &  27  & 2.6 &  \\
                      &  $D_s\bar D_s$ &  & &  6  &  &  \\
                      &  $D_s^*\bar D_s^*$ &  & &  11  &  &  \\
                      &  total             & $78 \pm 20$ & $107\pm 8$ & 57 &   &   \\
\hline
$\eta_{c2}'(4158)\  [2^1D_2]$  & $D\bar D^*$ & &  &  39 &   &   \\
                             & $D^*\bar D^*$ & &   &  33 &   &   \\
                             & $D_s\bar D_s^*$ & &   &  14 &   &   \\
                             &                &  &   & 85 &   &   \\
\hline
$\psi_2'(4158)\  [2^3D_2]$ & $D\bar D^*$ & &  &  24 &  &  \\
                          & $D^*\bar D^*$ & &  & 25  &  &   \\
                          & $D_s\bar D_s^*$ & &  & 20  &  &   \\
                          & total &  & & 69  &  &   \\
\hline
$\psi_3'(4167)\  [2^3D_3]$ & $D \bar D$ & &   & 18 &   &   \\
                         & $D\bar D^*$ & &  & 39 &   &   \\
                         & $D^*\bar D^*$ & &  & 52 &   &   \\
                         & $D_s\bar D_s$ & &  & 4 &   &   \\
                         & $D_s\bar D_s^*$ & &  & 1 &   &   \\
                         & total           & &  & 114 &   &   \\
\hline
$\psi(4040)\  [3^3S_1]$ & $D\bar D$ & &    &  0.1 &  2.3 &   \\
                      & $D\bar D^*$ & &  & 25 &  17 &    \\
                      & $D^*\bar D^*$ & &  & 25 &  5 &    \\
                      & $D_s\bar D_s$ & &  & 6 & 1.6 &    \\
                      & total         & $52 \pm 10$ & $88 \pm 5$  & 57  & 26 &   \\
\hline
$\eta_c''(4043)\  [3^1S_0]$ & $D\bar D^*$ &    & &  36 &  25  &  \\
                          & $D^*\bar D^*$ &   & & 25  & 6  &   \\
                          & total         &   & & 61 & 31 &   \\
\hline
$\psi(4415)\  [4^3S_1]$ & $D^*\bar D^*$ & &  & 12 &  &  \\
                      & $D\bar D_1$ & &    & 24 &  &  \\
                      & $D\bar D_2$ & &    & 18 &  &  \\
                      & $D_s\bar D_s^*$ & &    & 2 &  &  \\
                      & total & $43\pm 15$  & $119 \pm 15$ & 60 &  &  \\
\hline
\hline
\end{tabular}
\label{decaysTab}
\end{table*}

The $\rho K\rho$ and Cornell computations should agree in the event that similar 
wavefunctions and  meson masses were employed and the resulting width is not too
large.  The few cases where it is possible to make a comparison indicate rough agreement
between the computations, although  more extensive calculations should be
made. It appears that the predicted $\rho K\rho$  widths are always smaller (ranging from a
factor of 2 to a factor of 10) than those
of the $^3P_0$ model. Finally comparing $^3P_0$ and $\rho K\rho$ to (the rather sparse)
data indicates that the $^3P_0$ model tends to over-predict decay rates while the $\rho K\rho$
model under-predicts them. It is clear that more theoretical and experimental effort is
required before this situation can be clarified.

\begin{figure}[h]
  \includegraphics[width=7 true cm, angle=270]{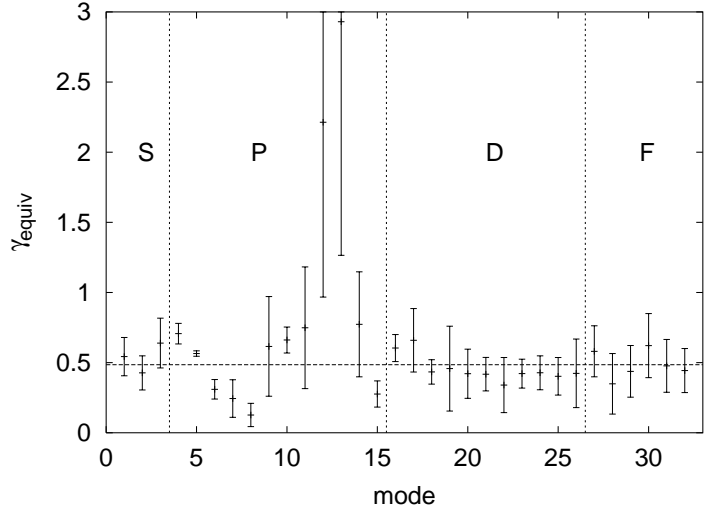}
  \caption{$^3P_0$ Couplings Required to Reproduce Data\protect\cite{CS}.}
  \label{3p0fitFig}
\end{figure}

Finally, we compare the $^3P_0$ model to all well measured rates in the PDG in Fig. \ref{3p0fitFig}
\footnote{The decay modes are as follows.
[1]  $b_1 \to \omega\pi$,
[2]  $\pi_2 \to f_2 \pi$,
[3]  $K_0 \to K\pi$,
[4]  $\rho \to \pi\pi$,
[5]  $\phi \to K\bar K$,
[6]  $\pi_2 \to \rho\pi$,
[7]  $\pi_2 \to K^*\bar K +cc$,
[8]  $\pi_2\to \omega\rho$,
[9]  $\phi(1680) \to K^*\bar K + cc$,
[10]  $K^* \to K\pi$,
[11]  ${K^*}' \to K\pi$,
[12]  ${K^*}' \to \rho K$,
[13]  ${K^*}' \to K^* \pi$,
[14]  $D^{*+} \to D^0 \pi^+$,
[15]  $\psi(3770) \to D\bar D$,
[16]  $f_2 \to \pi\pi$,
[17]  $f_2 \to K\bar K$,
[18]  $a_2 \to \rho \pi$,
[19]  $a_2 \to \eta \pi$,
[20]  $a_2 \to K\bar K$,
[21]  $f_2' \to K\bar K$,
[22]  $D_{s2} \to DK + D^*K + D_s\eta$,
[23]  $K_2\to K \pi$,
[24]  $K_2 \to K^*\pi$,
[25]  $K_2\to \rho K$,
[26]  $K_2\to \omega K$,
[27]  $\rho_3 \to \pi\pi$,
[28]  $\rho_3 \to \omega\pi$,
[29]  $\rho_3 \to K\bar K$,
[30]  $K_3 \to \rho K$,
[31]  $K_3 \to K^* \pi$,
[32]  $K_3 \to K\pi$.  }.
As can be seen, the computation of strong widths is not as reliable as the computation of masses; a typical
width error is 30\% and can reach factors of 2 or even 3.

\subsubsection{Critique}
\label{spectrumCritiqueSec}

{\bf Constituent Quark Models}

Most complaints about constituent quark models centre on one theme: quark models are
not QCD. But models do not seek to be QCD, rather they attempt to capture the dominant physics
relevant to the problem at hand; in this case, the structure and dynamics of hadrons.
In this regard, most quark models suffer from serious deficiencies which will be
become apparent as states higher in the spectrum are discovered. Namely:
(i) Fock sector mixing (or coupled channel effects) will become more important
as more continuum thresholds open up. Of course virtual meson loops also impact the
low lying spectrum, but these effects are more easily absorbed into model
parameters\cite{ess2}. (ii) Although gluons may be integrated out to good accuracy low
in the spectrum, they must manifest themselves as dynamical degrees of freedom somewhere
(roughly 1300 MeV above the lowest S-wave states in a  given flavour sector). (iii)
The use of chirally non-invariant effective interactions (for example, scalar confinement
and the $^3P_0$ model) cannot be accurate for systems with typical momentum scales
well above the chiral symmetry breaking and quark mass scales\cite{chiralRes}.

{\bf Short Range Interactions}

Although one may hope that short range quark interactions are described by 
one gluon exchange, this is far from obvious in the intermediate regime
relevant to hadrodynamics. In particular, spin-dependence in the quark interaction
may arise due to one gluon exchange, coupled channel effects, relativistic dynamics, and
instantons.  The last three remain serious challenges for model builders.

A good example of the difficulties which remain to be surmounted is the use
of the scalar confinement interaction discussed in Section \ref{spindepSec}.
Among its many problems are that it breaks chiral symmetry (and therefore
is only useful as an effective interaction low in the spectrum where explicit chiral
symmetry breaking is relevant)
and that it
cannot be used to describe baryons (quark-quark scalar interactions are of opposite sign to 
quark-antiquark scalar interactions, implying that baryons are dominated by an unphysical negative
long range interaction). 
Indeed the entire
analysis of Section \ref{spindepSec} was much too naive: QCD does {\it not} generate
quark interactions of the structure $\int J K J$. In fact, spin-dependent 
interactions arise from nonperturbative gluodynamics
involving virtual hybrid states. This can lead to complex behaviour -- including 
permitting a vector confinement potential to have an effective scalar spin-dependent
interaction. How this can be achieved is illustrated in Ref. \cite{ss2}.

{\bf Lattice}

Although lattice gauge theory is `in principle' equivalent to QCD, in practice
it has far to go.  A simple way to see this is by 
counting parameters -- `in principle' there should only be one parameter necessary
for all lattice computations, $\alpha_s$, or equivalently, $\Lambda_{QCD}$. In
practice there are many hidden parameters, including the lattice spacing, the
lattice volume, the structure of the interpolating field (which typically must
be carefully tuned to obtain a signal), Monte Carlo parameters (number of sweeps,
Metropolis parameters, etc), autocorrelation parameters, chiral tuning parameters, 
plateau parameters, etc. Of course the benefit of the lattice is that all of these
dependences may be categorised as systematic errors and eventually reduced; something
which cannot be said of models.

However, more serious concerns remain. The computations reported in Table \ref{spectrumTab}
employed the quenched approximation, which unfortunately is not an approximation, but a
truncation which renders QCD a sick theory. Furthermore, disconnected diagrams
are rarely included in spectrum computations. Although we look forward to the day when these
issues are overcome a warning is appropriate: unquenched lattice computations permit
hadron decay, both virtual and real. Accommodating these in a rigorous way can
only be achieved by computing observables such as scattering matrix elements. This
represents a fearsome technical challenge. It also
implies that full lattice computations will require all of the techniques of
(and be subject to all of the ambiguities associated with) experimental partial wave
analysis. Finally, it is worth stating that quenched lattice computations tend to succeed where
constituent quark models succeed and fail where they fail. The convergence is not
surprising: quark models fail when states have a complicated Fock structure and the lattice
has difficulty with such states because they represent multiple scale unquenched systems.

{\bf Decay Models}

The importance of continuum channels on the spectrum and hadronic properties has
already been mentioned. Of course attempts to describe such physics relies on 
a thorough understanding of the transition operator which mixes Fock sectors.
As discussed above, our understanding of these operators is essentially nonexistent.
This leads to a variety of issues when attempting to `unquench' the quark model.

First, as opposed to quark model lore, continuum thresholds affect states throughout
the spectrum. However, these effects may be largely renormalised away leaving residual
effects near thresholds.
Of course, the problem in hadronic physics is that continuum
channels tend to get dense above threshold.
Summing over these channels is nontrivial\footnote{Note that the Cornell group only sum over six continuum channels, $D\bar D$, $D\bar D^*$, $D^*\bar D^*$, $D_s\bar D_s$, $D_s\bar D_s^*$, 
$D_s^*\bar D_s^*$; the
sensitivity of their results to this truncation should  be tested.} -- indeed the 
sum may not converge.

Furthermore, one expects that when the continuum virtuality is much
greater than $\Lambda_{QCD}$ quark-hadron duality will be applicable and the sum
over hadronic channels should evolve into perturbative quark loop corrections to
the Wilson loop or quark model potential. Correctly incorporating this into constituent
quark models requires marrying QCD renormalisation with effective models
and is not a simple task.

Finally, pion and multipion loops can be expected to dominate the virtual continuum
component of hadronic states (where allowed) due to the light pion mass. This raises
the issue of correctly incorporating chiral dynamics into unquenched quark models.
The relationship of chiral symmetry breaking to the constituent quark model has been
discussed in Ref. \cite{ss6} and a variety of hadronic models which incorporate
chiral symmetry breaking exist\cite{chiralModels} but much remains to be achieved.

\vfill\eject
\section{The X(3872)}      

The $X(3872)$ is the poster boy of the new heavy hadrons -- it has been observed by four
experiments in three decay and two production channels and continues to refuse to 
fit into our expectations for charmonium.   Experimental properties of the $X$ will be reviewed
in the next subsection. Subsequent subsections will present model and effective field theory 
attempts at understanding the $X$. Finally, other heavy quark molecular states are considered.

\subsection{Experiment}
\label{XexptSect}

A summary of the discovery modes of the $X(3872)$ is
presented in Table \ref{XmodesTab}.

\begin{table}[h]
\caption{$X(3872)$ Discovery Modes.}
\label{XmodesTab}
\begin{tabular}{cclccl}
\hline
mass & width & production/decay mode & events & significance & experiment \\
\hline
\hline
$3872.0 \pm 0.6 \pm 0.5$  & $< 2.3$ 90\% C.L.  & $B^\pm \to K^\pm X \to K^\pm \pi^+ \pi^- J/\psi$   &  $25.6 \pm 6.8$ & $10 \sigma$     & Belle\cite{belleX}\\
$3871.3 \pm 0.7 \pm 0.4$  & resolution & $p\bar p \to  X \to \pi^+ \pi^- J/\psi$   &  $730 \pm 90$ & $11.6 \sigma$  & CDFII\cite{cdfX}\\
$M(J/\psi) + 774.9 \pm 3.1 \pm 3.0$ & resolution & $p\bar p \to X \to \pi^+\pi^-J/\psi$ & $522 \pm 100$ & $5.2 \sigma$  & D{\O}\cite{D0X} \\
$3873.4 \pm 1.4$  &  --  & $B^- \to K^- X \to K^- \pi^+ \pi^- J/\psi$   &  $25.4 \pm 8.7$ &$3.5 \sigma$ & BaBar\cite{BabarX}\\
\hline
\hline
\end{tabular}
\end{table}

The world average $X$ mass is 

\be
M(X) =  3871.9 \pm 0.5\ {\rm MeV}
\label{Xmass}
\ee
assuming independent systematic errors.

\begin{figure}[h]
  \includegraphics[width=7 true cm, angle=0]{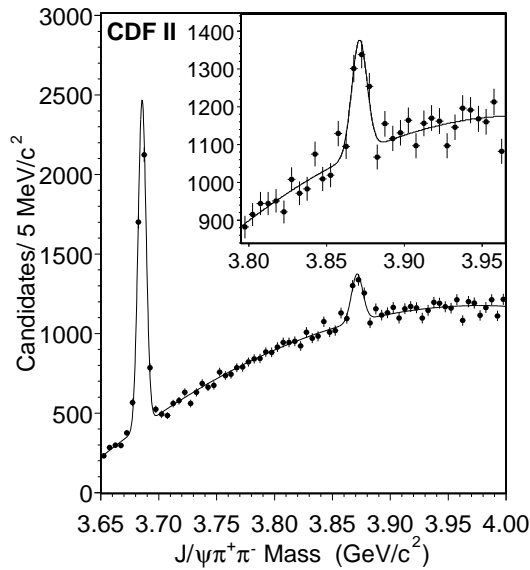}
  \caption{The $X$ in $\pi\pi J/\psi$\cite{cdfX}.}
  \label{cdfXFig}
\end{figure}

The clarity of the $X$ signal is exemplified in the $\pi\pi J/\psi$ invariant mass distribution as seen by CDF-II and  shown in Fig. \ref{cdfXFig}.

\begin{figure}[h]
  \includegraphics[width=7 true cm, angle=0]{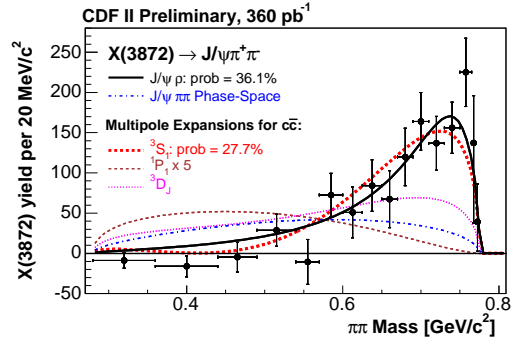}
  \caption{$\pi\pi$ Invariant Mass Distribution\cite{cdfNote}.}
  \label{cdfpipiFig}
\end{figure}

As will be discussed below, the invariant mass $\pi\pi$ distribution carries important information
about $X$ decays.  Thus it is of interest that
CDF-II have collected enough statistics to verify that the $\pi\pi$ invariant
mass distribution in $X$ decays is dominated by an intermediate $\rho$ meson. This is illustrated in Fig. \ref{cdfpipiFig}.

\begin{figure}[h]
\includegraphics[width=7 true cm, angle=0]{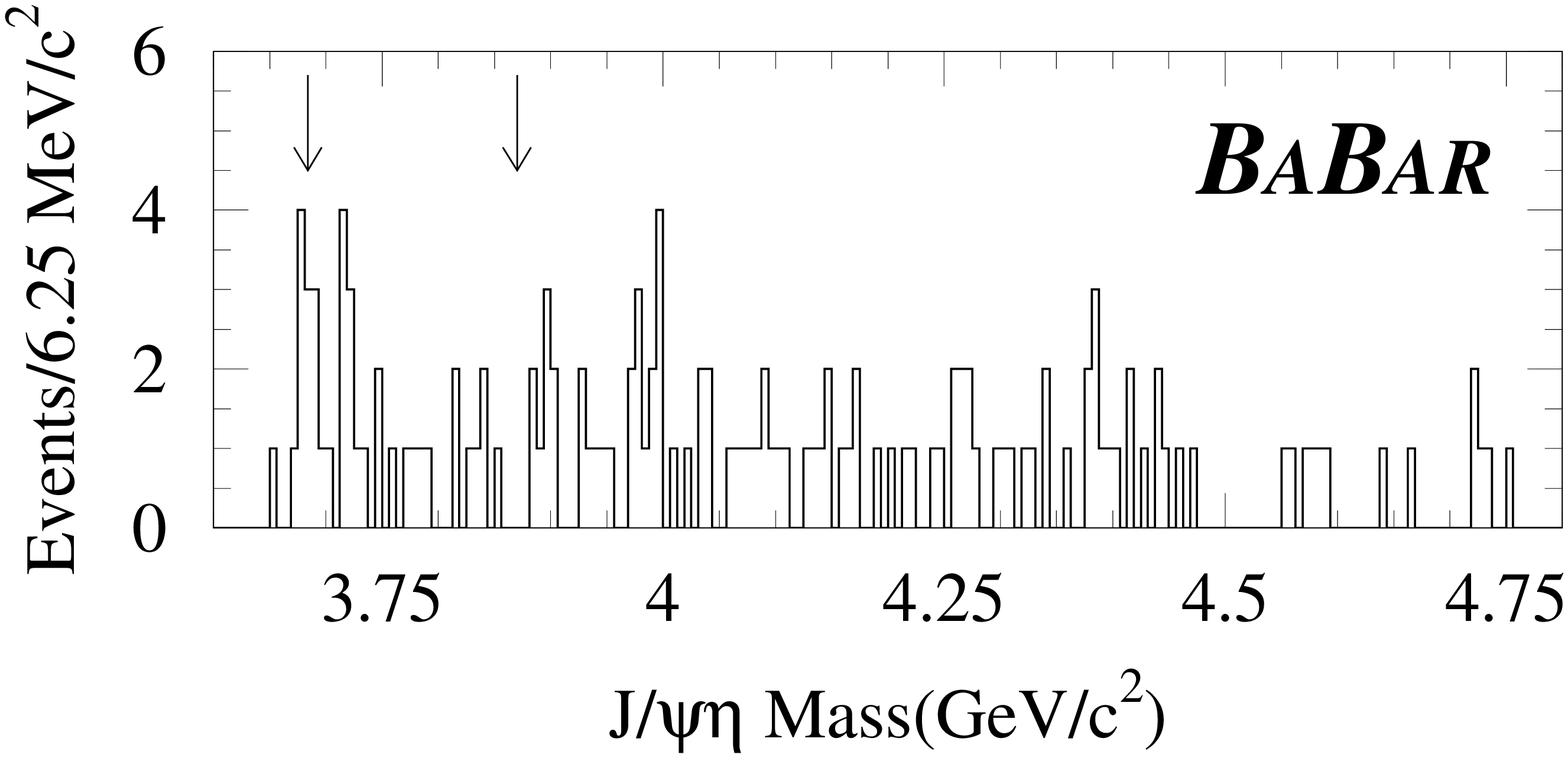}
\caption{Search for $X \to \eta J/\psi$\cite{Aubert:2004fc}.} 
\label{etapsiFig}
\end{figure}

BaBar\cite{BabarX} have 
set an upper limit on the $\eta J/\psi$ decay mode of the $X$\cite{Aubert:2004fc}:

\be
Br(B^\pm \to X K^\pm)\, Br(X \to \eta J/\psi) < 7.7 \cdot 10^{-6}\  90\% {\rm C.L.}
\ee
The relevant data are shown in Fig. \ref{etapsiFig}.

If the $X$ is an isovector state charged partners should exist and one expects that the branching
ratio for the charged $X$ should be twice that of the neutral $X$. BaBar\cite{BabarCh}
have searched for these and find the limits

\be
Br(B^0 \to K^+ X^-)\, Br(X^- \to \pi^-\pi^0 J/\psi) < 5.4 \cdot 10^{-6}
\ee
and

\be
Br(B^- \to \bar K^0 X^-)\, Br(X^- \to \pi^-\pi^0 J/\psi) < 22 \cdot 10^{-6}.
\ee

Spurred on by tetraquark models of the $X$ (see Section \ref{XmodelSec}), BaBar\cite{bernard}
have examined the mass difference of the $X$s produced in charged and neutral $B$
decays and determine this to be 

\be
\Delta M(X) = 2.7 \pm 1.3 \pm 0.2\ {\rm  MeV}.
\label{deltaX}
\ee
They have also determined that the ratio of neutral to charged production of the $X$ is
constrained to lie between 15\% and 134\%:

\be
0.15 < {Br(B^0\to K^0\pi\pi J/\psi)\over Br(B^+\to K^+\pi\pi J/\psi)} < 1.34
\ee
at 90\% C.L. This result has recently been improved by BaBar\cite{bernard}, who measure\footnote{This result
has been updated from the previous $0.61 \pm 0.36 \pm 0.06$\cite{bernard2}.}

\begin{equation}
{Br(B^0 \to X K^0)\over Br(B^+ \to X K^+)} = 0.50  \pm 0.30 \pm 0.05.
\label{b0b+}
\end{equation}

\begin{figure}[h]
  \includegraphics[width=7 true cm, angle=0]{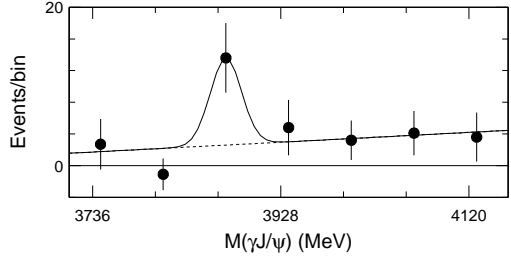}
  \caption{Observation of $X \to \gamma J/\psi$\cite{Abe:2005ix}.}
  \label{gammapsiFig}
\end{figure}

Motivated by the predictions of Ref. \cite{essX2}, Belle have searched for and found the $X$ 
in the channel $\gamma J/\psi$\cite{Abe:2005ix} thus establishing that the $X$ has
positive charge conjugation parity.
The signal is shown in Fig. \ref{gammapsiFig} and has a significance of 4 $\sigma$.

Belle also have measured the product of branching fractions

\be
Br(B \to KX)\, Br(X \to \gamma J/\psi) = (1.8 \pm 0.6 \pm 0.1) \cdot 10^{-6}
\ee
and the ratio of widths

\be 
{\Gamma(X \to \gamma J/\psi) \over  \Gamma(X \to \pi^+\pi^- J/\psi)} = 0.14 \pm 0.05.
\label{gammapsiEq}
\ee

\begin{figure}[h]
  \includegraphics[width=8 true cm, angle=0]{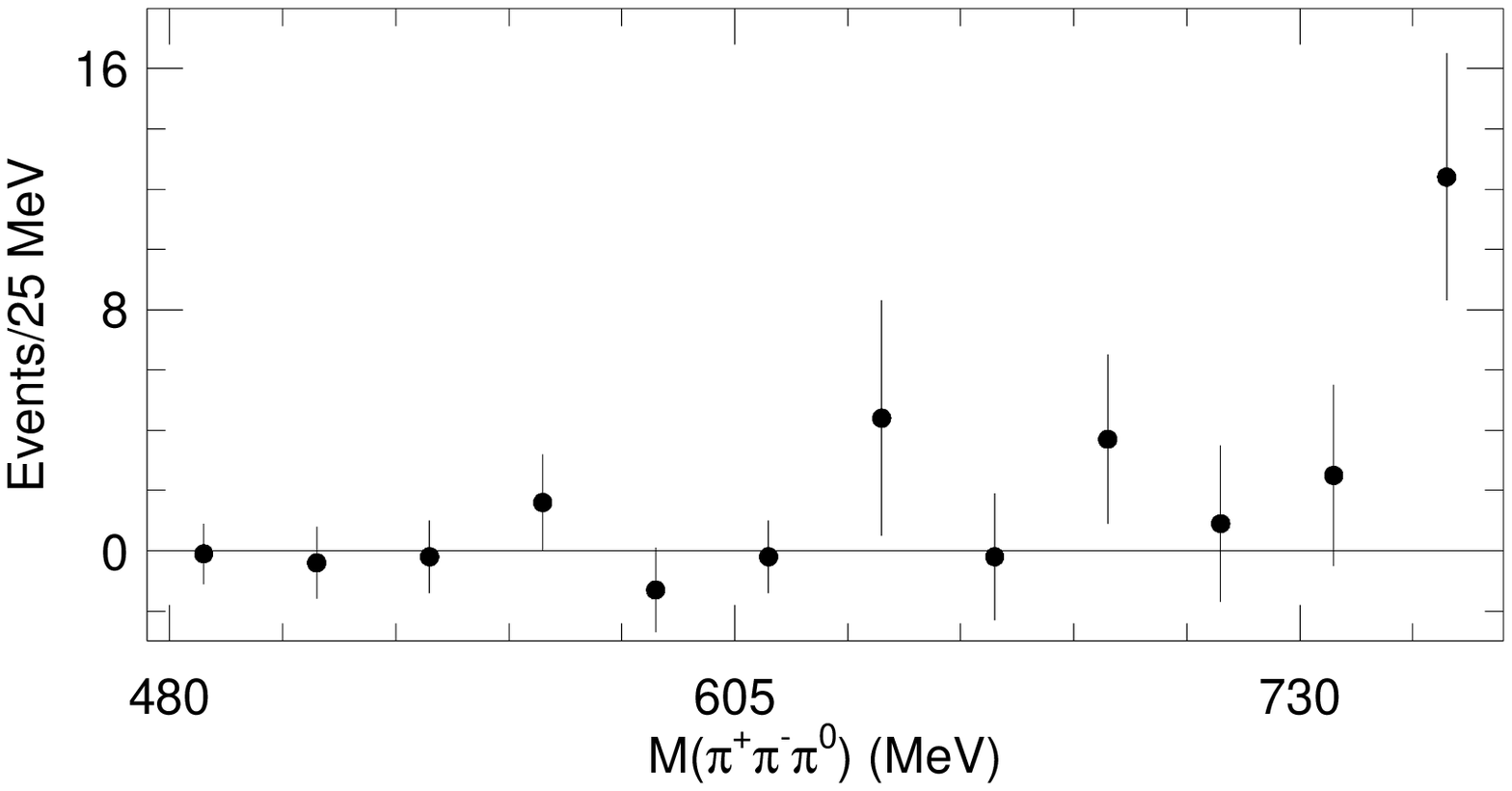}
  \caption{Observation of $X \to \pi\pi\pi J/\psi$\cite{Abe:2005ix}.}
   \label{X3piFig}
\end{figure}

The same paper reports the discovery of the $X$ decaying to $\pi^+\pi^- \pi^0 J/\psi$ with
a significance of 4 $\sigma$ (see Fig. \ref{X3piFig}) with a strength that is comparable to 
that of the $\pi\pi J/\psi$ mode:

\be
{Br(X \to \pi^+\pi^-\pi^0 J/\psi)\over Br(X \to \pi^+\pi^- J/\psi)} = 1.0 \pm 0.4 \pm 0.3.
\label{3pi-2pi}
\ee
This observation establishes strong isospin violation effects, which will be discussed further
in Section \ref{XmodelSec}.

The products of branching ratios have been measured by 
Belle\cite{Abe:2005iy} and BaBar\cite{bernard,bernard2}:

\begin{equation}
Br(B^+ \to X K^+) Br(X \to \pi\pi J/\psi) = (1.3 \pm 0.3) \cdot 10^{-5}
\label{BrBr}
\end{equation}
and

\begin{equation}
Br(B^+ \to X K^+) Br(X \to \pi\pi J/\psi) = (1.01 \pm 0.25 \pm 0.10) \cdot 10^{-5}.
\label{BrBr2}
\end{equation}

Finally, there are rumours that Belle observes the $X$ decaying to $D^0 \bar D^0 \pi^0$
at a  rate ten times larger than that of $\pi\pi J/\psi$\cite{belleConf}. The new decay mode 
is seen in $11.3 \pm 3.6$ events at 5.6 $\sigma$ with a product of branching fractions of
\be
Br(B \to KX)\, Br(X\to D\bar D\pi) = (2.2 \pm 0.7 \pm 0.4)\cdot 10^{-4}.
\label{DDpi}
\ee

Determining the quantum numbers of the $X$ is a high priority experimental task. 
BES\cite{bes} have searched for the $X$ in $e^+e^-$ collisions and find the bound

\be
\Gamma(e^+e^-)\, Br(X \to \pi\pi J/\psi) < 10\ {\rm eV}
\ee
at 90\% C.L. Thus the $X$ is unlikely to be a vector charmonium state. This conclusion is
confirmed by  Babar\cite{babarISR} who search for the $X$ in initial state radiation and find
an upper limit to the product of branching ratios:

\be
\Gamma(X\to e^+e^-) Br(X \to \pi\pi J/\psi) < 6.2 \ {\rm eV} 
\ee
at 90\% C.L.

Belle have measured angular correlations in the $\pi\pi J/\psi$ system and use
these to rule out possible $J^{PC} = 0^{++}$ or $0^{-+}$  values for the $X$. Furthermore
the dipion mass distribution strongly disfavours  $1^{-+}$ or $2^{-+}$ assignments
for the $X$\cite{Abe:2005ix}. Thus, it is likely that the $X$ is a $1^{++}$ charmonium 
state\cite{Abe:2005iy}.

\begin{figure}[h]
  \includegraphics[width=8 true cm, angle=0]{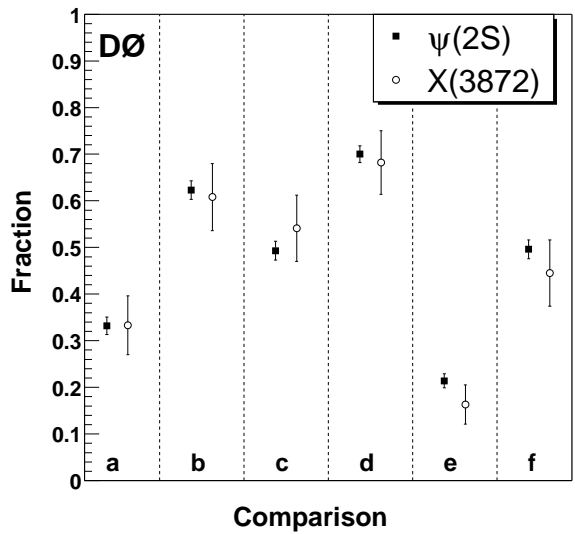}
  \caption{Comparison of event yield fractions for $X(3872)$ and $\psi(2S)$. (a) $p_T(J/\psi) > 15$ GeV, (b) $|y|(J/\psi) < 1$, (c) $\cos\theta_\pi < 0.4$, (d) effective proper decay length $< 0.01$ cm, (e) isolation-1, (f) $\cos\theta_\mu < 0.4$\cite{D0X}.}
   \label{D0FracsFig}
\end{figure}

CDF\cite{Bauer:2004bc} and D{\O}\cite{D0X} have examined the production characteristics of the $X$.
In particular, it is found that approximately 16\% of the $X$ sample at CDF comes from $B$
decay; the remainder comes from direct production or decay of short-lived particles.
The mix of production sources is similar to that observed for the $\psi(2S)$ charmonium 
state\cite{Bauer:2004bc} (see Fig. \ref{D0FracsFig}). Thus it appears that there is no 
penalty for producing the $X$ in
$p\bar p$ collisions.

In summary, a narrow state at 3871.9 MeV is seen in $B$ decays and $p\bar p$ collisions
decaying to $\pi\pi J/\psi$, $\pi\pi\pi J/\psi$, $\gamma J/\psi$, and $D^0\bar D^0 \pi^0$
which has production characteristics similar to the $\psi(2S)$.
The quantum numbers $J^{PC} = 1^{++}$ are strongly preferred by the data. Various attempts
to interpret this data are presented in the following sections.

\subsection{Models}
\label{XmodelSec}

\subsubsection{Charmonium}

The first place to seek an explanation of the $X(3872)$ is evidently in the charmonium
spectrum\cite{BG}. A glance at Table \ref{spectrumTab} reveals that the only viable candidates for
the $X$ are $2P$ or $1D$ states, with the $\chi_{c0}'$ being the most likely candidate based solely
on mass. Table \ref{decaysTab} indicates that $^3P_0$ widths are too large to accommodate a narrow
$2P$ or $1D$ charmonium (however, the  $\rho K\rho$ model yields smaller widths). Subsequent
experimental effort (described above) has focussed attention on $1^{++}$ charmonia, of which the
$\chi_{c1}'$ is the only possible candidate. But the predicted width of this state is  127 MeV
in the $^3P_0$ model and 16 MeV in the $\rho K \rho$ model -- much too large. The apparent
lack of viable charmonium candidates has led to much theoretical speculation.

Three options are available when the quark model fails us: (i) attempt to change the quark model;
(ii) seek supernumerary states; (iii) claim experimental or interpretational error. 
The first option is extremely difficult, as the average mass 
errors reported in Table \ref{errorTab} illustrate. In general one must find an effect which
is localised in the spectrum so as not to destroy the good agreement seen elsewhere. 

In the list of usual supernumerary suspects are hybrids, glueballs, diquark clusters, and molecular states.
All of these options have been suggested for the $X$. The prominent ones are discussed in the following.

\subsubsection{Tetraquarks}

\begin{figure}[h]
  \includegraphics[width=7 true cm, angle=0]{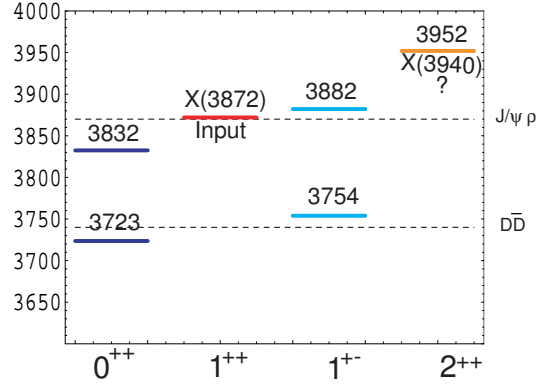}
  \caption{Spectrum of Tetraquark Charmonium States\protect\cite{maianiA}.}
  \label{maianiFig}
\end{figure}

Maiani {\it et al.}\cite{maianiA} have constructed a model of the $X(3872)$ (and several
other states as we shall see) which postulates that it is dominated by a diquark-diquark
structure. The results of a relativistic tetraquark model are reported in Ref. \cite{Ebert:2005nc}.
Chief assumptions in the model are that quarks group in colour triplet scalar
and vector clusters and that the interactions of these are dominated by a simple spin-spin
interaction\cite{jaffe}. Using the $X(3872)$ as input, Maiani {\it et al.} predict the spectrum of
new diquark charmonia shown in Fig. \ref{maianiFig}.

This spectrum assumes that the mass of the scalar and vector $[cq]$ diquarks are 1933 MeV
and constructs states as follows:

\begin{eqnarray}
|0^{++}\rangle &=& | [cq]_S [\bar c \bar q]_S; J=0\rangle \\
|{0^{++}}'\rangle &=& | [cq]_V [\bar c \bar q]_V; J=0\rangle \\
|1^{++}\rangle &=& {1\over \sqrt{2}}\left( | [cq]_S [\bar c \bar q]_V; J=1\rangle +
                                           | [cq]_V [\bar c \bar q]_S; J=1\rangle \right) \\
|1^{+-}\rangle &=& {1\over \sqrt{2}}\left( | [cq]_S [\bar c \bar q]_V; J=1\rangle -
                                           | [cq]_V [\bar c \bar q]_S; J=1\rangle \right) \\
|{1^{+-}}'\rangle &=& | [cq]_V [\bar c \bar q]_V;  J=1\rangle  \\
|2^{++}\rangle &=& | [cq]_V [\bar c \bar q]_V; J=2\rangle 
\label{tetraStates}
\end{eqnarray}

\noindent
Thus the model predicts that charged partners of the $X(3872)$ should exist, namely
$X^+ = [cu][\bar c \bar d]$ and $X^- = [cd][\bar c \bar u]$. Furthermore, the two neutral
$X$ states comprised of $[cu][\bar c \bar u]$ and $[cd][\bar c \bar d]$ should mix with angle
$\theta$ and hence produce two neutral $X$s with a mass difference of 

\be
M(X_h) - M(X_l) = {2(m_d - m_u)\over\cos(2\theta)} = {7 \pm 2 \over \cos(2\theta)} \, {\rm MeV} \approx (8 \pm 3) \, {\rm MeV}
\label{deltaX2}
\ee
(similar considerations apply to all the states in Fig. \ref{maianiFig}).
Furthermore, one of these states will populate $B^+$ decay while the other will populate $B^0$ decay.


Relative strengths of the production of tetraquarks may be simply estimated. However, I disagree
with the arguments of Ref. \cite{maianiA} where it is claimed that the amplitude for a strange 
quark to combine
with the spectator quark is comparable or less than  the amplitude for it to combine with a produced
quark. Rather this amplitude is suppressed by a factor of $1/N_c$ due to its colour structure. 
Thus one concludes 
that $B^0 \to K^- X^+$, $B^0 \to K^0 X_d$ ,
$B^+ \to K^+X_u$, and $B^+ \to K^0 X^+$ should all occur at the same rate.

The model makes strong predictions which may be compared with experiment. For example,
the ratio of neutral to charged production of the $X$ is predicted to be unity whereas 
BaBar measure (Eq. \ref{b0b+}) a ratio of 50\% with large errors. Unfortunately, many of the 
other predictions of the model are not as successful.  The predicted mass difference of the
neutral $X$s (Eq. \ref{deltaX2}) is 2 $\sigma$ removed from the measured mass 
difference (Eq. \ref{deltaX}) and no 
charged $X$ states, or any of the other states of Fig. \ref{maianiFig}, have been sighted.
Finally, the existence of diquark correlations in hadrons implies an extensive new exotic
and cryptoexotic spectroscopy for which very little evidence exists\footnote{The
fiascos associated with pentaquarks and baryonia will only be mentioned in this footnote.}. 

While the ideas surrounding diquarks and tetraquarks are interesting and even 
compelling\cite{Karliner:2006hf},
one must reserve judgement on their reality until strong evidence for them comes to light.

\subsubsection{Additional Models}

Li has suggested that the $X$ is a $c\bar c g$ hybrid meson\cite{li} which decays predominantly
via 
$X \to J/\psi gg \to J/\psi \pi\pi$
and has a significant $J/\psi \sigma$ branching fraction. The existence of the opposite 
$G$-parity $3\pi$ decay mode is explained as due to isospin violation. Its large magnitude 
(see Eq. \ref{3pi-2pi}) is left as a conundrum. 

Current lattice\cite{ccgLatt} and flux tube model\cite{ccgFTM} expectations for the mass of $c\bar c g$ hybrids are in
the range 4200-4400 MeV, thus there must exist significant systematic error in these models to
permit a charmonium hybrid at 3872 MeV. In the absence of countervailing data, one must therefore
regard hybrid models of the $X$ as unlikely.

Seth\cite{seth2} has proposed that the $X$ is a vector glueball with a small admixture of vector $c\bar c$.
This proposal has since been eliminated by the observation of the $\omega J/\psi$ (Fig. \ref{X3piFig}) and
$\gamma J/\psi$ (Fig. \ref{gammapsiFig}) decay modes.

On the basis of a simple chromomagnetic interaction model, 
H{\o}gaasen {\it et al.}\cite{Hogaasen:2005jv} postulate that the $X$ is a loose four quark agglomeration
which is a mixture of $cu\bar c \bar u$ and $c d \bar c \bar d$ flavour states. Thus another neutral $X$ is expected
along with two charged states $c u \bar c \bar d$ and $c d \bar c \bar u$. In addition, a broad  isospin
multiplet is predicted at 3742 MeV.

Kalashnikova has constructed a simple coupled channel model of the $c\bar c$ spectrum
which incorporates a central confining potential and $^3P_0$ model couplings to $D\bar D$,
$D\bar D^*$, $D^*\bar D^*$, $D_s\bar D_s$, $D_s\bar D_s^*$, and $D_s^*\bar D_s^*$ 
continua\cite{Kal}. Couplings are computed assuming SHO wavefunctions for the mesons
and a simple renormalisation procedure is carried out. An interesting result of this
computation is that the renormalised $\chi_{c1}'$ sits at roughly 4000 MeV, while a
virtual bound state appears just above $D\bar D^*$ threshold. It is natural to 
identify this with the $X(3872)$, and in fact, such an identification will be 
indistinguishable from a very weakly bound $D\bar D^*$ molecule (to be discussed in the
next section). Although this approach suffers from all of the problems mentioned
in Section \ref{spectrumCritiqueSec}, and in some cases can ruin agreement with
naive quark model results (such as vector meson leptonic widths), 
this interesting observation warrants further investigation. 

Finally, Bugg\cite{bugg} has reminded us that cusps occur in amplitudes at thresholds,
and these can manifest themselves as bumps in cross sections slightly above threshold.
The proximity of the $X$ to $D^0\bar D^{*0}$ threshold implies that the cusp scenario
should be treated seriously. In this regard, we note that typical enhancements due to cusps
are ${\cal O}(\Lambda_{QDC})$ in width, in contrast with the very narrow $X$. Furthermore, 
if the 
$X$ is simply due to the opening of a threshold, then the reaction
$B^+ \to K^0 D^+ \bar D^{*0}$
should exhibit a peak similar to the $X$ channel $B^+ \to K^+ D^0 \bar D^{*0}$. That it does not 
indicates that significant
final state interactions are affecting $B^+ \to K^+ D^0 \bar D^{*0}$. And of course 
these interactions
may be strong enough to create a hadronic resonance. Additional observations on distinguishing
bound states from cusps can be found in Ref. \cite{BK4}.

\subsubsection{Molecular Interpretation of the $X$}
\label{XMoleculeSect}

In principle nothing in QCD prevents the formation of nuclear-like bound states of mesons and
speculation on the existence of such states dates back thirty years\cite{oldMol,NAT}. The proximity
of the $X$ to $D^0\bar D^{0*}$ threshold immediately led to speculation that the $X$ is 
a $D\bar D^*$ resonance\cite{XmolT,XmolPC,CYW,PS,essX}. Indeed $D^0\bar D^{0*}$ threshold is
at 

\be
D^0\bar D^{0*} = 3871.2 \pm 1.0\ {\rm MeV}
\ee
while the world average $X$ mass is (Eq. \ref{Xmass}) $3872.0 \pm 1.8$  MeV, $0.7 \pm 1.2$ MeV
higher than threshold. The positive binding energy (with large error) is disconcerting for 
molecular interpretations, but, is no problem for tetraquark, cusp, or virtual 
state interpretations. In any event, the binding energy for a putative molecule is 
very small
and this forms the basis for an effective field theory approach to the $X$, which will 
be discussed
in the following section.

In general, if the $X$ is dominated by a loose $D^0\bar D^{0*}$ component then it
does not have good isospin which should give rise to a distinct decay pattern. The most
likely bound state is an S-wave which implies that the quantum numbers of the $X$ are
$J^{PC} = 1^{++}$. Mixing with nearby $\chi_{c1}$ states should induce weak
isospin violating decays in them as well. Finally, since the mesons are weakly bound they
decay as if they were free, thus the $X$ should decay via the modes 
$X \to D^0\bar D^0 \pi^0$ and $X \to D^o \bar D^0 \gamma$ with a relative rate of 62:38.

Making more detailed predictions forces one to employ a specific model because lattice
gauge theory is not sufficiently well developed to permit accurate examination of this
state 
(although a recent attempt has appeared\cite{lattX}; see Section \ref{spectrumCritiqueSec} for additional comments) and too many unknown 
couplings are present in effective field theory descriptions of the $X$.

The first step in constructing a molecular (or any) model of the $X$ is to determine the
relevant degrees of freedom. The $D^0\bar D^{*0}$ channel has already been identified as 
important. This in turn implies that the binding energy is very small which implies that
$c \bar c$ components in the $X$ are small (this is quantified below). Similarly excluding
hybrid components leaves meson-meson channels. Of these the four closest are listed in Table 
\ref{XchansTab}.

\begin{table}[h]
\caption{Continuum Channels near 3872 MeV.}
\begin{tabular}{llcc}
\hline
channel & threshold & $\Gamma(A)$ & $\Gamma(B)$ \\
\hline
\hline
$D^0 \bar D^{*0}$ & 3871.2  & --  & $< 2.1$ MeV \\
$D^+ D^{*-}$ &  3879.3  &  -- & $96 \pm 22$ keV \\
$\rho^0 J/\psi$ & 3867.9  & $150.3 \pm 1.6$ MeV & $91.0 \pm 3.2$ keV \\
$\omega J/\psi$ & 3879.5 & $8.49 \pm 0.08$ MeV & " \\
\hline
\hline
\end{tabular}
\label{XchansTab}
\end{table}

The next element in a microscopic description is dynamics. Two candidates come to mind:
pion exchange and quark-level interactions. Indeed, if quark dynamics are limited to 
quark/colour exchange interactions then it is natural to admit both interactions since pion
exchange is long range while quark exchange is a short range effect.

The quark model employed here assumes nonrelativistic quark dynamics mediated
by an instantaneous confining interaction and a short range spin-dependent interaction
motivated by one gluon exchange. The colour structure is taken to be the
quadratic form of  perturbation theory. This is an important assumption for
multiquark dynamics which has received support from recent lattice
computations for both confinement\cite{Bali} and multiquark interactions\cite{BB}.
The final form of the interaction is thus taken to be

\begin{equation}
\label{Vij}
\sum_{i<j}{\bm{\lambda}(i) \over 2}\cdot {\bm{\lambda}(j) \over 2} \left \{
{\alpha_s \over r_{ij}} - {3\over 4} br_{ij}
- {8 \pi \alpha_s \over
3 m_i m_j } \bm{S}_i \cdot \bm{S} _j \left ( {\sigma^3 \over
\pi^{3/2} } \right ) e^{-\sigma^2 r_{ij}^2}
\right \},
\end{equation}
where ${\bm{\lambda}}$ is a colour Gell-Mann matrix, $\alpha_s$ is
the strong coupling constant, $b$ is the string tension, $m_i$ and
$m_j$ are the interacting quark or antiquark masses, and $\sigma$ is a
range parameter in a regulated  spin-spin hyperfine
interaction.  The parameters used were $\alpha_s = 0.59$, $b = 0.162$ GeV$^2$,
$\sigma = 0.9$ GeV, and $0.335$, $0.55$, and $1.6$ GeV for up, strange, and
charm quark masses respectively. Relevant meson masses obtained from this model are
$\rho = 0.773$ GeV,
$J/\psi = 3.076$ GeV, $D = 1.869$ GeV, and $D^* = 2.018$ GeV, in good agreement
with experiment.

\begin{figure}[h]
 \includegraphics[width=7 true cm, angle=0]{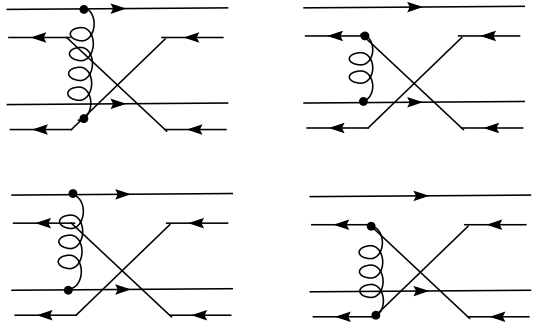}
 \caption{Quark Exchange Diagrams Contributing to Meson-Meson Scattering.}
 \label{qexFig}
 \end{figure}

Meson-meson interactions are obtained by computing the Born order scattering amplitude shown in
Fig. \ref{qexFig}
for a given process\cite{BS}. Because of the colour factors in Eq. \ref{Vij} this amplitude
necessarily involves an exchange of quarks between the interacting mesons. Thus the
leading order $D\bar D^*$ interaction  couples $D\bar D^*$ with hidden charm states
such as $\rho J/\psi$ and $\omega J/\psi$. This amplitude may be unitarised by
extracting an effective potential and iterating it in a Schr\"odinger equation\cite{BS}.
The method has been successfully applied to a variety of processes such
as $KN$ scattering\cite{KN}
and  $J/\psi$ reactions relevant to RHIC physics\cite{WSB}. It has even
proven surprisingly useful for relativistic (and chiral) reactions such as
$\pi\pi$ scattering\cite{BS}.

\begin{figure}[h]
  \includegraphics[width=6 true cm, angle=270]{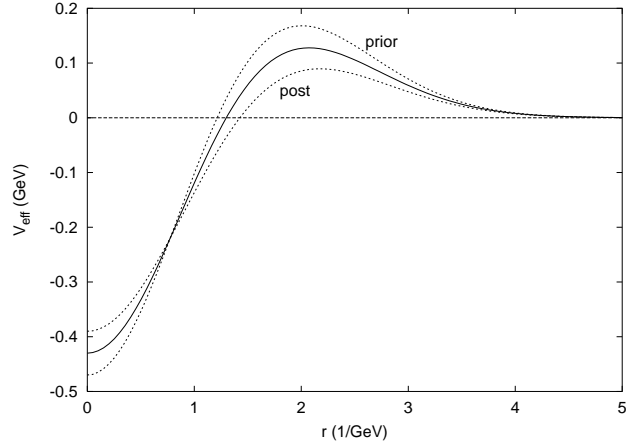}
  \caption{Effective Potential for $D\bar D^* \to \omega J/\psi$. The dashed lines indicate
different partitions of the four-quark Hamiltonian.}
  \label{veffFig}
\end{figure}

An effective
potential is extracted by equating the scattering amplitude to that obtained for
point-like mesons interacting via an arbitrary S-wave potential. 
The resulting potential
is shown in Fig. \ref{veffFig}. The distinctive
``mermaid potential''
seen here is due to destructive interference between diagrams in the quark level
amplitude. Thus details of the potential are sensitive to the assumed microscopic
interaction, however, its general shape and strength are quite robust.

Long range interactions are assumed to be dominated by one-pion-exchange. Ref. \cite{essX}
chose to 
follow the method of T\"ornqvist\cite{NAT} in constructing an effective
pion-induced interaction. This is based  on a microscopic quark-pion interaction
familiar from nuclear physics (in more modern language, the form of the interaction is dictated 
by spontaneous chiral symmetry breaking):

\begin{equation}
L = -{g\over \sqrt{2} f_\pi} \int d^3x \bar\psi(x) \gamma^\mu \gamma_5 \tau^a \psi(x) \partial_\mu \pi^a(x).
\label{piq}
\end{equation}
Here $f_\pi = 92$ MeV is the pion decay constant, $\tau$ is an SU(2) flavour generator,
and $g$ is a coupling to be determined. The effective potential is derived by projecting
the quark level interactions onto hadronic states in the nonrelativistic limit. In
the case of pseudoscalar-vector states one obtains\cite{NAT} (see Appendix B for more details):


\begin{equation}
V_{\pi} = - \gamma V_0 \left[ \pmatrix{1 & 0 \cr 0 & 1}C(r) + \pmatrix{0 & -\sqrt{2} \cr
-\sqrt{2} & 1 } T(r) \right]
\label{Vpi}
\end{equation}
where

\begin{equation}
C(r) = {\mu^2\over m_\pi^2} {{\rm e}^{-\mu r} \over m_\pi r},
\label{C}
\end{equation}
\begin{equation}
T(r) = C(r)\left( 1 + {3\over \mu r} + {3\over (\mu r)^2} \right),
\label{T}
\end{equation}
and
\begin{equation}
V_0 \equiv {m_\pi^3\over 24 \pi}{g^2 \over f_\pi^2} \approx 1.3 {\rm MeV}.
\end{equation}
The matrix elements refer to S- and D-wave components of the pseudoscalar-vector
state in analogy with the deuteron.
The strength of the interaction has been fixed by comparing to the $\pi NN$ coupling
constant via the
relationship $f_{\pi N}^2/4\pi = 25/18\cdot m_\pi^2 g^2/f_\pi^2$. This allows a
prediction of the $D^*$ decay width which is in good agreement with experiment\cite{NAT}.
The parameter $\mu$ is typically the pion mass, however, one can incorporate recoil
effects in the potential by setting $\mu^2 = m_\pi^2 - (m_V-m_{pS})^2$. 
Finally, the coupling $\gamma$ is a spin-flavour matrix element which takes on the
following values: $\gamma = 3$ for $I=0$, $C=+$; $\gamma = 1$ for $I=1$, $C=-$;
$\gamma = -1$ for $I=1$, $C=+$; and $\gamma = -3$ for $I=0$, $C=-$. Thus the isoscalar
positive charge parity channel is the most likely to form bound states and subsequent
discussion focusses on it.

The potential of Eq. \ref{T} is often regulated to represent the effect
of additional unknown short distance interactions.
The regulator scale, $\Lambda$
can be fixed by comparison with nuclear physics; for example $NN$ interactions
yield preferred values for $\Lambda$ in the range 0.8 GeV to 1.5 GeV depending on model
details. Alternatively,
reproducing the deuteron binding energy requires $\Lambda \approx 0.8$ GeV.
T\"ornqvist has employed an intermediate value of $\Lambda = 1.2$ GeV which is
appropriate for $D$ mesons and this is taken as the
canonical cutoff in the following.

Unfortunately,
$D$ mesons are sufficiently light that the $D\bar D^*$ system (just barely) does not
bind with canonical parameters. However, the combined pion and quark
induced effective interactions are sufficient to cause binding. 
Thus it is of interest to study the properties of possible bound states as a function of
their binding energy. This has been achieved by allowing the regulator scale to
vary between 1.2 and 2.3 GeV. Binding is seen to occur for $\Lambda$ larger
than approximately 1.23 GeV.

\begin{figure}[h]
 \includegraphics[width=7 true cm, angle=270]{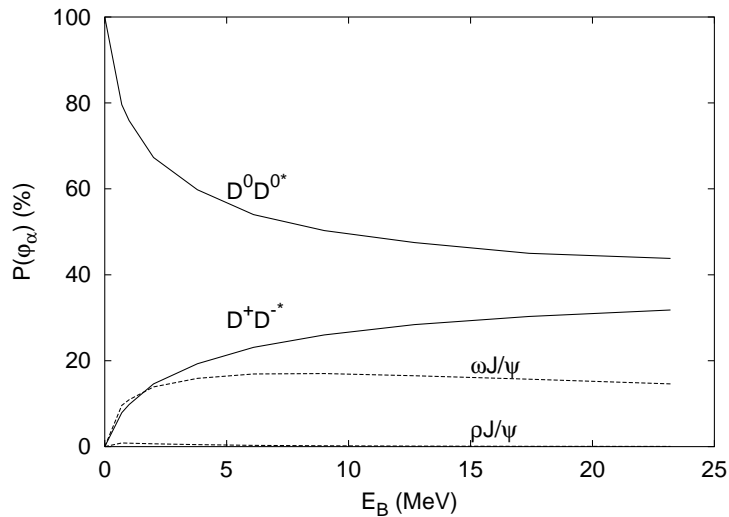}
 \caption{Wavefunction Components of the $X$\cite{essX}.} 
 \label{wfCompsFig}
 \end{figure}

Wavefunction probabilities are defined by

\be
Z_\alpha = \int |\varphi_\alpha|^2
\label{ZdefnEq}
\ee
where $\alpha$ is a channel index. These are 
shown as a function of binding energy in Fig. \ref{wfCompsFig}. It is clear
that the $D^0\bar D^{0*}$ component dominates the wavefunction, especially near
threshold.  However, the $D^+D^{-*}$ component rises rapidly in strength  with
isospin symmetry being recovered at surprisingly small binding energies (on the order of
30 MeV). Alternatively, the $\omega J/\psi$ component peaks at roughly 17\% at $E_B \approx
 9$ MeV.
The contribution of the $\rho J/\psi$
wavefunction  remains small, peaking at less than 1\% very close to threshold.

At the time of publication this model had two post-dictions: a charmonium resonance of mass 3872 MeV 
with a
decay mode of $\pi\pi J/\psi$. Predictions included the quantum numbers $J^{PC} = 1^{++}$; that the 
$\pi\pi$ invariant mass distribution should reflect its origins as a virtual $\rho$;  and  a 
$3\pi J/\psi$
decay mode of strength comparable to the $2\pi$ mode with an invariant mass distribution dominated
by the $\omega$. All of these predictions were subsequently confirmed experimentally 
(see Figs \ref{cdfpipiFig}, \ref{X3piFig}, and Eq. \ref{3pi-2pi}). Decays are discussed in 
Sect. \ref{XdecaySect}.
Further data and discussion of
the molecular interpretation of the $X$ appears in Sect. \ref{newDataSect}.


\subsection{Effective Field Theory}
\label{EFTSect}

The notion that the $X$ is dominated by a weakly bound $D^0\bar D^{*0}$ system
suggests that an effective field theory approach to the $X$ may be useful, in
analogy to the successful description of the deuteron\cite{W}. The utility of this
approach is further enhanced by an approximate heavy quark symmetry and the anomalously
small binding energy of $E_B = -0.5 \pm 0.9$ MeV, which should be compared to the 
natural scale of $m_\pi/(2 \mu_{DD^*}) \approx 20$ MeV\cite{voloshin,BK1}.

Weakly bound systems may be described in terms of a single parameter, the scattering
length $a$ -- a property known as `low energy universality'. Simple results
which apply to a weakly bound system include estimates of binding energy

\be
E_B = {1\over 2 \mu a^2},
\ee
and the weak binding wavefunction

\be
\psi(r) = {{\rm e}^{-r/a}\over \sqrt{2\pi a}\, r},
\ee
or
\be
\tilde\psi(k) = \sqrt{8\pi \over a} {1\over k^2 + a^{-2}}.
\ee

The scattering lengths can be quite large (compare to the deuteron, $a_D = 4.3$ fm). For example,
$E_B = 0.5$ MeV corresponds to $a = 6.4$ fm, while $E_B = 0.1$ MeV implies that $a = 14.4$ fm\footnote{A subtlety associated with these scales will be discussed below.}.
Universality ideas were first exploited by Voloshin\cite{voloshin} who estimated
the effects of interference on the processes $X \to D^0\bar D^0 \gamma$ and
$X \to D^0\bar D^0 \pi^0$.

Braaten and Kusunoki have followed this initial work with a series of papers
examining production and decay of the $X$. This began with the observation that
the probabilities for 
non-$D^0\bar D^{*0}$ components of the $X$, $X$ production, and $X$ decays  all
scale as $1/a$ and hence are strongly suppressed in the weak binding limit.

The analysis was performed with
the aid of an effective field theory which couples $D$, $D^*$ and $\chi$ mesons\cite{BK1}:

\begin{eqnarray}
{\cal H} &=& 
m_{D^0}  \left( D^\dagger D + \bar D^\dagger \bar D \right)
+ m_{D^{*0}} \left( {\bf D}^\dagger \cdot {\bf D} 
+ \bar {\bf D}^\dagger \cdot \bar {\bf D} \right)
\nonumber
\\
&&+ (m_{D^0}+m_{D^{*0}} + \nu_0) 
\mbox{\boldmath $\chi$}^\dagger \cdot \mbox{\boldmath $\chi$}  \nonumber
\\
 &-& \frac{1}{2 m_{D^0}} \left( D^\dagger \nabla^2 D 
+ \bar D^\dagger \nabla^2 \bar D \right) 
- \frac{1}{2m_{D^{*0}}} 
\left( {\bf D}^\dagger \cdot \nabla^2 {\bf D}
+ \bar {\bf D}^\dagger \cdot \nabla^2 \bar {\bf D} \right)
\nonumber
\\
&&- \mbox{$1\over2$} (m_{D^0}+m_{D^{*0}})^{-1} 
\mbox{\boldmath $\chi$}^\dagger \cdot \nabla^2 \mbox{\boldmath $\chi$} \nonumber
\\
&+& \lambda_0 \left( D \bar {\bf D} + \bar D {\bf D} \right)^\dagger
\cdot \left( D \bar {\bf D} + \bar D {\bf D} \right)
\nonumber
\\
&& + g_0 \left[ \mbox{\boldmath{$\chi$}}^\dagger \! \cdot \! 
(D \bar {\bf D} + \bar D {\bf D})
+ (D \bar {\bf D} + \bar D {\bf D})^\dagger
\! \cdot \! \mbox{\boldmath $\chi$} \right].
\label{HBK}
\end{eqnarray}
The bare $\chi$ binding energy is represented by $\nu_0$.
The four body interaction of strength $\lambda_0$ represents short range interactions between
$D$ and $D^*$ mesons while the term proportional to $g_0$ couples the $\chi$ to its constituent
$D\bar D^*$ channel via a pointlike interaction.

\begin{figure}
\includegraphics[width=4 true cm, angle=0]{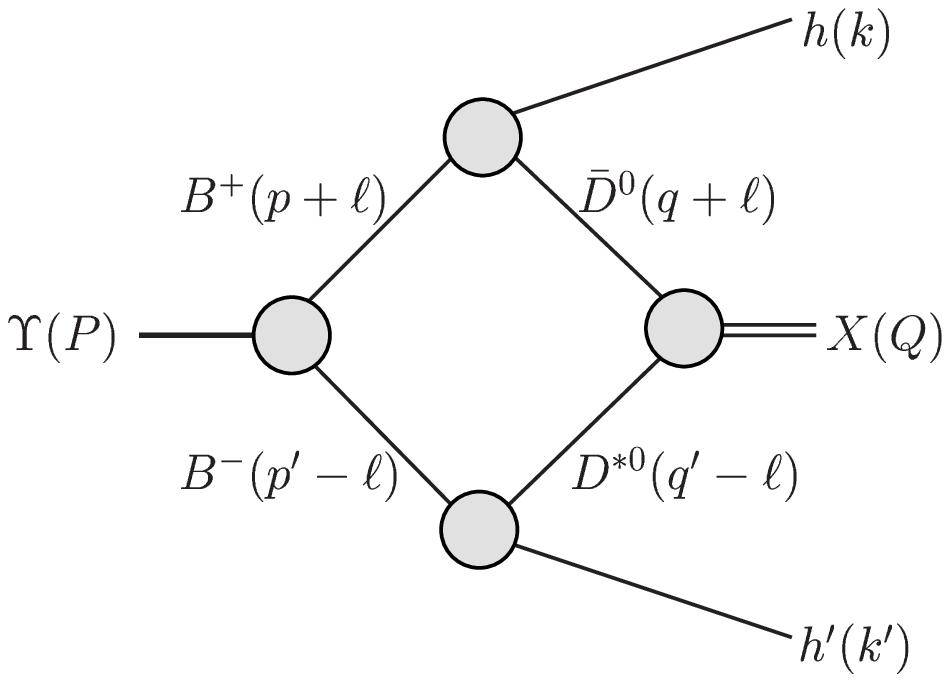}
\hskip 1 true cm
\includegraphics[width=4 true cm, angle=0]{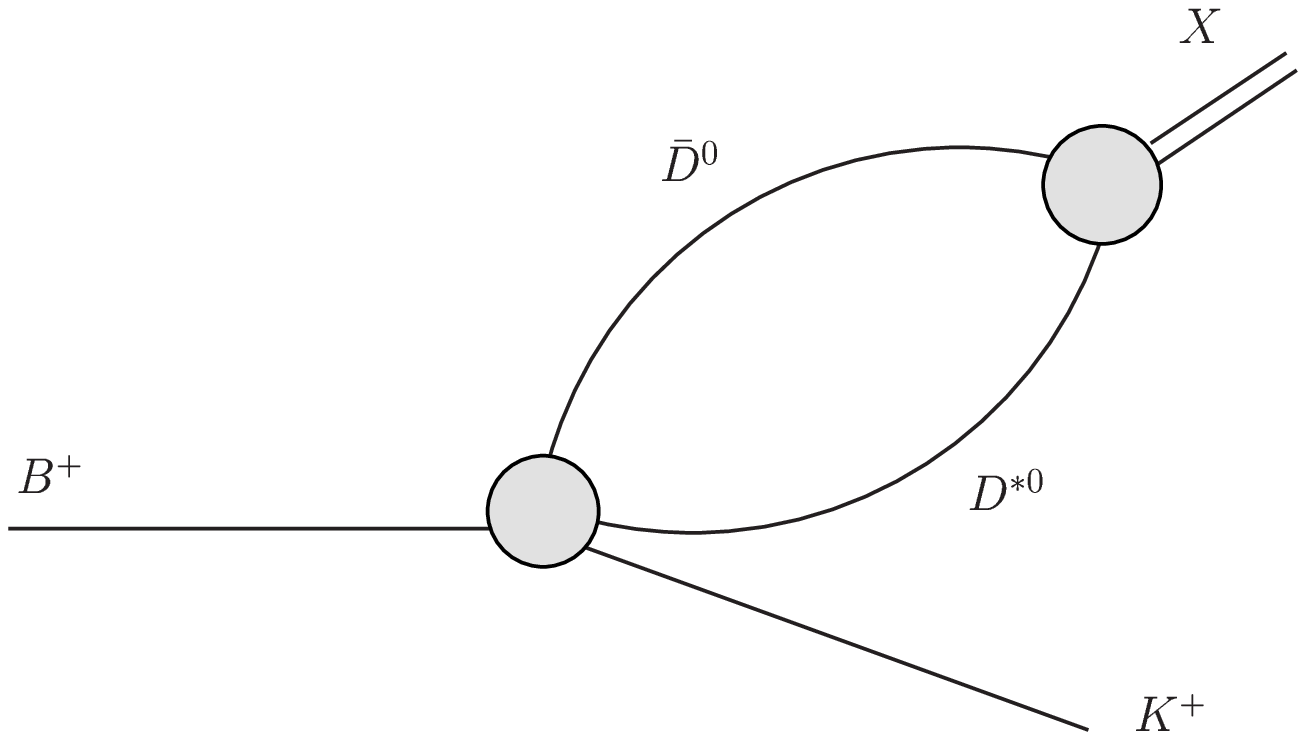}
\caption{$X$ Production from $\Upsilon(4S)$ Decays via Coalescence (left)\protect\cite{BK2} and from $B$ Decays (right)\protect\cite{BKN}.} 
\label{XcoalesceFig}
\end{figure}

Low energy universality also permits analysis of `coalescence' models of $X$ production
in $\Upsilon$ and $B$ decays (see Fig.~\ref{XcoalesceFig})
with the results\cite{BK2}

\be
\Gamma(\Upsilon(4S) \to X+\ {\rm hadrons}) \sim E_B \log E_B
\ee
and
\be
{\cal A}(B \to XK) = \left[{Z_{00}  m_X \over \pi^3 m_{D} m_{D^*} a}\right]^{1/2}  2 c \Lambda p_B\cdot \epsilon^*_X
\ee
for $X$ production in $B$ decays\cite{BKN}. Here $\Lambda$ is an ultraviolet cutoff, and $c$
is a coupling defined by ${\cal A}(B \to DD^*K) = c p_B\cdot \epsilon$. This expression
was analysed further in Ref. \cite{BK3} where the authors set the cutoff to be $\Lambda
= \Lambda_\pi \approx m_\pi$ which is a `cross over momentum' scale defined by the transition from
a phase space dominated invariant mass distribution to a resonance dominated invariant 
mass distribution.
The improved estimate is then

\be
Br(B^+ \to X K^+) = 2.7 \cdot 10^{-5} \left({\Lambda_\pi \over \pi}\right)^{1/2} \left({E_B \over 0.5 \ {\rm MeV}}\right)^{1/2}.
\ee
Braaten and Kusunoki  also obtained

\be 
{Br(B^0 \to X K^0)\over Br(B^+ \to X K^+)} = 0 - 0.08.
\ee
(The upper end of this estimate agrees with the lower end of the molecular estimate to appear in Eq. \ref{b0b+2}.)

Finally, Braaten and Kusunoki have used factorisation between universal long range
behaviour and unknown short range behaviour to predict the binding energy dependence
of decay rates\cite{BK4,BK5}. Using the results of Ref.~\cite{essX} to fix the short
range results, yields predictions such as

\be
\Gamma(X \to \pi\pi\pi J/\psi) = 222 \ {\rm keV}\, \left( {E_B + \Gamma_X^2/16E_B\over 1 \, {\rm MeV}}\right)^{1/2}.
\label{scaling3PiEq}
\ee

This method describes decays such as $X \to \pi\pi J/\psi$ via a virtual $\rho J/\psi$ channel and
associates that channel with the scattering length $a$. It should be noted, however, that the presence
of multiple channels generates multiple scattering lengths via the coupled channel weak binding 
relations

\begin{equation}
E_B(\alpha) = {1\over 2 \mu_\alpha a_\alpha^2}
\end{equation}
where $\alpha$ denotes an $X$ channel and $E_B(\alpha) = m_{tot}(\alpha) - m_X$.
Thus, although the probability of channel $\alpha$ scales as $1/a$, the dynamics associated
with that channel scale with $a_\alpha = 1/\sqrt{2 \mu_\alpha E_B(\alpha)}$.

\begin{figure}[h]
 \includegraphics[width=6 true cm,angle=270]{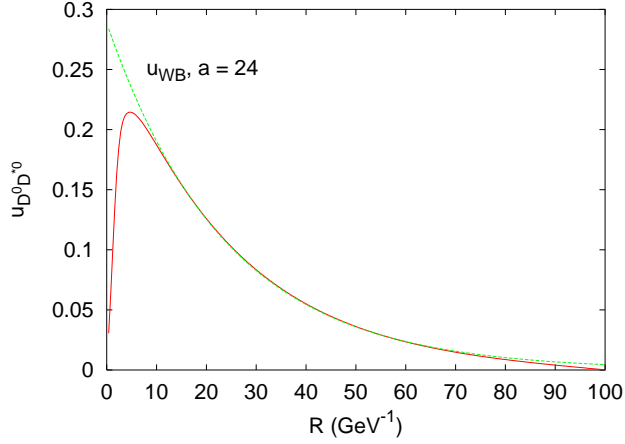}
 \caption{The full $D_0D^*_0$ wavefunction and the weak binding wavefunction.}
 \label{WBwfFig}
\end{figure}

These scales can be seen clearly in the $X$ molecule model of the previous section. 
Numerically integrating the coupled channel Schr\"odinger equation and fitting the subsequent
channel wavefunctions to the weak binding form (an example is shown in Fig. \ref{WBwfFig})
yields estimates of the channel scattering lengths.
For example, setting $\Lambda = 1.13$ GeV gives a binding energy of 0.9 MeV and hence 
$a_{D^0\bar D^{*0}} = 23.8$ GeV$^{-1}$ while the fit to the numerically obtained wavefunction
yields $a_{D^0\bar D^{*0}} = 24$ GeV$^{-1}$.
Similarly, for the $\rho J/\psi$ wavefunction one obtains a fit value of $a_{\rho\psi} = 9.5$ GeV$^{-1}$
while
using $E_B(\rho\psi) = 9.2$ MeV gives an expected value of $a_{\rho\psi} = 9.3$ GeV$^{-1}$. 
Similarly $a_{D^-D^{*+}} =7.6$ GeV$^{-1}$. A selection of $X$ properties as a function of $\Lambda$ are
given in Table \ref{XlambdaTab}. Thus many scales are present when multiple channels
contribute to the molecular structure of the $X$. As the binding energy decreases these additional 
channels disappear as $1/a_{D^0\bar D^{*0}}$ and the scale associated with each channel saturates
at $a_\alpha = [2 \mu_\alpha (m_{tot}(\alpha) - m_{tot}(D^0\bar D^{*0}))]^{-1/2}$. At larger
binding energies the interplay of these scales can complicate the description of $X$ dynamics and
care must be exercised.

\begin{table}[h]
\caption{$X$ Properties vs. Cutoff}
\begin{tabular}{c|ccccc}
\hline
$\Lambda$ (GeV) & $E_B$ (MeV) & $Z_{00}$ & $Z_{+-}$ & $Z_{\omega\psi}$ & $Z_{\rho\psi}$ \\
\hline
\hline
1.14 & 1.0 & 89\%  & 4.0\% & 5.0\% & 0.7\% \\
1.17 & 1.1 & 86\%  & 5.5\% & 6.4\% & 0.8\% \\
1.20 & 1.4 & 82\% &  7.5\% & 8.1\% & 0.8\% \\
1.23 & 1.7 & 78\% &  9.2\% & 9.3\% & 0.8\% \\
1.30 & 2.7 & 69\% &  14\%  & 12.7\% & 0.7\% \\
1.40 & 4.8 & 60\% &  20\%  & 15\% &   0.5\% \\
1.50 & 8.0 & 55\% & 24\%  & 16\%  & 0.3 \% \\
\hline
\hline
\end{tabular}
\label{XlambdaTab}
\end{table}

Notice that the weak binding wavefunction of Fig. \ref{WBwfFig} is quite accurate for distances greater
than 2 fm. Similarly, the Fourier transformed weak binding wavefunction is accurate for momenta less
than 200 MeV. Above this scale the wavefunction is quite small, hence the weak binding wavefunction
may serve as a useful approximation to the full wavefunction in favourable circumstances.

Finally, 
Al Fiky {\it et al.}\cite{alexey} have improved the effective field theory of Braaten and Kusunoki (Eq. \ref{HBK}) by 
incorporating heavy quark and chiral symmetries. 
Thus the theory is written in terms of a superfield doublet combining the 
pseudoscalar and vector mesons:
\be
H_a^{(Q)}=\frac{1+\not{v}}{2}\left[
D^{*(Q)}_{a\mu} \gamma^\mu - D_a^{(Q)} \gamma_5
\right], \qquad \overline{H}^{(Q) a} = \gamma^0 H_a^{(Q)\dagger} \gamma^0.
\ee
These fields have the usual transformation properties under heavy-quark
spin symmetry and SU(2)$_V$ flavor symmetry
and describe heavy mesons with a definite velocity $v$. One then has ${\cal L} = {\cal L}_2 + {\cal L}_4$ with 

\begin{eqnarray}
{\cal L}_2 ~&=& ~-i \mbox{Tr} \left[ \Hbar^{(Q)} v \cdot {\cal D} H^{(Q)} \right]
- \frac{1}{2 m^{\phantom{l}}_P} \mbox{Tr} \left[ \Hbar^{(Q)} {\cal D}^2 H^{(Q)} \right]
\nonumber \\
&+& ~\frac{\lambda_2}{m^{\phantom{l}}_P}  \mbox{Tr}
\left[ \Hbar^{(Q)} \sigma^{\mu\nu } H^{(Q)} \sigma_{\mu\nu} \right]
+ \frac{ig}{2} \mbox{Tr}  \Hbar^{(Q)} H^{(Q)} \gamma_\mu \gamma_5
\left[\xi^\dagger \partial^\mu \xi - \xi \partial^\mu \xi^\dagger
\right] 
\label{Lagr2}
\end{eqnarray}
and

\begin{eqnarray}
-{\cal L}_4&=& \frac{C_1}{4} \mbox{Tr} \left[ \Hbar^{(Q)} H^{(Q)} \gamma_\mu \right]
\mbox{Tr} \left[ H^{(\overline{Q})} \Hbar^{(\overline{Q})} \gamma^\mu \right]
+ \frac{C_2}{4} \mbox{Tr} \left[ \Hbar^{(Q)}  H^{(Q)} \gamma_\mu \gamma_5 \right]
\mbox{Tr} \left[ H^{(\overline{Q})} \Hbar^{(\overline{Q})} \gamma^\mu \gamma_5 \right].
\label{Lagr4}
\end{eqnarray}
The covariant derivative is given by
${\cal D}_{ab}^\mu=
\delta_{ab}\partial^\mu-(1/2)\left(\xi^\dagger \partial^\mu \xi +
\xi \partial^\mu \xi^\dagger
\right)_{ab}$ and $g$ is the $D^{*}D\pi$ coupling. The pion is realised nonlinearly as $\xi = \exp(i \pi\cdot\tau/\sqrt{2}f_\pi)$.
The third term in Eq.~(\ref{Lagr2}) is needed to account for the $D-D^*$ mass
difference $M(D^*)-M(D) = -2\lambda_2/M(D)$.

Evaluating the traces yields for the $D\overline{D^*}$ sector
\begin{eqnarray}
{\cal L}_{4,DD^*} = &-& C_1 D^{(Q)\dagger} D^{(Q)}
D^{*(\overline{Q})\dagger}_\mu D^{* (\overline{Q}) \mu}
- C_1 D^{*(Q)\dagger}_\mu D^{*(Q) \mu}
D^{(\overline{Q})\dagger} D^{(\overline{Q})} \nonumber \\
&+& C_2 D^{(Q)\dagger} D^{*(Q)}_\mu
D^{* (\overline{Q})\dagger \mu} D^{(\overline{Q})}
+ C_2 D^{* (Q)\dagger}_\mu D^{(Q)}
D^{(\overline{Q})\dagger} D^{* (\overline{Q}) \mu}
+\dots
\label{LocalLagr}
\end{eqnarray}

This lagrangian differs from that of Eq. \ref{HBK} where the interaction is described in terms
of a single parameter $\lambda_0$. This is related to the parameters above by $\lambda_0 = -C_1 = C_2$. The additional parameter in the lagrangian of Al Fiky {\it et al.} may be ascribed to distinguishing
effective pseudoscalar and scalar exchange interactions between heavy mesons.
Nevertheless, the binding energy of heavy pseudoscalar-vector
states depends on a single linear combination of $C_1$ and $C_2$. The mass of the $X$ may be used
to fix this linear combination and a 
prediction for a $B\bar B^*$ bound state at 10604 MeV is obtained\cite{alexey}. Predictions
for possible $D\bar D$, $B\bar B$, $B\bar D^*$, and $D \bar D^*$ bound states are not
possible until another linear combination of the couplings can be determined. Further discussion
of other molecules is contained in Sect. \ref{OtherMoleculesSect}.

The mass of the $B\bar B^*$ molecule was obtained with the aid of the heavy quark mass scaling
relation $C_i \sim 1/m_Q$ which implies that $C_i(m_b) = m_c/m_b C_i(m_c)$. Furthermore, the pion
exchange portion of the coupling constant (obtained upon integrating out the pions) is much
smaller than the contact portion\cite{alexey2}. Both of these observations are not in agreement
with the microscopic model quark+pion exchange model: pion exchange dominates the effective
interaction and the pointlike interaction is controlled by scales such as the string tension,
the light quark mass, and the pion mass.

The power and limitations of the effective field theory approach are easily seen. When
enough data exists to determine coupling constants reliable predictions are possible in a 
broad range of problems. These predictions are under-pinned by QCD symmetries or universality.
An example is the scaling of decay rates with binding energy shown in 
Eq. \ref{scaling3PiEq}. However, the method fails when data is sufficiently sparse that parameters
cannot be determined or multiple scales affect the dynamics.

\subsection{Microscopic Decay Models}
\label{XdecaySect}

\subsubsection{Strong Decays}
\label{XdecaysSect}

\begin{figure}[h]
\includegraphics[width=5 true cm, angle=0]{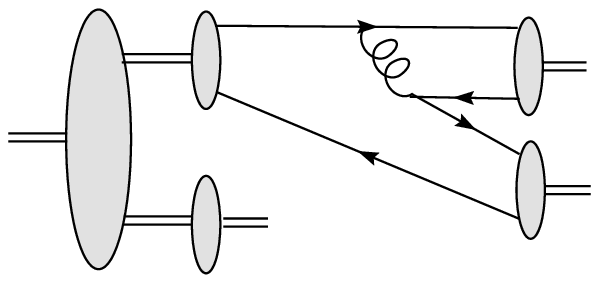}
\hskip 1 true cm
\includegraphics[width=5 true cm, angle=0]{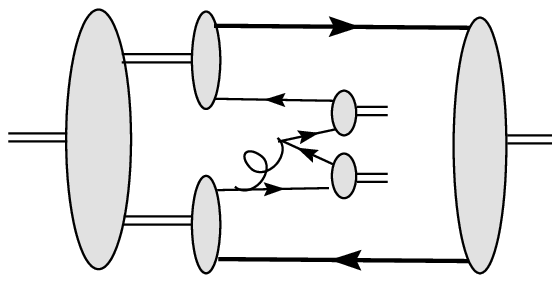}
\caption{$X$ Strong Decay via Dissociation (left) and $\pi\pi$ Production (right).}
\label{Xdecays2Fig}
\end{figure}

\begin{figure}[h]
\includegraphics[width=5 true cm, angle=0]{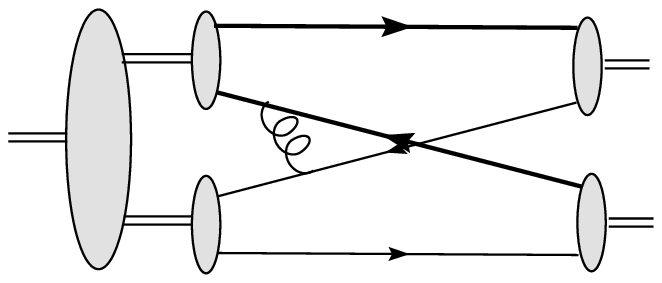}
\hskip 1 true cm
\includegraphics[width=5 true cm, angle=0]{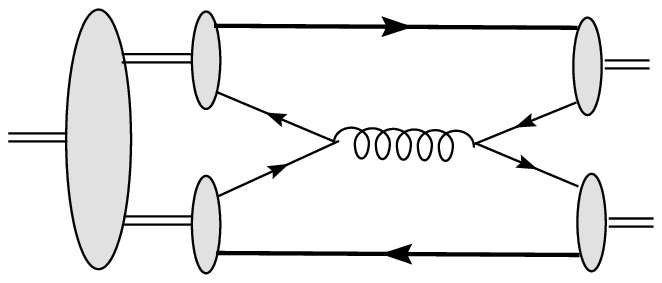}
\caption{$X$ Strong Decay via Rearrangement (left) and Annihilation (right).}
\label{XdecaysFig}
\end{figure}

The difficulties associated with predicting strong decays of mesons have already been discussed in
Section \ref{spectrumCritiqueSec}. These difficulties are worse in molecular decays where 
uncertainties are compounded by poorly constrained  structure and dynamics.  There is, however, 
no utility in being too easily dissuaded! In this spirit three- and two-body decay processes driven
by one gluon exchange are displayed in Figs. \ref{Xdecays2Fig} and \ref{XdecaysFig} respectively.
The latter replace constituent mesons with kinematically allowed decay product mesons, either by
rearrangement or by flavour changing processes. The former proceed via quark pair creation such
as drives $^3P_0$ or Cornell model meson decays. In the case of the $X$ molecular model of the
previous section, dissociation gives rise to decays such as $X \to \rho J/\psi \to \pi\pi J/\psi$ or
$X \to D\bar D^* \to D \bar D \gamma$.

In terms of the molecular $X$ model, the leading source of decays is dissociation. In the 
weak binding limit the decay rates are simple to estimate: to good approximation dissociation 
will proceed via the free space decay of the constituent mesons. Thus, for example, 

\be
\Gamma(X \to \pi\pi J/\psi) = Z_{\rho\psi} \Gamma(\rho \to \pi\pi)
\ee
and
\be
\Gamma(X \to D^0\bar D^0 \pi^0) = 2 Z_{00} \Gamma(D^{*0} \to D^0 \pi^0).
\ee
The factor of two in the latter expression accounts for the S-wave combination $\frac{1}{\sqrt{2}}(D^0\bar D^{*0} + \bar D^0 D^{*0})$ in the bound state wavefunction.
A more extensive list of predictions is given in Table \ref{XDecayTab}.

\begin{table}[h]
\caption{Some Decay Modes of the $X(3872)$ (keV).}
\begin{tabular}{c|ccccccccc}
\hline
$B_E$ (MeV)$^{\phantom{X}^{\phantom{X}}}$ & $D^0\bar D^0 \pi^0 \ \ $ & $D^0\bar D^0\gamma \ \ $ & $D^+D^-\pi^0 \ \ $ & $(D^+\bar D^0\pi^-$+c.c)$/\sqrt{2}$ & $D^+D^-\gamma \ \ $ & $\pi^+\pi^-J/\psi \ \ $ & $\pi^+\pi^-\gamma J/\psi \ $ & $\pi^+\pi^-\pi^0J/\psi \ \ $ & $\pi^0\gamma J/\psi \ \ $  \\
\hline
\hline
0.7  & 67 & 38 & 5.1  &  4.7  &  0.2  & 1290  & 12.9 &  720  &  70 \\
1.0  & 66 & 36 & 6.4  &  5.8  &  0.3  & 1215  & 12.1 &  820 &  80 \\
\hline
\hline
\end{tabular}
\label{XDecayTab}
\end{table}

Strong decays in the case of strong binding are considerably more complicated. The formalism which describes this is discussed in Appendix A.

\subsubsection{Radiative Decays}

\begin{figure}[h]
\includegraphics[width=4 true cm, angle=0]{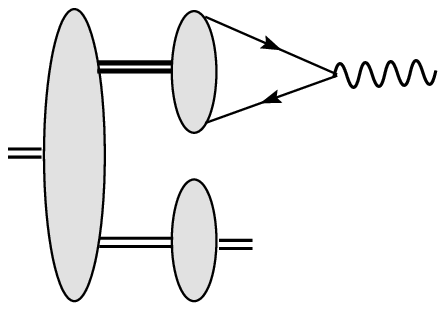}
\hskip 1 true cm
\includegraphics[width=4 true cm, angle=0]{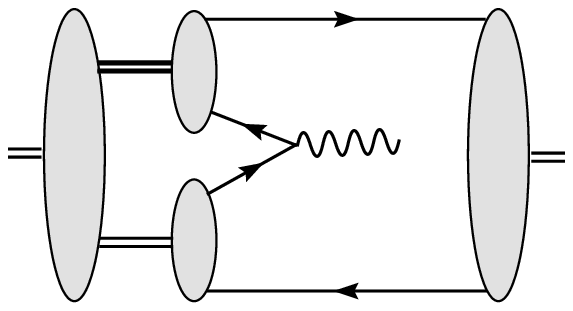}
\caption{$X$ Radiative Decay via Vector Meson Dominance (left) and Annihilation (right)
\protect\cite{essX2}.}
\label{Xdecays3Fig}
\end{figure}

Radiative decays of molecules provide an interesting contrast to the impulse approximation processes
which drive mesonic radiative decays. In the case of the $X$ radiative decay can occur via
vector meson dominance (in the $\omega J/\psi$ and $\rho J/\psi$ components) or annihilation\cite{essX2}
($D \bar D^*$ components), as illustrated in Fig. \ref{Xdecays3Fig}. The vector meson dominance
diagram will contribute to $X \to \gamma J/\psi$. 
 It is clear that the amplitude must be proportional to the light vector meson wavefunction at 
the origin
and to the $\hat\chi$ wavefunction for the channel in question. The specific result is

\begin{equation}
\Gamma_{\rm VMD} = {4\over 27} \alpha {q E_\psi \over m_\chi} |\psi_\omega(r=0)|^2  \left( Z_{\omega\psi}^{1/2}\phi_{\omega\psi}(q) + 3 Z_{\rho\psi}^{1/2} \phi_{\rho\psi}(q)\right)^2.
\label{vmd}
\end{equation}
Decays to $\gamma \psi'$ or $\gamma \psi_2$ can 
only proceed via annihilation and hence are very small. 

A variety of predictions are presented in Table \ref{E1X}.
Here, the fourth column summarises the results of Barnes and Godfrey\cite{BG} which are 
computed in the impulse, nonrelativistic, zero recoil, and
dipole approximations. The results for $X \to \gamma J/\psi$ are particularly sensitive to model 
details -- in fact this rate is identically zero for SHO wavefunctions. The sensitivity is examined in
columns five and six which present the results of two additional computations. The first,
labelled [A], employs the same approximations of Barnes and Godfrey but uses meson
wavefunctions computed with a simple, but accurate, Coulomb+linear+smeared hyperfine
potential. It is apparent that the $\gamma J/\psi$ rate is very sensitive to wavefunction
details. Furthermore, one may legitimately question the use of the zero recoil and dipole approximations
for the $\gamma J/\psi$ mode since the photon momentum is so large in this case. The
sixth column (model [B]) dispenses with these approximations, and one finds a relatively
large effect for $\gamma J/\psi$.

Table~\ref{E1X} makes it clear that computations of the $\gamma J/\psi$ radiative transition of
the $\chi'_{c1}$ are very sensitive to model details\footnote{There is an additional error induced by arbitrarily changing the quark model $\chi'$ mass to 3872 MeV. This is made clear through the observation
that the dipole formula for the width scales as $q^3$ whereas the momentum space formula scales as $q$.}
. The result of Barnes and Godfrey is
similar to that computed here for the molecular $\hat\chi$ state but is much smaller than models A and B. Furthermore, the rates for $\gamma \psi^{''}$ and $\gamma \psi_2$ are
very small for a molecular $X$ (at the order of eV) and quite small for a charmonium $X$.
Perhaps the most robust diagnostic is the $\gamma \psi'$ decay mode. For a molecular $\hat \chi$ this can only proceed via the annihilation diagram and hence is very small. Clearly a measurement of the $\gamma J/\psi$ and $\gamma \psi'$ decay modes of the $X(3872)$ will provide compelling clues to its internal structure. 

\begin{table}[h]
\caption{E1 Decays of the X(3872)\protect\cite{essX2}.}
\begin{tabular}{ccc|ccc|c}
\hline
mode & $m_f$ (MeV) & $q$ (MeV)  & $\Gamma[c\bar c]$ (keV) & $\Gamma[c\bar c]$ (keV) & $\Gamma[c\bar c]$ (keV) & $\Gamma[\hat \chi_{c1}]$ (keV)  \\
   &   &   & [B\&G]   &   [A]   &   [B]   &  \\
\hline
\hline
$\gamma J/\psi$ & 3097 & 697 & 11 & 71 & 139 & 8 \\$\gamma \psi'(2^3S_1)$ & 3686 & 182 & 64 & 95 & 94  & 0.03 \\$\gamma \psi^{''}(1^3D_1)$ & 3770 & 101 &  3.7 & 6.5 & 6.4 &   0 \\
$\gamma \psi_2(1^3D_2)$ & 3838 & 34 & 0.5 & 0.7 & 0.7 & 0 \\
\hline
\hline
\end{tabular}
\label{E1X}
\end{table}

\subsection{New Data and Further Analysis}
\label{newDataSect}

The apparent success of the molecular interpretation of the $X$
has been challenged by several recent experimental results.  These are reviewed here.

\begin{figure}[h]
  \includegraphics[width=3 true cm, angle=0]{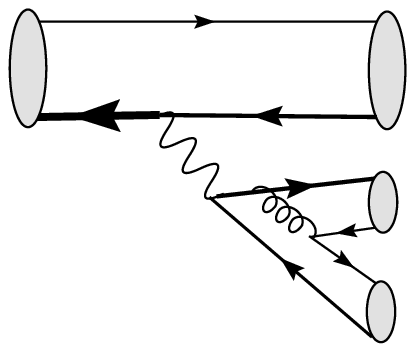}
  \hskip 2 true cm
  \includegraphics[width=3 true cm, angle=0]{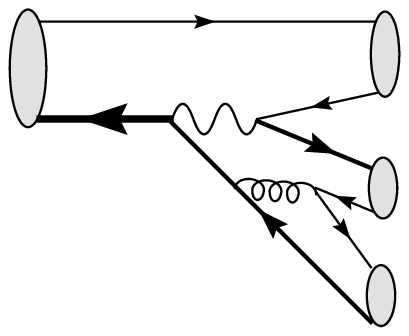}
  \caption{$B \to K X$. Colour enhanced amplitude (left). Colour suppressed amplitude (right).}
  \label{BdecaysFig}
\end{figure}

$\bullet\ $ The ratio of production of the $X$ in neutral and charged $B$ decays (Eq. \ref{b0b+}) lies in the
range 0.25 - 0.97 with 64\% C.L.  and hence is rather large compared to expectations based on the
molecular picture. This ratio may be estimated with the diagrams shown in Fig. \ref{BdecaysFig}. 
The colour leading diagram (left panel) permits the processes $B^+ \to K^+ D^0\bar D^{*0}$ which
will contribute to $K^+X$, $B^+ \to K^0 \bar D^0 D^{*+}$ which tests $X^+$ production and cusp
effects, $B^0 \to K^0 D^-D^{*+}$ which accesses the $X$ through the suppressed charged component,
and
$B^0 \to K^+ D^- D^{*0}$ which tests $X^-$ production and cusp effects.  The colour suppressed 
diagram (right panel) permits the following decays: $B^+ \to K^+ D^0 \bar D^{*0}$, $B^+ \to K^+ D^+D^{*-}$,
$B^0 \to K^0 D^0 \bar D^{*0}$, and $B^0 \to K^0 D^+D^{*-}$. Since the diagrams involve identical mesons
(at the strong level) and only differ by topology one may predict the ratio

\be
{Br(B^0 \to K^0 X)\over Br(B^+ \to K^+X)} = {|4 Z_{+-}^{1/2} + Z_{00}^{1/2}|^2 \over
|4 Z_{00}^{1/2} + Z_{+-}^{1/2}|^2 } \approx 0.06 - 0.29.
\label{b0b+2}
\ee
The lower estimate assumes that there is no relative phase between the topologies (it is 
difficult to imagine why there should be one) and that the $X$ is completely dominated by the
neutral $D\bar D^*$ component. The higher estimate assumes that the neutral component is 90\% of
the $X$ while the charged is 10\%\footnote{Assuming a negative relative phase raises this estimate
to 0.25 -- 0.51.}. Given the large errors in the experimental result of Eq. \ref{b0b+},
it is too early to rule out the molecular interpretation of this state. An accurate determination
of this  ratio will measure the relative importance of the suppressed $D^+D^{*-}$  component of the $X$.

$\bullet\ $ Eqs. \ref{BrBr}, \ref{BrBr2}, and \ref{DDpi} imply that

\be
R = {Br(X \to D^0\bar D^0 \pi^0) \over Br(X \to \pi^+\pi^- J/\psi)} = 22 \pm 13.
\label{DDpi-pipipsi}
\ee

This should be contrasted with an earlier measurement from Belle
which determined\cite{Abe:2003zv}  

\be
Br(B^+ \to X K^+) Br(X \to D^0\bar D^0 \pi^0) < 6 \cdot 10^{-5}
\label{belleDDpi}
\ee
at 90\% C.L. Eqs. \ref{BrBr} and \ref{BrBr2} then imply that $R < 6$. A reliable determination of this ratio will be crucial to understanding the $X$ and further experimental effort
is encouraged.

In the molecular picture $R$ is driven by decays $D^{*0} \to D^0 \pi^0$ and
$\rho \to \pi\pi$ and hence is given (in the weak binding limit)  by 

\be
R = {2 Z_{00} \Gamma(D^{*0})\over Z_{\rho\psi} \Gamma(\rho)} .
\label{DDpipiThy}
\ee
With the pion and quark
dynamics assumed in Section \ref{XMoleculeSect} one obtains $Z_{00} \approx 0.8$ and $Z_{\rho \psi} \approx 0.01$ at a binding energy of 0.7 MeV. Thus $R \approx 0.08$, clearly in conflict with experiment. 

If the $X$ is indeed a 
weakly bound molecule, Eq. \ref{DDpipiThy} must be accurate and hence the only way to
reconcile this prediction with the data is if the $\rho J/\psi$ component has been severely
overestimated. The required probability is $Z_{\rho \psi} \approx 0.007\%$.

$\bullet\ $ The large rate for $D^0\bar D^0 \pi^0$ implies that 
$Br(X \to \pi\pi J/\psi) < 0.05$. With Eqs. \ref{BrBr}, \ref{BrBr2}, this implies that 

\be
Br(B \to K X) > 2 \cdot 10^{-4}
\ee
which is comparable to the branching ratio for $\chi_{c1}$: $Br(B \to \chi_{c1}K) = 4.0(1.0)\cdot 10^{-4}$\cite{pdg}.  However, both of these results are very close to bounds established by 
BaBar\cite{Aubert:2005vi,bernard,bernard2}:

\be
Br(B^\pm \to K^\pm X) < 3.2 \cdot 10^{-4} (90\% {\rm C.L.})
\ee
and
\be
Br(X \to \pi\pi J/\psi) > 4.2\%.
\ee
These results, and the conflict with the earlier Belle result (Eq. \ref{belleDDpi}), can be most easily
interpreteted by assuming that the rate to $D^0\bar D^0 \pi^0$ has been overestimated in 
Eq. \ref{DDpi-pipipsi}. 

\begin{figure}[h]
  \includegraphics[width=5 true cm, angle=0]{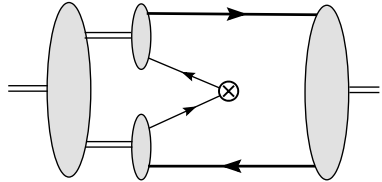}
  \caption{$X - \chi_{c1}$ Mixing.}
  \label{XXFig}
\end{figure}

If a large $D^0\bar D^0 \pi^0$ decay mode is confirmed, the
simplest interpretation is that the
$X$ contains substantial $c\bar c$ -- in keeping with the production characteristics discussed at the
end of Sect. \ref{XexptSect}. More specifically the $X$ must mix strongly with
the $\chi_{c1}'$. This may be estimated with the aid of the $^3P_0$ model and Fig. \ref{XXFig}.
The net result is a mixing matrix element 

\be
a_\chi = \sqrt{2} Z_{00}^{1/2}\, \int d^3 k \,  \psi_X(k) {\cal A}(-k)
\ee
where ${\cal A}$ is the $^3P_0$ amplitude for $\chi_{c1}' \to D^0\bar D^{*0}$. Here 
$\psi_X(k) = (\pi\sqrt{a})^{-1} (k^2 + a^{-2})^{-1}$ is the weak binding $X$ wavefunction.

Evaluating the integral yields surprisingly large mixing matrix elements, which are summarised in 
Table \ref{XXTab}. As can be seen, the mixing amplitude is diminished as the $X$ scattering length 
gets larger. Surprisingly, mixing with the excited $\chi_{c1}'$ is not strongly suppressed with 
respect to the ground state $\chi_{c1}$. Most surprisingly, the mixing matrix element can be quite
large, so large that the relatively close $\chi_{c1}'(3930)$ has a very (nonperturbatively) large
mixing amplitude with the $X$. This unexpected result explains the large $KX$ branching ratio in
$B$ decays and the $X$ production characteristics in $p\bar p$ collisions. The complete phenomenological
implications of this result have not been explored\cite{HS}.

\begin{table}
\caption{$X-\chi_{c1}$ Mixing.}
\label{XXTab}
\begin{tabular}{cccccc}
\hline
state  &  $E_B$ (MeV) &  $a$ (fm)  &  $Z_{00}$  & $a_\chi$ (MeV) & prob \\
\hline
\hline
$\chi_{c1}$ & 0.1  &  14.4 & 93\%  &  94 & 5\% \\
       & 0.5  &   6.4 & 83\%  & 120 & 10\% \\
$\chi_{c1}'$ & 0.1  & 14.4  & 93\% & 60 & 100\% \\
             & 0.5  & 6.4   & 83\% & 80 & $> 100\%$ \\
\hline
\hline
\end{tabular}
\end{table}

$\bullet\ $ Belle have found the $\gamma J/\psi$ decay mode of the $X$ and report a 
large branching of 14\% of the $\pi\pi J/\psi$ mode (see Eq. \ref{gammapsiEq}). 
In the model of Sect. \ref{XdecaySect} one expects this to be 0.4\%  at a binding energy of 0.7 MeV -- 
much too small to explain the data. Note, however,  that the
predicted rate for $\chi_{c1}' \to \gamma J/\psi$ is 10 -- 140 keV (Table \ref{E1X}); the upper
end of this prediction implies a branching ratio of 14\% if the $\pi\pi J/\psi$ mode
has a width of roughly 1 MeV.
But this implies that the $X$ has a very large $\chi_{c1}'$ component or that the predicted
$\pi\pi J/\psi$ rate is too large. In view of the previous discussion, both of these 
possibilities appear likely.

\subsection{Summary}

It is possible that all of the new experimental data may be explained if the coupling of
the $X$ to the $\chi_{c1}'$ is correctly incorporated in the coupled channel formalism. This
would increase the $\gamma J/\psi$ branching fraction, permit the large $D^0\bar D^0 \pi^0$ decay
mode, and allow the charmonium-like production characteristics. A detailed phenomenology of this
scenario remains to be constructed.

Note that cusp, tetraquark, renormalised $\chi_{c1}'$, virtual state, and molecular interpretations 
all assume that the $X$ is dominantly (or heavily modified by)  a $c\bar c u\bar u$ 
Fock space component. The chief difference among the models is the spatial, colour, and
spin  configuration of the state, which is of course, driven by different assumed 
dynamics. Thus distinguishing these models is a qualitative issue and choosing between them
is question of efficiency of description and consistency of adjunct predictions.

Finally, we comment on two misapprehensions in $X$ dynamics that have appeared in the 
literature. First, 
the $3\pi J/\psi$ decay mode proceeds via the $\omega J/\psi$ component
of the $X$.  It is sometimes thought that the decay occurs via the natural line
width of the $\omega$.  Since the width of $\omega$ is only 8 MeV and the $\omega J/\psi$
system is 7.5 MeV above the $X$ mass, the probability of the $X(\omega J/\psi)$ fluctuating into
$3\pi J/\psi$ is very small. Thus the large relative branching fraction for the $3\pi$ mode
cannot be correct. This argument is incorrect: the $\omega J/\psi$
component of the $X$ is off-shell due to the dynamics leading to binding and the
$\omega$ need not fluctuate to a ``lower mass'' to permit decay.

Second, Suzuki\cite{Suzuki:2005ha} has argued that the one-pion-exchange potential as described in 
Sect. \ref{XMoleculeSect} is not valid because the process $D^* \to D\pi$ can occur on-shell.
This argument is incorrect. The four point function that describes $D\bar D^*$ scattering
does indeed exhibit a pole (as opposed to the $NN$ interaction) due to 
set of measure zero in the phase space available to the $D\bar D\pi$ system. The pole
induces a small imaginary part to the scattering amplitude, which describes the decay
$X \to D\bar D\pi$. Of course, this decay may be separately estimated in perturbation
theory as in Sect. \ref{XdecaySect}. The real part of the scattering amplitude remains unchanged
and its analysis may proceed as for $NN$ scattering. The procedure is described in more detail in
Appendix B.

\subsection{Other Molecules}
\label{OtherMoleculesSect}

Other heavy molecular states can be generated from the one pion exchange mechanism discussed
above. Effective potentials in the various channels are discussed in Appendix B. Isovector 
channels never bind; some properties of isoscalar channels which bind are listed in Table
\ref{MoleculesTab}. 
These results have been computed with $\Lambda = 1.23$ GeV and generally agree with 
T{\"o}rnqvist\cite{NAT} except that I find only one $D^*\bar D^*$
bound state with $J^{PC} = 0^{++}$.

\begin{table}[h]
\caption{Isoscalar Heavy Quark Molecules}
\begin{tabular}{ccccc}
\hline
state & $J^{PC}$ & channels & mass (MeV) & $E_B$ \\
\hline
\hline
$D^*\bar D^*$ & $0^{++}$ & $^1S_0$, $^5D_0$ & 4019 & 1.0 \\
$B \bar B^*$ & $0^{-+}$ & $^3P_0$ & 10543 & 61 \\
$B \bar B^*$ & $1^{++}$ & $^3S_1$, $^3D_1$ & 10561 & 43 \\
$B^*\bar B^*$ & $0^{++}$ & $^1S_0$, $^5D_0$ & 10579 & 71\\
$B^*\bar B^*$ & $0^{-+}$ & $^3P_0$ & 10588 & 62\\
$B^*\bar B^*$ & $1^{+-}$ & $^3S_1$, $^3D_1$ & 10606 & 44 \\
$B^*\bar B^*$ & $2^{++}$ & $^1D_2$, $^5S_2$, $^5D_2$, $^5G_2$ & 10600 & 50 \\
\hline
\hline
\end{tabular}
\label{MoleculesTab}
\end{table}

Note that binding energies in the $BB$ channels are quite large. In particular, the $1^{++}(B\bar B^*)$
has a binding energy of 43 MeV, in contrast to the binding energy of 0.2 MeV predicted by Al Fiky
{\it et al.}\cite{alexey,alexey2} (see Sect. \ref{EFTSect}). The latter is predicated upon an effective field theory
description of $H\bar H^*$ interactions and the scaling relation $C_i \sim 1/m_Q$. This is quite
different from one pion exchange and quark exchange interactions which do not scale with $m_Q$. 
Thus there is a conflict between these approaches (where in principle, none should exist) and further
effort is required to resolve this discrepancy. Certainly, one expects that the $B\bar B^*$ system
should be more strongly bound than $D\bar D^*$ since the kinetic energy is greatly reduced  --
in agreement with the pion exchange model predictions.


The molecular $\hat \chi_{c0}$ (the $J^{PC}=0^{++}$ $D^*\bar D^*$ state)  is interesting 
because it may be accessible in B factories and at
Fermilab. This state  can decay to $\omega J/\psi$, $\eta \eta_c$, and
$\eta' \eta_c$. A rough computation of the partial widths yields

\be
\Gamma(\hat \chi_{c0} \to \omega J/\psi) \approx 200 \ {\rm keV}
\ee
and
\be
\Gamma(\hat \chi_{c0} \to \eta \eta_c) \approx 600 \ {\rm keV}.
\ee
Finally, unlike the case of the $X$, no nearby hidden flavour meson-meson channels
exist for these states. Thus their short range structure is expected to be substantially simpler
than that of the $X(3872)$.

%
%
%
%
%
%

\vfill\eject
\section{Other New Charmonia}  

\subsection{$h_c$}
\label{hcSec}

\begin{figure}[h]
  \includegraphics[width=8 true cm,angle=0]{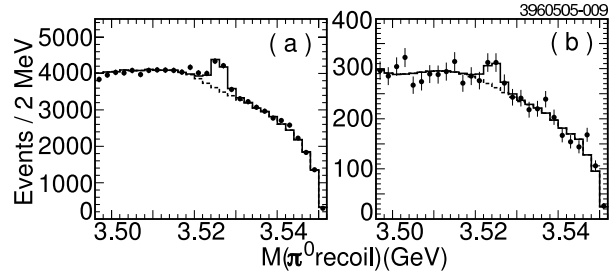}
\caption{Inclusive Production of the $h_c$ in $\psi(2S) \to \pi^0 X$. Monte Carlo (left);
Data (right)\protect\cite{cleoB}.}
\label{hc}
\end{figure}

The $h_c$ has been observed by CLEO\cite{cleoB} in the reaction
$\psi(2S) \to \pi^0 h_c \to (\gamma\gamma)(\gamma \eta_c)$ with a mass of $3524.4 \pm 0.6 \pm 0.4$ MeV
 at a significance greater than 5 $\sigma$. The observation has been made in 
exclusive measurements where the $\eta_c$ has been reconstructed, and in inclusive measurements
(see Fig \ref{hc}).
The mass may be compared to the quark model 
predictions listed in Table \ref{spectrumTab} which range from 3516 to 3526 MeV (or 3474 MeV for quenched lattice computations). Furthermore, the measured product of branching fractions

\be
Br(\psi(2S) \to \pi^0 h_c) Br(h_c \to \gamma\eta_c) = (4.0 \pm 0.8 \pm 0.7) \cdot 10^{-4}
\ee
is in close agreement with  a theoretical estimate of 
$3.8 \cdot 10^{-4}$ of  Ref.~\cite{GR}.
(This reference may also be consulted for an extensive discussion  of spin splittings in
quark models.) Thus it appears that the $h_c$ is conforming to expectations.

Nevertheless, as discussed in Section \ref{spindepSec}, the $h_c$ is an important diagnostic
for the spin-dependent structure of the quark-quark interaction. The fact that it lies
within 2 MeV of the spin-averaged mass of the $\chi_{cJ}$ multiplet:
$M_{c.o.g.} = {1\over 9}(\chi_0 + 5 \chi_1 + 9 \chi_2) = 3525.36$ implies that the 
long range spin-spin interaction $V_4$ is very small, in agreement with the effective scalar confinement
picture.

\subsection{$\eta_c'$}

As discussed above, hyperfine splittings are a useful diagnostic of spin-dependent
confinement interactions. An old result from Crystal Ball\cite{cball} caused some consternation
since it implied a rather large spin splitting in the radial $\psi(2S)-\eta_c(2S)$ system. The
situation has recently changed with new measurements from Belle, CLEO, and BaBar. 

Belle\cite{BelleA} have
observed an $\eta_c'$ candidate in $39 \pm 11$ events in the $K_SK\pi$ system  of the
decay $B \to KK_SK^-\pi^+$. The measured
mass of the resonance is $3654 \pm 6 \pm 8$ MeV and the width was determined to be
$\Gamma = 15^{+24}_{-15}$ MeV ($\Gamma < 55$ MeV at 90\% C.L.).

Belle also measured the ratio of branching fractions

\be
{Br(B \to K \eta_c') Br(\eta_c' \to K_SK^-\pi^+)\over
Br(B \to K \eta_c) Br(\eta_c \to K_SK^-\pi^+)} = 0.38 \pm \pm 0.12 \pm 0.05.
\label{etacRatEq}
\ee
Assuming equal branching ratios for $\eta_c$ and $\eta_c'$ to $K_SK\pi$ (this assumption is 
discussed in Ref. \cite{Chao:1996sf})
gives 
$Br(B \to K\eta_c')/Br(B\to K \eta_c) = 0.38$. This may be compared to the factorisation expression:

\be
{Br(B \to K\eta_c') \over Br(B \to K\eta_c)} = {p_f(\eta_c') \over p_f(\eta_c)}\cdot 
{f_{\eta_c'}^2 \over f_{\eta_c}^2} \cdot {|\xi(w(\eta_c'))|^2 \over |\xi(w(\eta_c))|^2} 
\cdot { \left[(m_B + m_K)^2 - m_{\eta_c'}^2\right]^2 \over
\left[(m_B + m_K)^2 - m_{\eta_c}^2\right]^2}
\label{etacRatThyEq}
\ee
where $w = v \cdot v' = (m_B^2 + m_K^2  - m_\eta^2)/(2 m_B m_K)$.

A quark model computation gives\cite{LS2} 

\be
f_{\eta_c} = 424\ {\rm MeV}
\ee
and
\be
f_{\eta_c'} =  243 \ {\rm MeV}.
\ee
And, using a simple model for the Isgur-Wise form factor $\xi(w) = (2/(1+w))^2$ gives

\be
{Br(B \to K\eta_c') \over Br(B\to K \eta_c)} = 0.38
\ee
in agreement with the data and the assumed equality of branching ratios.

\begin{figure}[h]
  \includegraphics[width=7 true cm, angle=0]{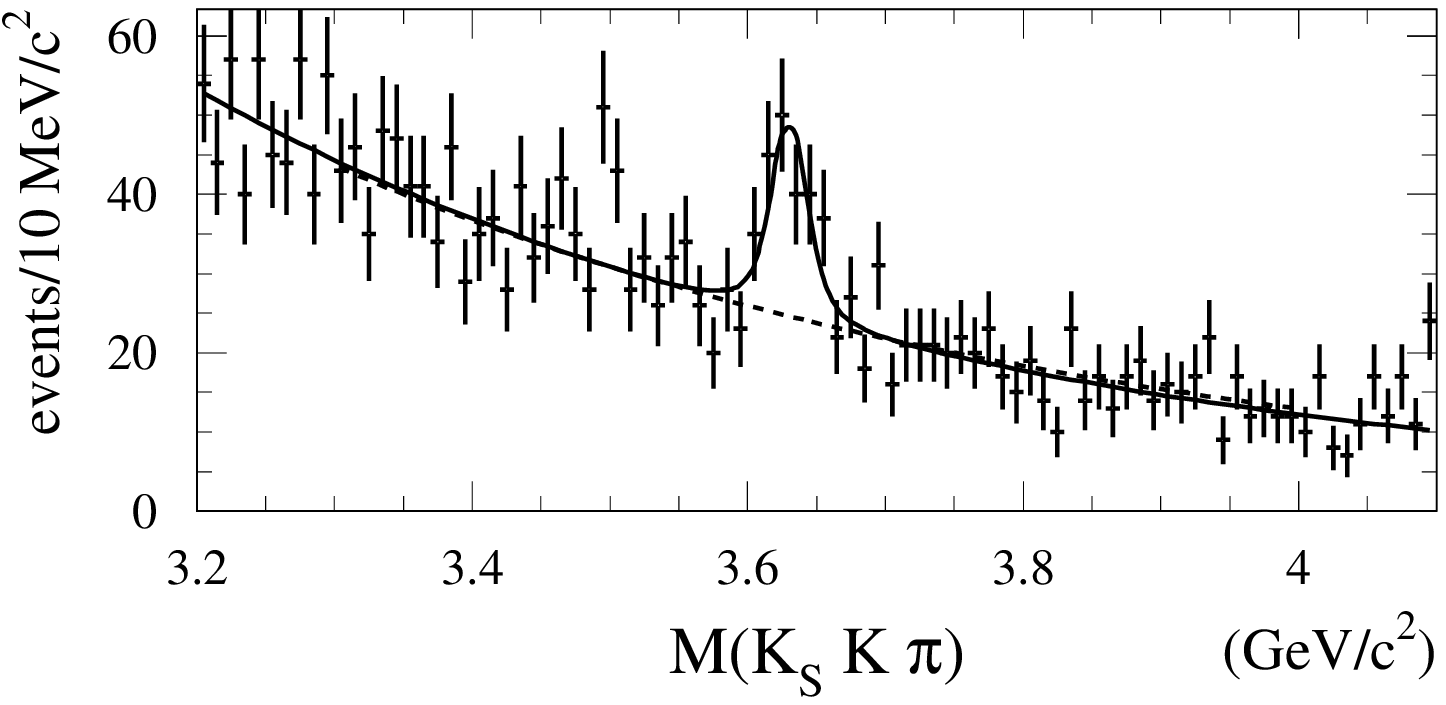}
  \hskip 1 true cm
  \includegraphics[width=7 true cm, angle=0]{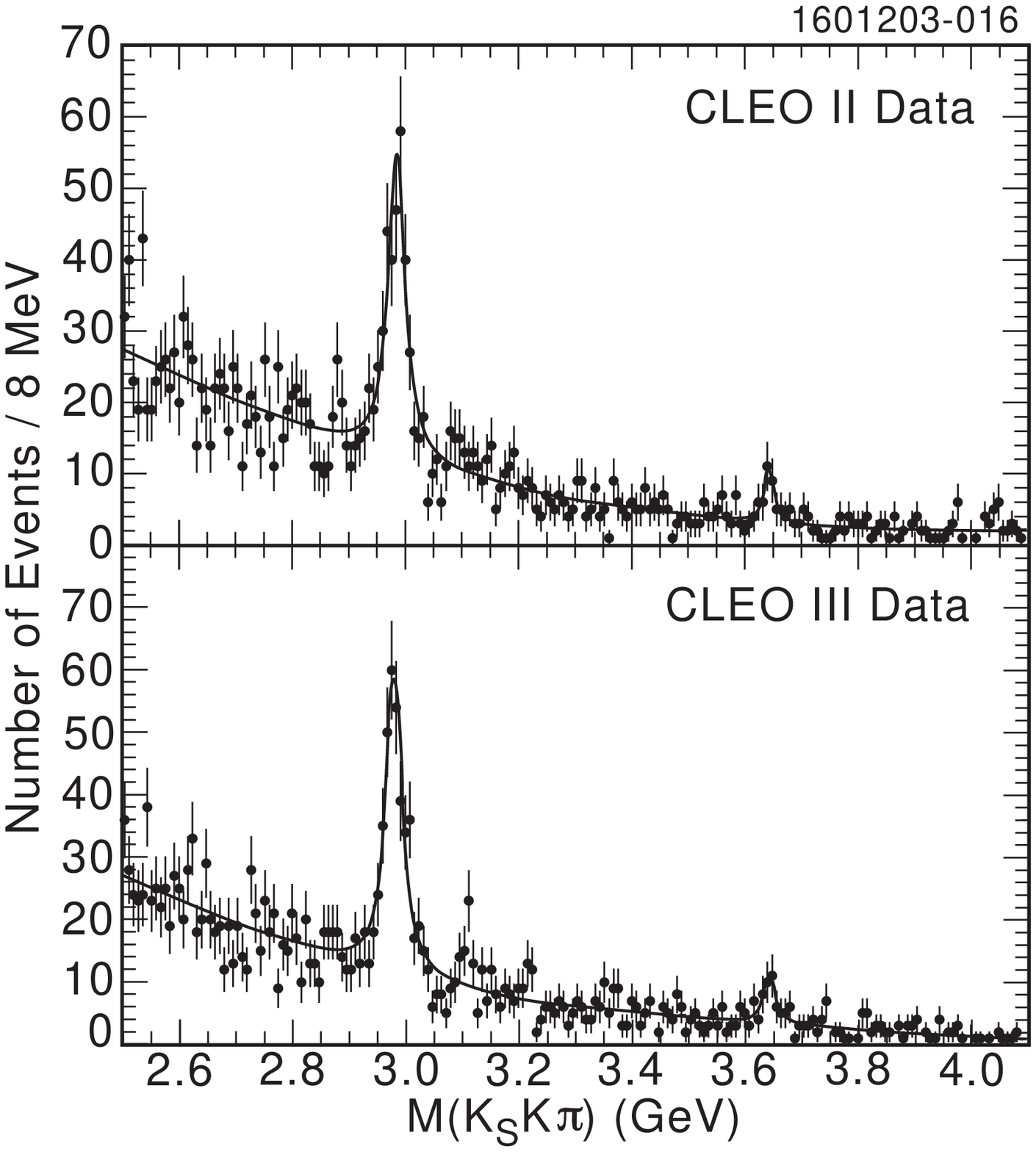}
  \caption{Fit $\eta_c(2S)$ by BaBar\protect\cite{Aubert:2003pt}(left) and CLEO\protect\cite{cleoC}(right). }
 \label{etaFig}
\end{figure}

More recently, the CLEO collaboration\cite{cleoC} have observed the $\eta_c'$  in 
$\gamma\gamma \to \eta_c' \to K_SK^\pm \pi^\mp$ (Fig. \ref{etaFig}; right panel) and find a Breit-Wigner mass and width of
$3642.9 \pm 3.1 \pm 1.5$ MeV  and  
$\Gamma = 6.3 \pm 12.4 \pm 4.0$ MeV respectively. Assuming equal $K_SK\pi$ branching fractions as
above, they also obtain 

\be
\Gamma(\eta_c' \to \gamma\gamma) = 1.3 \pm 0.6 \ {\rm  keV}.
\ee

Finally, BaBar\cite{Aubert:2003pt} have observed an $\eta_c'$ candidate in $112 \pm 24$ events in 
$\gamma\gamma \to \eta_c' \to K_SK\pi$ (Fig. \ref{etaFig}; left panel). They fit a mass of 
$3630.8 \pm 3.4 \pm 1.0$ MeV and obtain a width of $\Gamma = 17.0 \pm 8.3 \pm 2.5$ MeV. 
The angular distribution of the decay products and of the total transverse momentum both suggest 
an assignment of  $J^{PC} = 0^{-+}$
or $J>2$, with the latter being unlikely for such a light state.


As mentioned above, the $\eta_c'$ is of some interest because its splitting with the $\psi'$ is driven by
the hyperfine interaction and hence probes this interaction in a new region. Specifically, the ground
state vector-pseudoscalar splitting is

\be
M(J/\psi) - M(\eta_c) = 117\ {\rm MeV}
\ee
whereas the excited splitting is now measured to be

\be
M(\psi') - M(\eta_c') = 48 \ {\rm MeV}.
\ee
Theoretical expectations for the former range from 108 to 123 MeV in simple (or `relativised') 
quark models (see Table \ref{spectrumTab})  and thus are within expectations. 
Alternatively, the latter is predicted to be 67 MeV in the quark model of Eichten,
 Lane, and Quigg\cite{ELQ}. However, the authors note that including
unquenching effects due to open charm meson loops lowers this splitting to 46 MeV  which is
taken as evidence in favour of their `unquenched' quark model. However, the simple quark models 
mentioned above\cite{BGS} find splittings of 42 - 53 MeV,
indicating that it is too early to make definitive conclusions about loop effects.
It is worth noting, furthermore, that attempts to unquench the 
quark model are fraught with technical difficulty (see Section \ref{spectrumCritiqueSec}) and 
a great deal of
effort is required before we can be confident in the results of any model.

\subsection{$X(3940)$}

The $X(3940)$ is seen by Belle\cite{belleC} recoiling against $J/\psi$ in $e^+e^-$ collisions. The
state has a Breit-Wigner mass of $3943 \pm 6 \pm 6$ MeV and a width of less than 52 MeV at 
90\% C.L.\cite{trabelsi}. The $X$ is
seen to decay to $D\bar D^*$ and not to $\omega J/\psi$ or $D\bar D$. 

\begin{figure}[h]
  \includegraphics[width=8 true cm, angle=0]{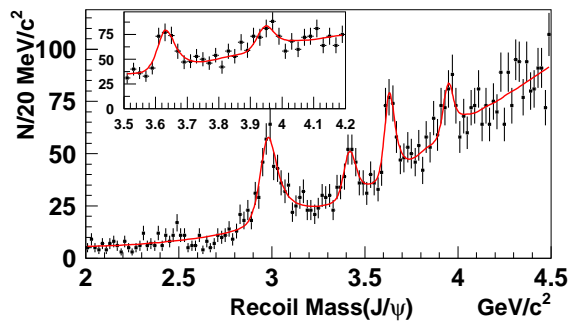}
  \caption{$X(3940)$ Signal in $e^+e^- \to J/\psi +$ charm. Ref. \cite{belleC}.}
  \label{X3940Fig}
\end{figure}

It is natural to attempt a $2P$ $c\bar c$ assignment for this state since the expected mass 
of the $2^3P_J$ multiplet is 3850 -- 3980 MeV (see Table \ref{spectrumTab}) 
and the expected widths are 20 - 130 MeV (see Table \ref{decaysTab}).
Indeed, 
if the $D\bar D^*$ mode is dominant it suggests that the $X(3940)$ is the $\chi_{c1}'$.
There is, however, a problem with this assignment. Fig. \ref{X3940Fig} shows the 
recoil mass spectrum up to 4.5 GeV. The bumps visible in the figure are a 3.3 $\sigma$ fluctuation
at 2500 MeV\footnote{Could this be the tensor glueball?}, the $\eta_c$ at 2980 MeV, a structure usually associated with the $\chi_{c0}(3415)$
(but probably due to $\chi_{c0}$, $\chi_{c1}$, and $\chi_{c2}$), the $\eta_c'$ at 3638, and the
$X(3940)$.  Thus there is no clear evidence for the $\chi_{c1}$,
 and one may suspect that the $\chi_{c1}'$ should also
not be a strong signal in this reaction. This has led to speculation that the $X(3940)$ is the
radially excited $\eta_c''$\cite{Olsen2,rosner}.
Unfortunately this interpretation also has its problems as the expected mass of the 
$\eta_c''$ is 4040 -- 4060 MeV, approximately 100 MeV too high.  An angular analysis of the decay
products of the $X$ will help resolve this issue.

An interesting addendum to the issues surrounding the $X(3940)$ concerns attempts to 
compute the production of the various charmonium states. Table \ref{recoilTab} displays
the predicted cross sections from two NRQCD computations and experimental results from Belle
and BaBar.  It appears that NRQCD is under-predicting $\eta_cJ/\psi$ production cross sections by an
order of magnitude.  However Bondar and Chernyak\cite{Bondar:2004sv} have argued that NRQCD is
not applicable to processes such as $e^+e^- \to J/\psi \eta_c$ because the charm quark is not
sufficiently heavy. Using general methods for hard exclusive processes and a model $J/\psi$ light
cone wavefunction gives the result labelled as BC in Table \ref{recoilTab}. The last line refers
to a light cone computation\cite{BLL} which also produces much larger cross sections than NRQCD.
The authors also not that NLO corrections can be large.

\begin{table}[h]
\caption{Cross Sections (fb) for $e^+e^- \to J/\psi H$ at $\sqrt{s} = 10.6$ GeV.}
\begin{tabular}{lccc}
\hline 
 $H$  & $\eta_c$  & $\chi_{c0}$ & $\eta_c'$ \\
\hline
\hline
BaBar\cite{bernard} &  $17.6 \pm 2.8 \pm 2.1$ & $10.3 \pm 2.5 \pm 1.8$ & $16.4 \pm 3.7 \pm 3.0$ \\
Belle\cite{belleD} &  $25.6 \pm 2.8 \pm 3.4$ & $6.4 \pm 1.7 \pm 1.0$ & $16.5 \pm 3.0 \pm 2.4$ \\
\hline
BL\cite{BL} &  $2.31 \pm 1.09$ & $2.28 \pm 1.03$ & $0.96 \pm 0.45$ \\
LHC\cite{LHC} & 5.5                 & 6.9             & 3.7            \\
BC\cite{Bondar:2004sv} & $\sim 33$ &  &  \\
BLL\cite{BLL} & 26.7 & & 26.6 \\
\hline
\hline
\end{tabular}
\label{recoilTab}
\end{table}

\subsection{$Y(3940)$}

\begin{figure}[h]
  \includegraphics[width=12 true cm, angle=0]{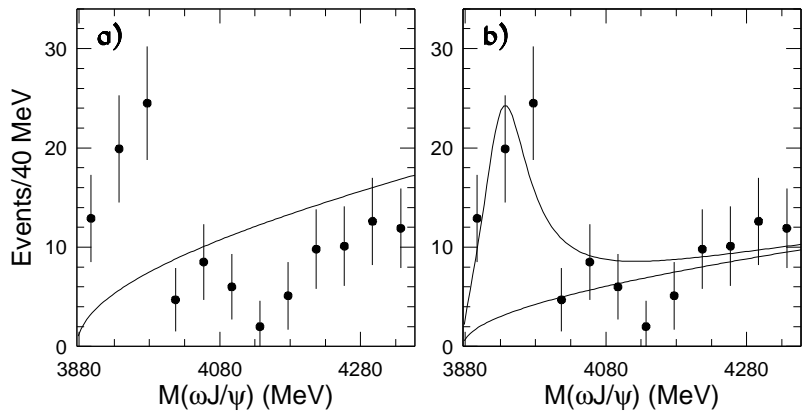}
  \caption{Threshold Enhancement in the $\omega J/\psi$ System (left) and the Breit 
          Wigner Fit (right)\protect\cite{belleB}.}
  \label{Y3940Fig}
\end{figure}

Belle claim the discovery\cite{belleB} of a second resonance at 3940 MeV, this time in the 
$\omega J/\psi$ subsystem of the process
$B \to K\pi\pi\pi J/\psi$. The S-wave Breit Wigner mass and width are (see Fig. \ref{Y3940Fig})  determined to
be $3943 \pm 11 \pm 13$ MeV and $87 \pm 22 \pm 26$ MeV respectively. A total of $58 \pm 11$ events
are claimed for the $Y$ with a significance of 8 $\sigma$.
The state has not been seen in the decay modes $Y \to D\bar D$ or $D\bar D^*$.  Because it does
not share the production or decay channels of the $X(3940)$ it is possible that these states are
distinct.

As for the $X(3940)$, the mass and width of the $Y$ suggest a radially excited P-wave charmonium. However,
the $\omega J/\psi$ decay mode is peculiar.
In more detail, Belle measure 

\be
Br(B \to KY) Br(Y \to \omega J/\psi) = (7.1 \pm 1.3 \pm 3.1)\cdot 10^{-5}.
\ee
One expects that the branching ratio of a radially excited $\chi_{cJ}$ is less than that of the 
ground state $\chi_{cJ}$:  $Br(B \to K \chi_{cJ}') < Br(B\to K\chi_{cJ}) = 4(1)\cdot 10^{-4}$\cite{pdg}.
Thus 
 $Br(Y \to \omega J/\psi) > 17\%$, which is unusual for a canonical $c\bar c $ state
above open charm threshold.
 
Thus the $Y$ is something of an enigma, driving the observation of the Belle collaboration that it
is perhaps a hybrid.  The unusual decay mode and branching fraction are consistent with the claims that
a hybrid should have strongly suppressed $D\bar D$, $D\bar D^*$ and $D^*\bar D^*$ decay modes\cite{hybridDecay}. However, a mass of less than 4000 MeV is in conflict with quenched lattice computations of the
low lying hybrid spectrum\cite{ccgLatt}.
Thus more data are required before the exotic nature of the $Y$ can be claimed with any certainty.

\subsection{$Z(3930)$}

\begin{figure}[h]
  \includegraphics[width=6 true cm, angle=0]{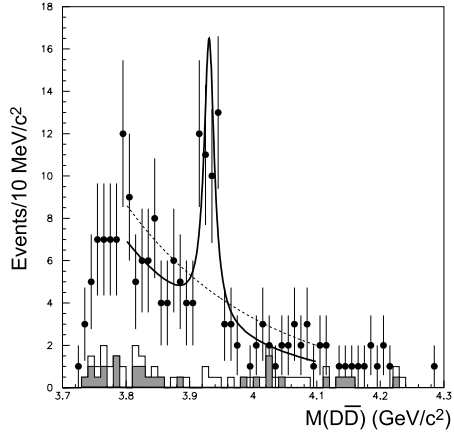}
  \caption{Observation of the $Z(3930)$ in $\gamma\gamma \to D\bar D$\protect\cite{belleZ}.}
  \label{z3930Fig}
\end{figure}

The $Z(3930)$ 
was observed  by the Belle collaboration\cite{belleZ} in $\gamma\gamma \to D\bar D$ with a mass of $3931 \pm 4 \pm 2$ MeV and a width of $20 \pm 8 \pm 3$ MeV at a claimed significance of 5.5 $\sigma$ (see 
Fig.~\ref{z3930Fig}).  The
$D\bar D$ helicity distribution was determined to be consistent with $J=2$.  Due to its
mass and decay mode, the
$Z$ may be considered a $\chi(2P)$ candidate along with the $X(3940)$ and $Y(3940)$. In this case
identification as the $\chi_{c2}'$ is natural.

The predicted 
mass of the $\chi_{c2}'$ is 3970 -- 3980 MeV (Table \ref{spectrumTab}) and the 
predicted width is 60 MeV (Table \ref{decaysTab}).
However, 
setting the mass to the measured 3931 MeV restricts phase space sufficiently that the
predicted strong width drops to 35 MeV, reasonably close to the measurement. The 
predicted branching fraction to $D\bar D$ is 70\%. Furthermore, the largest radiative 
transition is 
$\chi_{c2}' \to \psi' \gamma$ with a rate of $180 \pm 30$ keV\cite{BGS}.

If this assignment is correct it implies a 40 MeV error in the quark model predictions,
somewhat large compared to the average errors shown in Table \ref{errorTab}, and this
is perhaps an indication that the expected breakdown in the model is occurring. 
Furthermore, if the $X(3940)$ is the $\chi_{c1}'$ then these excited $P$ waves exhibit
an inverted spin splitting of -10 MeV rather than the expected  +45 MeV. The 
perturbative $\chi_{c2} - \chi_{c1}$ splitting is given by 
$\langle 16\alpha_s/(5 r^3) -  b/r\rangle/m^2$, so an inverted splitting is possible, but
indicates a serious lack of understanding of the relevant dynamics. If this scenario
is confirmed then the $\chi_{c0}'$ is expected at $M(\chi_{c1}') - 1/2 \langle 8 \alpha_s/r^3 - b/r\rangle/m^2$. This is difficult to estimate in view of the uncertainty in the dynamics. A reasonable guess
may be a 
$\chi_{c0}'$ mass of 3945 MeV.
Alternatively, if
the mass scale has only been misunderstood, then a $\chi_{c0}'$ is expected at approximately 3810 MeV -- it is possible that a glimpse of this state is already visible 
in Fig. \ref{z3930Fig}.  

\begin{figure}[h]
  \includegraphics[width=5 true cm, angle=270]{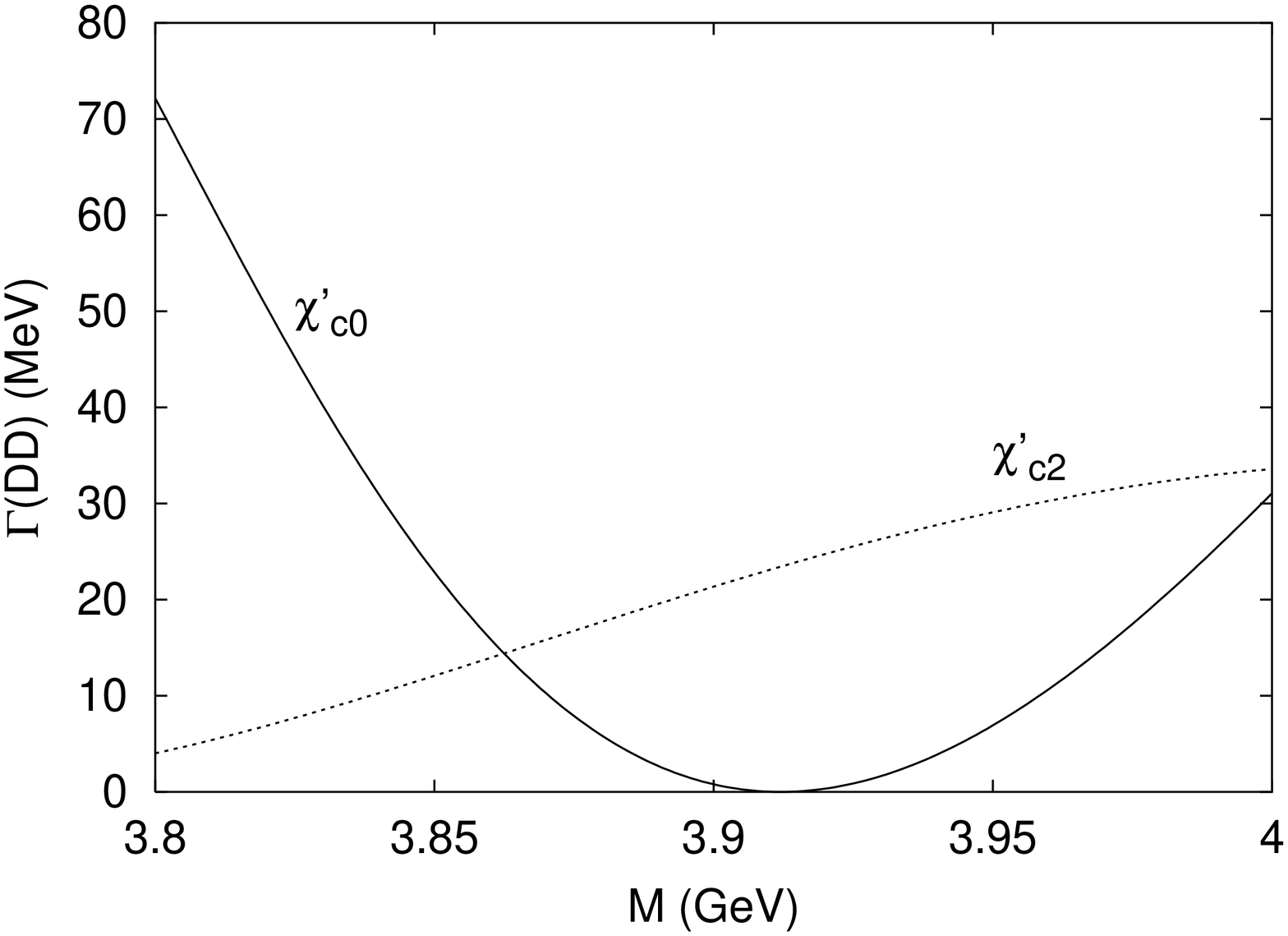}
  \hskip 1 true cm
  \includegraphics[width=5 true cm, angle=270]{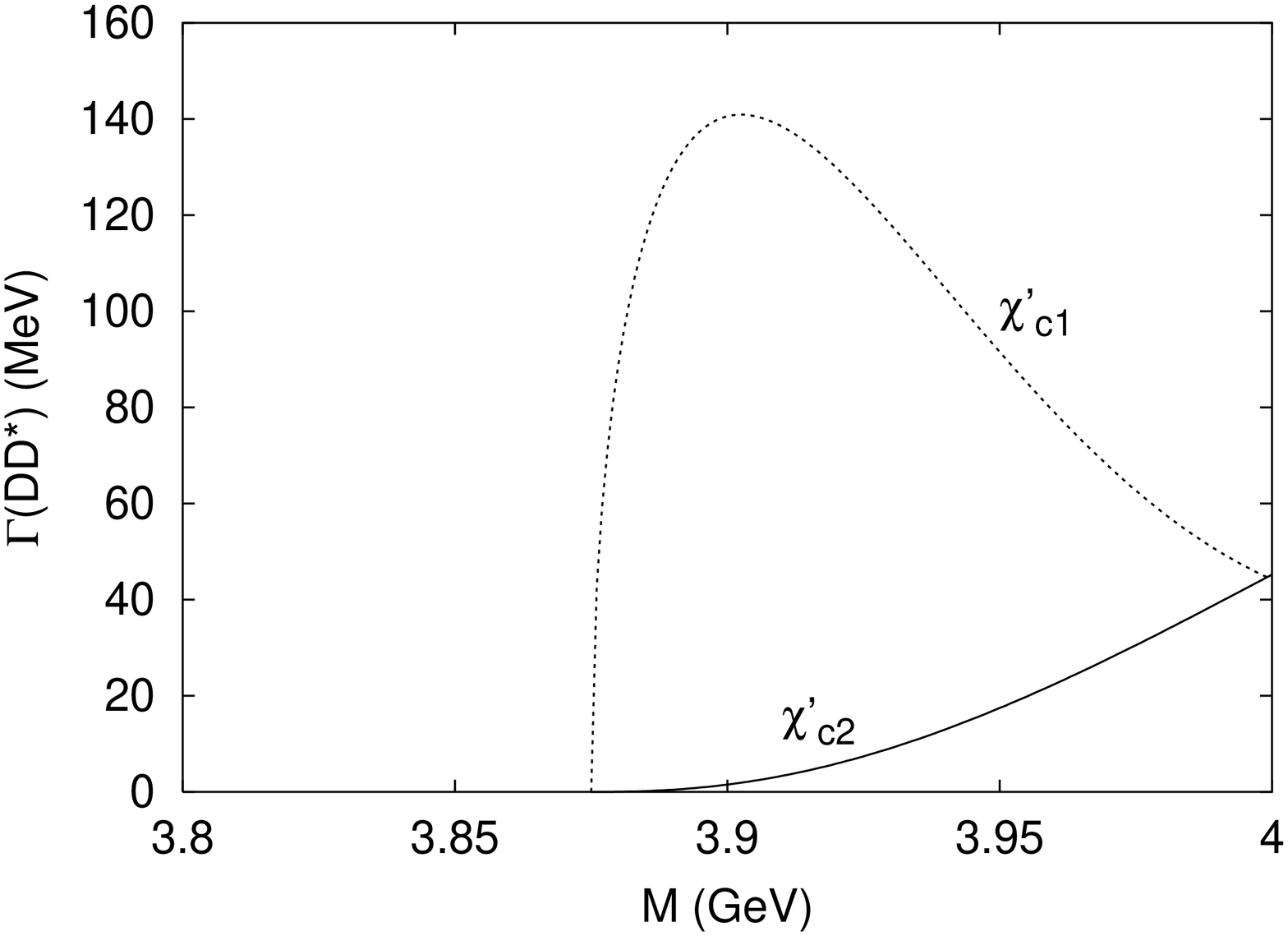}
  \caption{$D\bar D$ and $D\bar D^*$  Partial Widths vs. $\chi_{cJ}'$ Mass.}
  \label{chiWidthsFig}
\end{figure}

In view of the uncertainty in the masses of the $\chi_{cJ}'$ it is worth examining the
partial widths of the $\chi_{c0}'$ and $\chi_{c2}'$ as a function of the resonance
mass. The results are shown in Fig. \ref{chiWidthsFig} and indicate that the
$\chi_{c0}'$ can be very narrow near 3920 MeV. In this case
it may not be a surprise that this state is not strongly visible in Fig. \ref{z3930Fig}.

The production rate in $\gamma\gamma$ has been determined to be

\be
\Gamma_{\gamma\gamma} Br(Z \to D\bar D) = 0.23 \pm 0.06 \pm 0.04 \ {\rm  keV}.
\ee
The $^3P_0$ model results predict a $D\bar D$ branching ratio of 70\% for the
$\chi_{c2}'$ and hence $\Gamma(\gamma\gamma) = 0.33$ keV. This rate has
been estimated by Barnes\cite{barnes} to be 0.64 keV -- given the uncertainties, this 
is reasonably close to the data. 

Searching for $D\bar D^*$ decay modes of the $Z$  will be an important diagnostic for
this state. Measuring helicity angle distributions for the $X$, $Y$, and $Z$ will
also be important for determining quantum numbers. Finally, increasing statistics
for all these states will be useful.

\subsection{$Y(4260)$}

%

\begin{figure}[h]
  \includegraphics[width=7 true cm, angle=0]{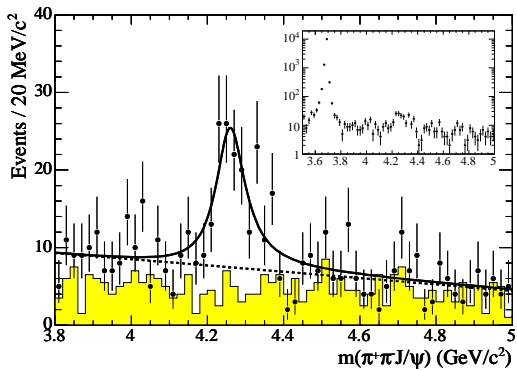}
  \caption{The $Y(4260)$ Observed in $e^+e^- \to \gamma_{ISR}J/\psi\pi\pi$\protect\cite{babarB}.}
  \label{Y4260Fig}
\end{figure}

\begin{figure}[h]
  \includegraphics[width=7 true cm, angle=0]{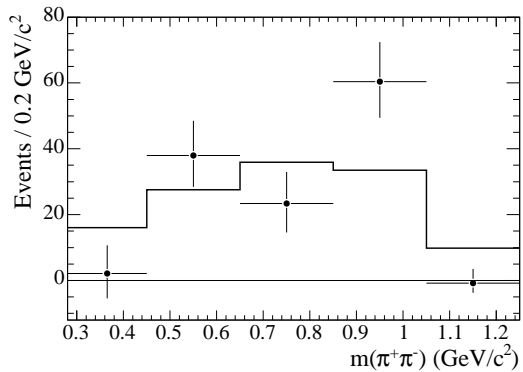}
  \caption{$\pi\pi$ Invariant Mass Distribution\protect\cite{babarB}. The line corresponds to an S-wave phase space model.}
  \label{Y4260pipiFig}
\end{figure}

The $Y(4260)$ is an intriguing recent addition to the charmonium spectrum. It was
discovered by BaBar\cite{babarB} as an enhancement in the $\pi\pi J/\psi$ subsystem of the reaction
$e^+e^- \to \gamma_{\rm ISR} J/\psi \pi\pi$ (see Fig. \ref{Y4260Fig}). The measured  mass is $4259 \pm 8 \pm 4$ MeV and the  width
is  $88 \pm 23 \pm 5$ MeV; with a total of $125 \pm 23$ events in the signal.
The $\pi\pi$ mass distribution peaks near 1 GeV (see Fig. \ref{Y4260pipiFig}), suggesting an isobar $Y \to J/\psi f_0(980) \to J/\psi \pi\pi$ decay (although there is substantial strength at lower invariant mass suggesting a $\sigma$ 
contribution). The reaction rate has also been determined to be

\be
\Gamma(Y\to e^+e^-)\, Br(Y \to J/\psi \pi\pi) = 5.5 \pm 1.0 ^{+0.8}_{-0.7} \ {\rm eV}.
\ee
The $Y$ is not seen in $e^+e^- \to {\rm hadrons}$, in particular in the final states $D \bar D$ or 
$D_s\bar D_s$, indicating that 
the branching fraction of the $Y$ to $J/\psi \pi\pi$ is large compared to that for the
$\psi(3770)$. 

BaBar have also searched for the $Y$ in $B$ decays and report\cite{Aubert:2005zh} 
``an excess of events" around 4.3 GeV at a significance of 3 $\sigma$. The
measured branching ratio is

\be
Br(B^- \to K^-Y) Br(Y \to J/\psi \pi\pi) = (2.0 \pm 0.7 \pm 0.2) \cdot 10^{-5}.
\ee

If the $J/\psi \pi\pi$ mode dominates the width of the $Y$ this implies a $B$ decay 
branching ratio of approximately $10^{-5}$, two orders of magnitude smaller that the
branching fractions for $KJ/\psi$ or  $K\psi'$. Similarly, $\Gamma(Y \to e^+e^-) \sim 5$ eV,
substantially smaller than typical vector leptonic widths.

Evidently the $Y$ is a vector
with $c \bar c$ flavour content. Of course the low lying charmonium vectors are well 
known (see Table \ref{spectrumTab}) and the first
mesonic charmonium vector available is the $3^3D_1$ state.
However quark model estimates of its mass
place it at 4460 MeV, much too heavy for the $Y$. Of course this statement relies on the robustness of 
the quark model. Thus, for example, 
Llanes-Estrada\cite{filipe} has argued that the $Y$ is the $\psi(4S)$ based on 
the spectrum
of a relativistic model. This displaces the $\psi(4415)$ which moves to the $\psi(5S)$ position.

A variety of other explanations for the $Y$ have been proposed. Among these are the suggestion
that the $Y$ is a  $\Lambda_c \bar \Lambda_c$ baryonium\cite{Qiao:2005av}.
Maiani {\it et al.}\cite{maiani} claim the $Y$ is the first orbital excitation of a 
tetraquark $[cs]_S[\bar c \bar s]_S$
state which decays predominantly to $D_s\bar D_s$.  A nonet of vector $Y$ states is predicted with
masses $Y(I=0) = 3910$ MeV and $Y(I=1/2) = 4100$ MeV.
Alternatively, Liu, Zeng, and Li suggest that the $Y$ is a $\chi_{c1}\rho$ molecule bound by $\sigma$
exchange\cite{Liu:2005ay}.
Thus two isotriplet partner states are predicted and possible isotriplet $\chi_{c0}\rho$ and $\chi_{c2}\rho$
states may exist as well.

Of the charmonium states which we know must exist, the most natural explanation is as a 
$c\bar c$ hybrid\cite{Zhu,CP,Kou}. A hybrid assignment agrees with the large $J/\psi\pi\pi$
width\footnote{Although it remains to be established that a large $J/\psi\pi\pi$ mode
is inconsistent with canonical charmonium.}, small leptonic coupling, and small $B$ decay 
branching fraction.
As stated in the Introduction, the
lightest charmonium hybrid is expected at 4400 MeV -- somewhat high, but perhaps acceptable
given our lack of experience in this sector.  The observed decay to a closed charm final state 
is simply explained: all hybrid decay models forbid decays to identical mesons, and many forbid
decays to pairs of S-wave mesons\cite{hybridDecay}. Thus the  allowed channels for a vector hybrid are $D\bar D^*$ (which
is suppressed), $D_2 \bar D$ (a suppressed $D$-wave mode), $D_1\bar D$, and $D_1'\bar D$.  The latter
modes are above threshold and thus are very small.  Ref. \cite{Kou} reminds us of these facts in the
context of a constituent glue model\footnote{Detailed predictions of constituent glue models of
hybrids should be regarded with care since they do not agree with level orderings of adiabatic
hybrid surfaces as computed on the lattice unless specific gluodynamics are assumed\cite{ss3}.}.
Thus the decays of a putative vector hybrid at 4260 MeV will be dominated by closed charm channels. 
The UKQCD collaboration\cite{ukqcd} has found a large coupling of $b\bar b$ hybrids 
to closed flavour modes such as $\chi_b \sigma$. Such a large coupling will be required to accommodate
the relatively large total width of the $Y$. Understanding the mechanisms behind this coupling will
then be a major task for phenomenologists.

If the $Y$ is indeed a vector $c\bar c$ hybrid, it should also decay to $J/\psi \sigma$, $J/\psi K \bar K$, $\chi_{c0}\omega$, 
$h_c \eta$, and $h_c \eta'$ and these modes should be sought. Furthermore, one may expect nearby pseudoscalar and exotic
$J^{PC} = 1^{-+}$ hybrids which will display their own unique decay patterns\cite{CP}.

The proximity of the $Y$ to $D\bar D_1$ threshold (at 4285 MeV) suggests that the resonance could
actually be a cusp at $D\bar D_1$ threshold combined with rescattering into the
$J/\psi f_0 \to J/\psi \pi\pi$ channel. Confirming the phase motion of the $Y$ is thus an important
(although very difficult) task. Searching for a $D\bar D \pi\pi$ decay mode with the $D\pi\pi$
subsystem
dominated by a virtual $D_1$ would also be an important diagnostic. Furthermore, if rescattering
drives the $Y$ enhancement then one may expect similar enhancements at $D\bar D_{s1}$ (and 
similar) thresholds. These may possibly be seen in the $J/\psi K\bar K$ channel.

In line with the previous discussion, it is tempting to speculate on molecular interpretations 
for this state.
In particular the 
$DD_1$ channel is an S-wave threshold and is thus a likely one for forming a molecule.
However, 
if the system does bind, it does so with a novel mechanism since pion exchange
does not lead to a diagonal interaction in this channel (unlike the case of the $X(3872)$).
It is possible that off-diagonal interactions may provide the required novelty. For example, 
$D\bar D_1 \to D^*\bar D^* \to D\bar D_1$ via iterated pion exchange; however, it seems 
unlikely that such a mechanism can lead to binding. Countering this, is a recent 
lattice computation that claims to see a four-quark state at $4238 \pm 31$ MeV and predicts a
$c\bar c s \bar s$ at $4450 \pm 100$ MeV\cite{lattY}.

Lastly, the reader may be disturbed that the $Y$ appears in ISR but is not seen $e^+e^-$
annihilation. This issue has been considered in Ref. \cite{babarB} where the $Y$ cross section
in $e^+e^-$ was estimated to be 50 pb. This corresponds to 0.34\% of the signal at $\sqrt{s} = 4.25$ GeV,
which is not experimentally discernible. Thus the ISR production mechanism serves as a useful 
filter for some vector channels, such as $\pi\pi J/\psi$.

\vfill\eject
\section{New Open Charm States}  

The new revolution in charmonia was contemporaneous with BaBar's discovery of an enigmatic
open charm state, the $D_s(2317)$\cite{babarDS}. 
We shall proceed with a review of theoretical expectations for heavy-light mesons
before moving on to the properties and interpretations of the new open charm states.

\subsection{Heavy Quark Symmetry and the Quark Model}

\subsubsection{Notation}

Generic $Q\bar q$ states will be denoted as $D$ in the following. Properties of the  six lowest lying $D$s are listed in Table \ref{DTab}. Note that charge conjugation is no longer a symmetry in the 
heavy-light sector; thus the $^3P_1$ and $^1P_1$ states mix to form the physical $D_1$ and $D_1'$.
Quark model assignments for the remaining states are as in the quarkonium sector.

In the $m_Q \to \infty$ limit the heavy quark acts as a static colour source and it is useful to 
describe the system in the $jj$ coupling scheme. Thus the light quark degrees of freedom are written
as $j_q = \ell_q + S_q$ where $\ell_q$ ($S_q$) is the angular momentum (spin) of the light quark. 
The total spin-parity is then given by $J^P = \frac{1}{2}^- \otimes j_q^p$.

The heavy-light system can be very different from quarkonium. 
For example, quarkonia contain a hierarchy of scales, $m_Q$, $\langle p \rangle = \alpha_s m_Q$,
$E = \alpha_s^2 m_Q$, etc, whereas a heavy-light system has two scales, $m_Q$ and $\Lambda_{QCD}$. This 
can have profound implications, for example, hyperfine splitting for heavy quarkonium must scale as
$\delta M_{hyp} \sim m_Q$ whereas it scales as $\delta M_{hyp} \sim \Lambda_{QCD}^2/m_Q$ for heavy-light
systems.

\begin{table}[h]
\caption{Low-Lying $Q\bar q$ States}
\begin{tabular}{l|cc|cccc}
\hline
property & $D$ &  $D^*$    & $D_0$  &  $D_1$  & $D_1'$  & $D_2$ \\
\hline
\hline
$J^P$         & $0^-$ & $1^-$ & $0^+$ & $1^+$ & $1^+$ & $2^+$ \\
$^{(2S+1)}L_J$  & $^1S_0$ & $^3S_1-{}^3D_1$ & $^3P_0$ & $^3P_1-{}^1P_1$ & $^3P_1-{}^1P_1$ & $^3P_2$ \\
$\ell_q$ &  0 & 0 & 1 & 1 & 1 & 1 \\
$j_q^p$  &  $\frac{1}{2}^-$  &  $\frac{1}{2}^-$ & $\frac{1}{2}^+$  &  $\frac{1}{2}^+$ & 
$\frac{3}{2}^+$  &  $\frac{3}{2}^+$ \\
decay & --  & -- &  $(DP)_S$  & $(D^*P)_S$  & $(D^*P)_D$ & $(DP)_D$ \\
\hline
\hline
\end{tabular}
\label{DTab}
\end{table}

\subsubsection{Spin Splitting and Multiplet Structure}

The different bases impose different multiplet structures on the $D$ spectrum. In the absence of
spin splitting, the spectroscopic $LS$ basis has an S-wave ($D$,$D^*$) doublet and a P-wave
quadruplet. Alternatively, the heavy quark basis contains three doublets labelled by $j_q^p = 
\frac{1}{2}^-$, $\frac{1}{2}^+$, and $\frac{3}{2}^+$. 

The spin splittings induced in the LS basis may be determined by evaluating matrix 
elements of the interactions of Eq. \ref{VSD}. In the following only terms of order $1/m_q^2$ or 
$1/m_q m_Q$ are retained. The perturbative hyperfine splittings is

\be 
\delta M(D) = -\frac{3}{4}{\langle V_4 \rangle \over 3 m_Q m_q}
\ee

\be 
\delta M(D^*) = +\frac{1}{4}{\langle V_4 \rangle \over 3 m_Q m_q}
\ee

The P-waves are split by tensor and spin-orbit interactions according to

\be
\delta M(D_0) = -{\langle V_{SO}\rangle \over m_q^2}
\ee

\be
\delta M(D_2) = \frac{1}{2}{\langle V_{SO}\rangle \over m_q^2}
\ee

\be
\delta M(^3P_1, {}^1P_1) = -\frac{1}{2}{\langle V_{SO}\rangle \over m_q^2} \pmatrix{1 & -\sqrt{2} \cr
                                                                                     -\sqrt{2} & 0 }
\ee
Here $V_{SO}$ is given by

\be
V_{SO} = { \epsilon' \over 2 r} + {V_1' \over  r} + \frac{1}{2} V_5(m_q,m_Q).
\ee
The sign of the off-diagonal term in the P-wave splitting is convention dependent. Here I have chosen
to work with $q\bar Q$ mesons and to combine angular momentum and spin in the order $L\times S$. 

Tensor splittings  are necessarily of order $1/m_Q m_q$ and are given by

\be
\delta M(D_0) = - 4 {\langle V_3 \rangle \over 12 m_Q m_q}
\ee
                                                     
\be
\delta M(D_2) = - \frac{2}{5} {\langle V_3 \rangle \over 12  m_Q m_q}
\ee

\be
\delta M(^3P_1) = +2 {\langle V_3 \rangle \over 12 m_Q m_q}
\ee

Diagonalising the spin-orbit splittings in the heavy quark limit yields

\be
M(D_2) = M(D_1') = M_0 + {\langle V_{SO}\rangle \over 2 m_q^2}
\ee
and
\be
M(D_0) = M(D_1) = M_0 - {\langle V_{SO}\rangle \over m_q^2}
\ee
in agreement with the $jj$ multiplets. The physical $D_1$ and $D_1'$ states are defined in terms
of a mixing angle:

\begin{eqnarray}
&|D_1\rangle  &= +\cos(\phi) |^1\P_1\rangle + \sin(\phi) | ^3\P_1\rangle \nonumber \\
&|D_1'\rangle  &= -\sin(\phi) |^1\P_1\rangle + \cos(\phi) | ^3\P_1\rangle.
\label{D1}
\end{eqnarray}

In the heavy-quark limit a particular mixing angle follows
from the quark mass dependence of the spin-orbit and
tensor terms, which is $\phi_{HQ} = -54.7^o$ $(35.3^o)$
if the expectation of the heavy-quark spin-orbit interaction
is positive (negative) \cite{GK}. Since the former implies that
the $2^+$ state is greater in mass than the $0^+$ state, and this agrees with
experiment, we employ $\phi = -54.7^o$ in the following.
This implies

\begin{eqnarray}
|D_1 \rangle_{HQ} &=&  +{1\over\sqrt{3}}| ^1P_1\rangle - \sqrt{2\over 3}|^3P_1\rangle \nonumber \\
|D_1' \rangle_{HQ} &=&  +\sqrt{2\over 3}| ^1P_1\rangle + {1\over \sqrt{3}}|^3P_1\rangle.
\label{axmix}
\end{eqnarray}

\subsubsection{Decays}

The P-wave $D$ mesons are sufficiently light that the only isospin conserving decays are to $DP$ or
$D^*P$ where $P$ is a pseudoscalar meson ($\pi$ for $D$ decays and $K$ for $D_s$ decays).
Conservation of angular momentum implies that the $D_0$ can only couple to $DP$ in an S-wave and one
may thus expect the $D_0$ to be broad. Similarly the $D_2$ only couples to $DP$ in a P-wave and
is narrow.

\begin{figure}[h]
  \includegraphics[width=5 true cm, angle=0]{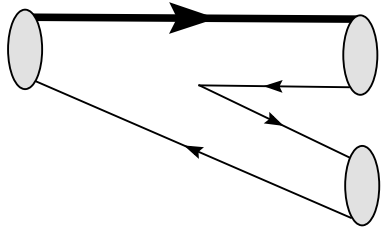}
  \caption{P-wave Decay in the Heavy Quark Limit.}
  \label{DdecayFig}
\end{figure}

In general a $D_1$ or $D_1'$ may decay to $D^*P$ in S- and D-wave. However, the conservation
of heavy quark spin implies that the $D_1$ only decays in an S-wave while the $D_1'$ decays
in a D-wave. Thus the $j_q = \frac{1}{2}$ doublet is broad and the $j_q = \frac{3}{2}$ doublet
is narrow. Fig. \ref{DdecayFig} illustrates this selection rule\cite{GK,HQdecays}.
We consider an initial $D_1^{(\prime)}$ decaying into a $D^*$ and a light 
pseudoscalar meson. Conservation
of angular momentum and heavy quark spin implies that

\be
j_q(D_1^{(\prime)}) = j_q(D^*) + L_{D^*P}.
\ee
Setting $j_q(D^*) = \frac{1}{2}$ and $j_q(D_1) = \frac{1}{2}$ then implies that the relative angular
momentum in the final state is $L_{D^*P} = 0$. Alternatively for the $D_1'$ one has $j_q(D_1') = 
\frac{3}{2}$ and hence $L_{D^*P} = 2$.

It is interesting to examine this behaviour in the $^3P_0$ model. One finds the following relation
between $^3P_1$ and $^1P_1$ decay amplitudes\cite{CS,GK}

\begin{eqnarray}
{\cal A}({}^1P_1 \to V P)_S &=& -{1\over \sqrt{2}} {\cal A}({}^3P_1 \to V P)_S \nonumber \\
{\cal A}({}^1P_1 \to V P)_D &=& \sqrt{2} {\cal A}({}^3P_1 \to V P)_D 
\label{AmpRatsEq}
\end{eqnarray}
Eq. \ref{D1} then implies 

\begin{eqnarray}
{\cal A}(D_1\to V P)_S  &=&  \Big(-{\cos\phi\over \sqrt{2}} + \sin\phi\Big) {\cal A}_S \\
{\cal A}(D_1\to V P)_D  &=&  ( \sqrt{2} \cos\phi + \sin\phi) {\cal A}_D \\
{\cal A}(D_1'\to V P)_S &=&  \Big({\sin\phi\over \sqrt{2}} + \cos\phi\Big) {\cal A}_S \\
{\cal A}(D_1'\to V P)_D &=&  (-\sqrt{2} \sin\phi + \cos\phi) {\cal A}_D
\end{eqnarray}
Substituting the heavy quark mixing angle $\cos\phi = 1/\sqrt{3}$ yields

\begin{eqnarray}
{\cal A}(D_1\to V P)_S  &=&  -\sqrt{3\over 2} {\cal A}_S \\
{\cal A}(D_1\to V P)_D  &=&  0 \\
{\cal A}(D_1'\to V P)_S &=&  0 \\
{\cal A}(D_1'\to V P)_D &=&  \sqrt{3} {\cal A}_D
\end{eqnarray}
in agreement with the heavy quark selection rule.  The agreement is likely quite general
because relations \ref{AmpRatsEq} hold in the Cornell model as well. This may be true because
the nonrelativistic structure of the $^3P_0$ and Cornell transition operators are very similar:

\be
H_{3P0} \sim \sum_k b^\dagger_k \sigma\cdot k d^\dagger_{-k}
\ee
while
\be
H_{\rm Cornell} \sim \mathbb{I}\cdot V(q) \sum_k b^\dagger_{k+q} \sigma\cdot q d^\dagger_{-k}.
\ee

Additional selection rules may be derived\cite{CS}. For example the following relations hold in the
$^3P_0$ model

\begin{eqnarray}
{\cal A}(2{}^3S_1 \to {}^1P_1 P)_S &=& -{1\over \sqrt{2}} {\cal A}(2{}^3S_1 \to {}^3P_1 P)_S \\
{\cal A}(2{}^3S_1 \to {}^1P_1 P)_D &=& \sqrt{2} {\cal A}(2{}^3S_1 \to {}^3P_1 P)_D \\
{\cal A}({}^3D_1 \to {}^1P_1 P)_S &=& \sqrt{2} {\cal A}({}^3D_1 \to {}^3P_1 P)_S
\label{ratios}
\end{eqnarray}
Thus the $D_1 \pi$ ($D_1' \pi$) mode will be large (small) in $D(2{}^3S_1)$ decays
and the $D_1 \pi$ ($D_1' \pi$) mode
will be small (large) in $D(^3D_1)$ decays.

\subsubsection{Spin Splitting with Continuum Mixing}

Although the conclusions reached above seem to be in accord with data (see Sect. \ref{DSect}),
the logic used -- namely 
first perform perturbation theory in the spin orbit interaction,
then use the perturbed states to compute decay widths -- is not correct. In general the coupling
to continuum channels also perturbs the energy and structure of the $D_1$s; hence spin orbit
and loop effects should be considered at the same time. In this case the perturbative mass splittings
 are given  by

\be
\delta M(D_0) = - {\langle V_{SO}\rangle\over m_q^2} + \int {d^3k \over (2\pi)^3} {|\omega_0(k)|^2 \over M_0 - E_{DP}(k)}
\ee

\be
\delta M(D_2) = + {\langle V_{SO}\rangle\over 2 m_q^2} + \int {d^3k \over (2\pi)^3} {|\omega_2(k)|^2 \over M_0 - E_{DP}(k)}
\ee
where $\omega_J(k) = \langle D_J | V | DP;\vec  k\rangle$ and $V$ is, for example, the $^3P_0$ 
transition operator.

The $D_1$ and $D_1'$ eigenstates are determined by diagonalising a Hamiltonian describing the 
interactions of the $^1P_1$, $^3P_1$, $(DP)_S$ and $(DP)_D$ channels.  Integrating out the
continuum channels and examining the perturbative mass shifts yields an effective 2x2 Hamiltonian
to be solved for the physical $D_1$ states:

\be
\delta H = - \frac{1}{2}{\langle V_{SO}\rangle\over m_q^2} \pmatrix{1 & -\sqrt{2} \cr -\sqrt{2} & 0 } + 
\int {d^3k\over (2\pi)^3} {1\over M_0 - E_{D^*P}} \pmatrix{
            (2|\tilde\omega_0(k)|^2 + \frac{1}{2} |\tilde\omega_2(k)|^2) &
            (-\sqrt{2}|\tilde\omega_0(k)|^2 + \frac{1}{\sqrt{2}} |\tilde\omega_2(k)|^2) \cr
            (-\sqrt{2}|\tilde\omega_0(k)|^2 + \frac{1}{\sqrt{2}} |\tilde\omega_2(k)|^2) &
            (|\tilde\omega_0(k)|^2 +  |\tilde\omega_2(k)|^2) }.
\ee
where $\tilde\omega_L(k) = \langle ^1P_1|V|D^*P;k\rangle_L$. 
The continuum contribution to $\delta H$ may be written as

\be
\delta H_{cont} = \int {d^3k\over (2\pi)^3} { |\tilde\omega_0(k)|^2 + |\tilde\omega_2(k)|^2 \over M_0 - E_{D^*P} }
\pmatrix{ 1 & 0 \cr 0 & 1} +
\int {d^3k\over (2\pi)^3} { |\tilde\omega_0(k)|^2 -\frac{1}{2} |\tilde\omega_2(k)|^2 \over M_0 - E_{D^*P} }
\pmatrix{ 1 & -\sqrt{2} \cr -\sqrt{2} & 0} 
\ee
showing that the eigenvalues and mixing angle remain unchanged from the naive mixing case.

The full perturbative mass shifts are

\be
\delta M(D_1) = -{\langle V_{SO}\rangle\over m_q^2} + 3\int {d^3k \over (2\pi)^3} {|\tilde\omega_0(k)|^2 
\over M_0 - E_{D^*P}}
\ee
and
\be
\delta M(D_1') = \frac{1}{2} {\langle V_{SO}\rangle\over m_q^2} + \frac{3}{2}\int {d^3k \over (2\pi)^3} {|\tilde\omega_2(k)|^2 
\over M_0 - E_{D^*P}}
\ee

Finally, in the heavy quark limit $E_{DP} = E_{D^*P}$ and $\omega_0 = -\sqrt{3}\, \tilde\omega_0$.
Thus the $(D_0,D_1)$ doublet is recovered. However, $\omega_2 = - \sqrt{3/5}\, \tilde\omega_2$ and
the heavy doublet $(D_1',D_2)$ is {\it not} recovered. This is not a great surprise -- while it was
established that the heavy doublet was narrow, there is no reason to believe that the widths of
the $D_1'$ and the $D_2$ are identical.

Furthermore,
it should be recalled that all possible virtual continua contribute
to the mass splitting. These continua do not necessarily obey the amplitude relations given in 
Eq. \ref{AmpRatsEq} and the naive heavy quark mixing angle and width relations need not hold. 
Also, the full computation should not be made in perturbation theory because some of the
widths, and hence mass shifts, can be very large.  In this case the concept of the mixing angle is
not valid because the continuum cannot be represented as an effective potential but must be
included in the Fock space expansion of the state

\be
|D_1 \rangle = Z_1^{1/2} |{}^1P_1\rangle + Z_2^{1/2} |{}^3P_1\rangle + \int d^3k Z(k)^{1/2} |D^*P;k\rangle.
\ee
Alternatively, the Hilbert space becomes infinite dimensional rather than simply two dimensional.

\subsubsection{D-waves}

Mixing in the $^3D_2 - {}^1D_2$ system may be addressed in the heavy quark
limit of the constituent quark model as above. The mixing angle may be defined as

\begin{eqnarray}
|D_2^*\rangle &=& + \cos(\phi_D) |^1D_2\rangle + \sin(\phi_D) |^3D_2\rangle \nonumber \\
|{D_2^*}'\rangle &=& -\sin(\phi_D) |^1D_2\rangle + \cos(\phi_D) |^3D_2\rangle
\end{eqnarray} 

Proceeding as with P-waves one finds a $^3D_2-{}^1D_2$ mixing matrix of

\be
M(^3D_2,{}^1D_2) = M_0 - {\langle V_{SO}\rangle \over 2} \pmatrix{ 1 & \sqrt{6} \cr \sqrt{6} & 0 }.
\ee
Diagonalising the spin-orbit interaction yields the results

\begin{eqnarray}
M(^3D_1) &=&  M(D_2^*) = M_0 - {3\over 2} \langle H_{SO}^q\rangle_D \\ \nonumber
M(^3D_3) &=&  M({D_2^*}') = M_0 + \langle H_{SO}^q\rangle_D ,
\end{eqnarray}
and a $^3D_2- {}^1D_2$ mixing angle of $\phi_D = -50.76^o$.
Thus, the heavy quark $D$-wave states are

\begin{eqnarray}
| D_2^* \rangle_{HQ} &=&  \sqrt{2\over 5}| ^1D_2\rangle - \sqrt{3\over 5}|^3D_2\rangle \nonumber \\
| {D_2^*}' \rangle_{HQ} &=&  \sqrt{3\over 5}| ^1D_2\rangle + \sqrt{2\over 5}|^3D_2\rangle.
\label{D2mix}
\end{eqnarray}

As with $P$-waves, the $^3P_0$ strong decay model makes specific predictions for
$D$-wave heavy
quark decay amplitudes which may be useful in interpreting the spectroscopy. Some of
these are:

\begin{eqnarray}
{\cal A}({}^1D_2 \to V P)_P &=& -\sqrt{2\over 3} {\cal A}({}^3D_2 \to VP)_P \\
{\cal A}({}^1D_2 \to V P)_F &=& +\sqrt{3\over 2} {\cal A}({}^3D_2 \to VP)_F \\
{\cal A}({}^1D_2 \to {}^1P_1 P)&=& 0
\end{eqnarray}
The second of these is an example of the $^3P_0$ selection rule forbidding such
transitions among $q\bar q$ spin singlets.

We note that, in analogy with the $P$-waves, the amplitude ratios above imply that the $D_2^*$
decays strongly in $P$ wave while the ${D_2^*}'$ decays only in $F$ wave and is thus
narrower than the $D_2^*$. As with $P$-waves, this conclusion agrees with spin
conservation in the heavy quark limit. Unfortunately, the ability to distinguish
the states is weakened by the many
other decay modes which exist for these states.
Nevertheless,
the $D_2^*$ and ${D_2^*}'$
have total widths which depend strongly on $\phi_D$ and measurement of any (or several)
of the larger decay modes will provide (over) constrained tests of the model and
measurements
of the mixing angle.

\subsection{Chiral Symmetry Breaking and Restoration in Heavy-Light Mesons}

\subsubsection{Effective Field Theory}

Ten years prior to the discovery of the $D_s(2317)$, Nowak {\it et al.}\cite{NRZ} and 
Bardeen and Hill\cite{BH} investigated
the consequences of an effective lagrangian description of heavy-light mesons, called the chiral
doublet model.  Nowak {\it et al.}
assumed heavy quark symmetry and broken chiral symmetry and thus arrived at the lagrangian of
Eq. \ref{Lagr2}. Alternatively, Bardeen and Hill assumed that chiral symmetry was manifest 
and thus wrote an analogous lagrangian with the linear realisation of chiral symmetry.
The properties of heavy-light mesons have also been examined in the context of a Nambu--Jona-Lasino
model with chiral and heavy quark symmetries by Ebert {\it et al.}\cite{EbertNJL}.

The combination of heavy quark and chiral symmetries imply that the bare low-lying
heavy-light states are degenerate multiplets, $H$ and $G$ (or $H'$) where 

\be
H = (0^-,1^-)
\ee
and 
\be
G = (0^+,1^+).
\ee
Note that this axial is the $j_q^p = \frac{1}{2}^+$ state. Heavy quark symmetry imposes the
doublet structure shown, while chiral symmetry implies that they are degenerate. This structure
is modified in two ways: the heavy quark symmetry is explicitly broken by the finite heavy quark
mass and chiral symmetry is spontaneously broken by nonperturbative vacuum gluodynamics 
(explicit chiral symmetry breaking is neglected). The subsequent mass splittings raise the $G$
multiplet above the $H$ by roughly the chiral symmetry breaking scale and induce {\it identical
hyperfine splittings} in the $G$ and $H$ multiplets. Thus one has the prediction

\be
M(D_1) - M(D^*) = M(D_0) - M(D) \approx \Lambda_{QCD}.
\label{chiralDoubletEq}
\ee

Both sets of authors have subsequently applied the formalism to the newly discovered $D_s$
states\cite{chiralDS}. As we shall see, Eq. \ref{chiralDoubletEq} is quite accurate:

\be
M(D_{s1}) - M(D_s^*) = 347.2 \pm 1.5 \ {\rm MeV}, \qquad\qquad M(D_{s0}) - M(D) = 349.1 \pm 1.0 \ {\rm MeV}.
\ee
Additional predictions are\cite{chiralDS}

\be
M(D_0) = 2215 \ {\rm MeV}
\ee
and
\be
M(D_1) = 2357 \ {\rm MeV}.
\ee
Experimentally these are $2352 \pm 50$ MeV and $2421.8 \pm 1.3$ MeV respectively\cite{pdg}. 

In the $B$ system the predictions are\cite{chiralDS}

\be
M(B_0) = 5627 \pm 35 \ {\rm MeV}
\ee
and
\be
M(B_1) = 5674 \pm 35 \ {\rm MeV}.
\ee
The PDG reports several narrow and broad resonances of undetermined quantum numbers at 5698 MeV, in 
rough agreement with these predictions.

The chiral nature of the lagrangian also permits computations of strong transitions involving Goldstone
bosons.  Indeed, 
parity doubling implies that the transition strength governing  $(0^+,1^+) \to (0^-,1^-) \pi$ is related
to the scalar-pseudoscalar mass difference by a Goldberger-Treiman relationship

\be
g_\pi f_\pi = M(0^+) - M(0^-)
\ee
where $f_\pi$ is the pion decay constant. Unfortunately, these decays are kinematically forbidden for
the $D_s$ system  and they can only proceed via isospin breaking $D_s \to D \eta \to D \pi^0$ transitions.
Similarly couplings to multiple pions may be computed; for example 
Bardeen {\it et al.}\cite{chiralDS} predict 
\be
\Gamma(D_{s1} \to D_s\pi\pi) = 4.2 \ {\rm keV}.
\ee

It is worth stressing that the relationship shown in Eq. \ref{chiralDoubletEq}  is very difficult to 
obtain in the 
constituent quark model. In particular, the $D_0$ and $D_1$ are P-wave excitations with
{\it small hyperfine splittings}. Thus the two sides of the equation are not related in any simple way.
This issue will be examined in detail in the following section.

\subsubsection{An Explicit Chiral Doublet Model}
\label{DoubletModelSect}

The speculation that chiral symmetry is relevant to heavy-light mesons is a vital part of
the preceding arguments. While general arguments have been made that this should be true high in the
meson or baryon spectra\cite{chiral2}, it is not clear that chiral symmetry is relevant to 
low lying states.
Thus it is useful to examine the 
issue of chiral symmetry restoration in the meson spectrum with the
aid of a simple relativistic Nambu--Jona-Lasinio quark model. The model is defined by the Hamiltonian

\be
H = \int d^3x\, \psi^\dagger (-i \alpha \cdot \partial + \beta m) \psi + \frac{1}{2}\int d^3x d^3y\, \rho^a(x) V(x-y) \rho^a(y)
\ee
where $\beta$ is the Dirac matrix $\gamma_0$, $\vec \alpha = \beta \vec \gamma$, and $\rho^a$ is the quark charge density current. 

Chiral symmetry breaking occurs for a wide range of interactions, $V$ and is embodied in the gap equation\cite{LS}:

\begin{equation}
Z_\psi(\Lambda) p s_p = \left[m(\Lambda) + \delta m(\Lambda)\right]c_p +
{C_F \over 2} \int {d^3k \over (2\pi)^3} \tilde V(p,k)\left[ s_k c_p - \hat p \cdot \hat k\, c_k s_p\right].
\label{qgap}
\end{equation}
The functions $s_k$ and $c_k$ are defined in terms of the Bogoliubov angle $\phi(k)$
as $s_k = \sin\phi(k)$ and $c_k = \cos \phi(k)$. The constant $Z_\psi$ and the UV cutoff $\Lambda$ 
are required for renormalisation. However, this will not be discussed further since it is irrelevant to
the following.

The quark gap equation is to be solved for the unknown Bogoliubov angle, which
then specifies the quark
vacuum and the quark field mode expansion via spinors of the form

\begin{equation}
u_s(k) = \sqrt{1 + s_k \over 2} \left( \begin{array}{c}
                                     \chi_s  \\
                                      {c_k\over 1 + s_k}  {\bm{\sigma} \cdot \hat k} \chi_s
                                 \end{array}
                                  \right).
\end{equation}
Comparing the quark spinor
to the canonical spinor (nonrelativistic normalisation is used) permits a simple
interpretation of the Bogoliubov angle through the relationship $\mu(k) = k \tan\varphi(k)$ where $\mu$ may
be interpreted as a dynamical momentum-dependent quark mass. Similarly $\mu(0)$
may be interpreted as a constituent quark mass.

Chiral and heavy-quark symmetry can be studied by projecting the Hamiltonian on pseudoscalar,
scalar, vector, and axial vector $Q\bar q$ channels. For example, the pseudoscalar and scalar
Tamm-Dancoff Salpeter  equations are

\be
M_{0^-}\phi(k) = (E_k + \bar E_k)\phi(k) + \frac{1}{2}\int {d^3 p \over (2\pi)^3} \left[a \hat p\cdot \hat k + b\right]\tilde V(p-k) \, \phi(p)
\ee
and
\be
M_{0^+}\phi(k) = (E_k + \bar E_k)\phi(k) + \frac{1}{2}\int {d^3 p \over (2\pi)^3} \left[b \hat p\cdot \hat k + a\right]\tilde V(p-k) \, \phi(p).
\ee
The factors
in the interaction are given by

\be
a = \sqrt{1-s_q}\sqrt{1-s_k}\sqrt{1+\bar s_q}\sqrt{1+\bar s_k} + \sqrt{1+s_q}\sqrt{1+s_k}\sqrt{1-\bar s_q}\sqrt{1-\bar s_k}
\ee
and
\be
b = \sqrt{1+s_q}\sqrt{1+s_k}\sqrt{1+\bar s_q}\sqrt{1+\bar s_k} + 
\sqrt{1-s_q}\sqrt{1-s_k}\sqrt{1-\bar s_q}\sqrt{1-\bar s_k}.
\ee
The function $s_k$ is related to the dynamical quark mass via
$s_k = \mu(k)/\sqrt{k^2 + \mu^2(k)}$.

The structure of these equations simplifies substantially in the nonrelativistic limit, 
$s_k, \bar s_k \to 1$.
In this case the pseudoscalar interaction reduces to $2\int V \phi$ while the scalar interaction reduces
to $2 \int \hat p \cdot \hat k V \phi$. Thus the pseudoscalar interaction is S-wave, while the scalar
interaction is P-wave and one expects the scalar mass to be larger than the pseudoscalar.  Of course
these conclusions are in agreement with constituent quark models. 

Alternatively,  these equations reduce to 

\be
M_{0^\pm}\phi(k) = (E_k + \bar E_k)\phi(k) + \int {d^3 p \over (2\pi)^3} \left[\hat p\cdot \hat k + 1\right]\tilde V(p-k) \, \phi(p)
\ee
in the heavy quark ($s_k \to 1$) and chiral ($\bar s_k \to 0$) limits.  A similar results applies to the
axial and vector equations.
Thus the expected multiplet structure is recovered. 


The  effects of spontaneous chiral symmetry breaking may be examined by 
setting $\bar s_k = \mu(k)/\bar E(k)$. In this case the $(D_0,D_1)$ doublet is split from
the $(D,D^*)$ doublet. Hyperfine interactions arising from a finite heavy quark mass then
split the doublets. Bardeen and Hill assume that this splitting is identical in both 
doublets, leading to Eq. \ref{chiralDoubletEq}. This is consistent with the analysis here
if the chiral splitting is small, or equivalently, $\langle s_k\rangle \ll 1$. However,
experience with light mesons indicates that
$\langle s_k\rangle \approx 1$ (or $\langle k \rangle << \mu(0)$).  
If this is also true in heavy-light mesons (and we shall argue that it is) then the 
chirally split $D_0$ and $D$ are dominantly P- and S-wave respectively, and hence have
very different hyperfine splittings and the conclusions of \cite{chiralDS, NRZ, BH} are incorrect.
In general one expects $\langle k \rangle \sim \mu(0) \sim \Lambda_{QCD}$ so that 
$\langle s_k \rangle = {\cal O}(1)$. Alternatively, the function $\mu$ is 
universal (it is determined by the structure of the vacuum) and hence the ratio

\be
{\langle k \rangle|_{B^*}\over \langle k \rangle|_\rho} \sim \Big({\mu(B^*)\over \mu(\rho)}\Big)^{1/3} \sim  (2)^{1/3} \sim 1.
\ee

In summary, the scenario of Nowak {\it et al.} and Bardeen {\it et al.}  relies on effective chiral symmetry restoration in low-lying heavy-light
mesons but there is no reason to believe that this actually occurs. Of course this is a theoretical
belief which flies in the face of the data (see the subsequent discussion); however, one must allow
the possibility that other effects may exist which can explain the same data.

Finally, 
Jaffe {\it et al.} claim\cite{Jaffe:2005sq} that the notion of the irrelevance of chiral symmetry
breaking in the spectrum cannot be correct under any circumstances and speculate that 
hints of parity restoration 
in the baryon spectrum are instead due to $U_A(1)$ restoration. These observations are at odds with
those of Ref. \cite{chiral2} and the simple model presented here.

\subsection{$D_s(2317)$ and $D_s(2460)$}
\label{DSect}

\subsubsection{Experiment}

\begin{figure}[h]
  \includegraphics[width=5 true cm, angle=0]{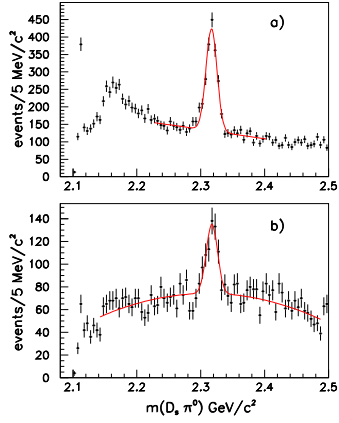}
  \caption{$D_s\pi$ Mass Distribution for (a) $D_s \to KK\pi$ and (b) $D_s \to KK\pi\pi$.}
  \label{babarDSFig}
\end{figure}

BaBar's announcement\cite{babarDS} of a narrow low mass state seen in the $D_s\pi^0$ decay mode
has generated renewed enthusiasm for open charm spectroscopy -- in large because of the curious
properties of the new state\cite{Colangelo:2004vu}.
 Fig. \ref{babarDSFig} shows fits to the $D_s\pi$ mass distributions 
in the $KK\pi\pi$ and $KK\pi\pi\pi$ modes. The resulting resonance masses are

\be
M(D_s) = 2316.8 \pm 0.4 \pm 3\ {\rm MeV}
\ee

\be
M(D_s) = 2317.6 \pm 1.3  \ \rm{MeV}
\ee
respectively.
The respective widths were determined to be $8.6 \pm 0.4$ MeV and $8.8 \pm 1.1$ MeV. Presumably
these are consistent with detector resolution since BaBar merely conclude that the width
of the $D_s(2317)$ is less than 10 MeV (no confidence interval is reported).

Evidently the new state is a $D_s$ meson which is decaying via
an isospin violating mode. Since its mass is below the well-known 
$D_{s1}(2536)$ and $D_{s2}(2573)$, it suggests that the $D_s(2317)$ is a $J^P = 0^+$ state. This 
assignment disallows the decay mode $D_s(2317) \to D_s \gamma$, which is consistent with the 
lack of signal reported in this channel in Ref. \cite{babarDS}. Similarly, the $D_s(2317)$ was
not seen in the $D_s\gamma\gamma$ decay mode.

\begin{figure}[h]
  \includegraphics[width=5 true cm, angle=0]{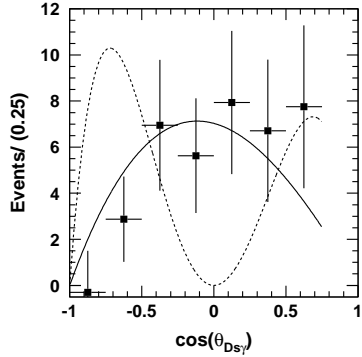}
  \caption{The $D_s(2460) \to D_s\gamma$ Helicity Distribution. Solid and dashed lines are Monte Carlo
predictions for the $J=1$ and $J=2$ hypotheses respectively\protect\cite{belleDS}.}
  \label{DShelFig}
\end{figure}

A search for the $D_s(2317)$ was also carried out in the $D_s \pi^0 \gamma$ mode. No signal
at 2317 MeV was observed; although a peak at 2460 MeV was commented on. The new state was
subsequently
confirmed by CLEO\cite{cleoDS} in the $D_s^*\pi^0$ mode.
Further data taking by BaBar\cite{babarDS2} yielded the following improved mass measurements:

\be
M(Ds(2317)) = 2318.9 \pm 0.3 \pm 0.9\ {\rm MeV}
\ee
and for the $D_s(2460)$\cite{babarDS2460}:

\be
M(D_s(2460) \to D_s\gamma)  = 2457.2 \pm 1.6 \pm 1.3\ {\rm MeV},
\ee
\be
M(D_s(2460) \to D_s\pi^0\gamma)  = 2459.1 \pm 1.3 \pm 1.2\ {\rm MeV},
\ee
\be
M(D_s(2460) \to D_s\pi^+\pi^-)  = 2460.1 \pm 0.3 \pm 1.2\ {\rm MeV}.
\ee
The combined result of the latter three measurements is

\be
M(D_s(2460))  = 2459.4 \pm 0.3 \pm 1.0\ {\rm MeV}
\ee

As seen in Fig. \ref{DShelFig}, the $D_s\gamma$ helicity distribution strongly prefers a $J=1$ 
assignment for the $D_s(2460)$\cite{belleDS}. Thus the low-lying $D_s(P)$ quadruplet
is complete.

The following ratios of branching fractions were also obtained:

\be
{Br(D_s(2460) \to D_s\gamma)\over Br(D_s(2460) \to D_s\pi^0\gamma)} = 0.375 \pm 0.054 \pm 0.057,
\ee

\be
{Br(D_s(2460) \to D_s\pi^+\pi^-)\over Br(D_s(2460) \to D_s\pi^0\gamma)} = 0.082 \pm 0.018 \pm 0.011,
\ee

\be
{Br(D_s(2460) \to D_s\gamma) \over Br(D_s(2460) \to D_s\pi^0)} < 0.17,
\ee

\be
{Br(D_s(2460) \to D_s\pi^0) \over Br(D_s(2460) \to D_s\pi^0\gamma)} < 0.11,
\ee
and

\be
{Br(D_s(2460) \to D_s(2317)\gamma) \over Br(D_s(2460) \to D_s\pi^0\gamma)} < 0.23.
\ee
The latter three are 95\% C.L.

Recently, 
Belle\cite{belleBD, Abe:2005zy} has reported the observations of the new $D_s$'s in $B$ decays.
The D$_{s0}^*(2317)$ is seen in the process 
B$^0 \to {\rm D}_{s0}^{*+}(2317) {\rm K}^-$, 
D$_{s0}^{*+}(2317)\to {\rm D}_s^+ \pi^0$, and the D$_{s1}^+(2460)$ 
may have been seen (at a much lower rate) in a similar chain to 
${\rm D}_s^+ \gamma$. The reported branching fractions are

\begin{equation}
Br({\rm B}^0 \to {\rm D}_{s0}^{*+}(2317) {\rm K}^-)
\cdot
Br({\rm D}_{s0}^{*+}(2317) \to {\rm D}_s^+ \pi^0)
= (4.4 \pm 0.8 \pm 0.6 \pm 1.1) \cdot 10^{-5},
\label{B_2317_BF}
\end{equation}
\begin{equation}
Br({\rm B}^0 \to {\rm D}_{s1}^{*+}(2460) {\rm K}^-)
\cdot
Br({\rm D}_{s1}^{*+}(2460) \to {\rm D}_s^+ \gamma)
= \Big(0.53 \pm 0.20 \pm 0.6 {+0.16 \atop -0.15}\Big) \cdot 10^{-5}.
\label{B_2460_BF1}
\end{equation}
One may compare these rates to those for $B \to D_sK$ with a 
branching fraction of roughly $4 \cdot 10^{-5}$. 
These processes can occur via W exchange or final state interactions. The tree diagram can
also contribute to a four-quark component of the $D_s$. Assuming that the $D_s\pi^0$ mode
dominates the width of the $D_s(2317)$ then implies that
the ratio

\be
{Br(\bar B^0 \to D_s (\pi,K,D)^-) \over Br(\bar B^0 \to D_{s0} (\pi,K,D)^-) } \approx 1.
\ee
This is in agreement with expectations 
for $c\bar s$ states\cite{ChenLi}. Alternatively this ratio is expected to be 10 for multiquark states.

Finally, the reported yields for possible charge partners of the $D_s(2317)$ are
$-28 \pm 25$ for 
$D_s^0(2317)$  and $-39 \pm 16$ for $D_s^{++}(2317)$. Thus tetraquark models (to be discussed below) 
of the $D_s(2317)$ appear to be ruled out.
Also we note that angular analysis\cite{drut} of $D_{s1}(2536) \to D^{*+}K_S$ indicates the 
presence of a large S-wave amplitude. This observation counters the notion that the narrow $j_q = 3/2$ 
states decay in D-wave.

\subsubsection{Models}

\begin{figure}[h]
  \includegraphics[width=7 true cm, angle=0]{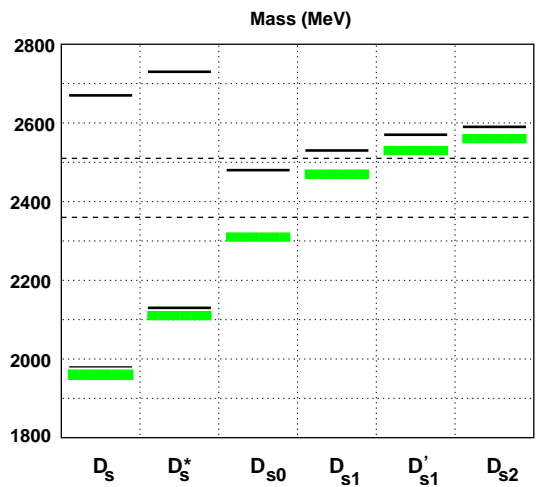}
  \caption{The Low Lying $D_s$ Spectrum. Lines are theory from Ref.\protect\cite{GI}, boxes are experiment. 
The dashed lines show the $DK$ and $D^*K$ thresholds\protect\cite{CS}.}
\label{DsSpectFig}
\end{figure}

The $D_s(2317)$ lies approximately 160 MeV below the majority of model computations. A selection
of such model predictions and a summary of the experimental masses are given in Table \ref{DSTab}.
One sees good agreement between data and models for the $j_q = 3/2$ multiplet and poor agreement 
for the $j_q = 1/2$ multiplet. This is also illustrated in Fig. \ref{DsSpectFig}.  The poor
agreement calls into question the $c\bar s$ interpretation of the new $D_s$ states; however this
interpretation has received support from light cone sum rule decay computations\cite{Colangelov}.
A comprehensive
analysis of the new states in the $c\bar s$ model is given by Cahn and Jackson\cite{CJ}.

The majority of models listed in Table \ref{DSTab} predict a $D_{s0}$ between 2400 and 2500 MeV.
The model of Matsuki and Morii\cite{Matsuki:2005vs} is unique in predicting the 
$D_{s0}$ and $D{s1}$ masses approximately correctly. The model employs a 
Foldy-Wouthuysen nonrelativistic 
reduction of an assumed scalar confining interaction and vector Coulomb interaction\footnote{The full vector current is used. 
Also, the authors found it was not possible to extract $^3D_1$ and $^3D_2$ masses, although recent
computations have resolved this issue\cite{MMprivate}.}. It is difficult to isolate the cause of the
good agreement -- it may due to relativistic light quark kinematics or the novel assumed spin
structure of the static interaction.  Further predictions of this model are discussed in Sect.
\ref{PBSect}. Finally, Godfrey discusses using radiative transitions to test models\cite{God2}.

\begin{table}[h]
\caption{Theory and Experimental $D_s$ Masses}
\begin{tabular}{lllll}
\hline
$D_{s0}$ &  $D_{s1}$  & $D_{s1}'$  &  $D_{s2}$   & ref \\
\hline
\hline
2480       &   2550   &     2550  &  2590         & GK\cite{GK} \\
2380       &   2510   &     2520  &   2520        & ZOR\cite{ZOR}  \\
2388       &   2521   &    2536   &   2573        & GJ\cite{GJ} \\
2508       &   2569   &    2515   &   2560        & EGF\cite{EGF} \\
2455       &   2502  &     2522   &    2586       & LNR\cite{LNR} \\
2487       &   2605  &     2535   &   2581        & DE\cite{DE} \\
2339       &   2496  &     2487   &   2540        & MM\cite{MM} \\
\hline
$2317.3 \pm 0.4 \pm 0.8$ &   $2458.0 \pm 1.0 \pm 1.0$  & &    &     BaBar\cite{babarDS} \\
$2317.2 \pm 0.5 \pm 0.9$ &   $2456.5 \pm 1.3\pm 1.3$   &  &  &   Belle\cite{belleDS} \\
$2318.5 \pm 1.2 \pm 1.1$ &   $2463.6 \pm 1.7\pm 1.2$  &  &   &    CLEO\cite{cleoDS} \\
$2317.4 \pm 0.9$      & $2459.3\pm 1.3$ &  $2535.35 \pm 0.34$  &  $2572.4\pm 1.5$ & RPP\cite{pdg} \\
\hline
\hline
\end{tabular}
\label{DSTab}
\end{table}

{\bf Molecules}

Such a strong deviation from expectations point to
exotic structure, such as $DK$ molecules, or to a deep misunderstanding
of heavy-light hadrons.

As can be seen in Fig. \ref{DsSpectFig}, the $D_{s0}$ lies just below $DK$ threshold, while the
$D_{s1}$ lies just below $DK^*$. This suggests that mixing with these continua may be important 
components of the new $D_s$ states. Indeed, Barnes, Close, and Lipkin have suggested\cite{bcl} that
the $D_{s0}$ and $D_{s1}$ may be $DK^{(*)}$ molecules. They note that such an assignment
naturally 
explains the anomalous masses of the $D_s(2317)$ and $D_s(2460)$.  Furthermore, since the $D^0K^+$
channel is roughly 9 MeV higher than the $D^+K^0$ channel, isospin breaking is induced, thereby 
explaining
the isospin breaking discovery mode of the $D_s(2317)$. The same observation holds for the $DK^*$
channel, suggesting that isospin breaking may be important in the $D_s(2460)$ as well. 

The  molecular
assignment for the
$D_{s0}$  can be tested by measuring the radiative decay $D_{s0} \to D_s^* \gamma$. A simple quark
model computation gives 1 keV for this E1 transition\cite{CS}.
Similarly the $D_s(2460)$ transitions are 
$D_{s1} \to D_s\gamma = 7.3 \, \sin^2\phi$ keV and
$D_{s1} \to D_s^*\gamma = 4.2 \, \sin^2\phi$ keV. Using the heavy quark mixing angle $\sin^2\phi = 2/3$
yields predicted rates of 4.5 keV and 2.8 keV respectively. 

Alternatively, the radiative decay of molecular $D_{sJ}$ proceeds via the annihilation diagram shown
in Fig. \ref{Xdecays3Fig} (right panel). A simple estimate of these rates gives\cite{CS}

\begin{eqnarray}
\Gamma(D_{s0}(DK) \to D_s^* \gamma) \approx 35\ {\rm keV} \\
\Gamma(D_{s1}(DK^*) \to D_s^* \gamma) \approx 45\ {\rm keV} \\
\Gamma(D_{s1}(DK^*) \to D_s \gamma) \approx 50\ {\rm keV} 
\end{eqnarray}
Thus it is possible that such a measurement will serve to distinguish the structure of the
anomalous $D_s$ states.  There is also the intriguing possibility of a radiative transition between
molecular $D_{s1}(DK^*)$ and $D_{s0}(DK)$ states which would be driven by the virtual transition
$K^* \to K \gamma$. This rate is estimated to be 17 keV in Ref. \cite{CS}, which may be compared
to the analogous $c\bar s$ rate of $(0.3 \cos\phi + 1.4 \sin\phi)^2 \approx 1$ keV. 
Recent computations of strong decay rates are presented in Ref. \cite{Liu:2006jx}.

Finally, if the $D_{s0}$ and $D_{s1}$ really are molecular states, extra $D_{sJ}$ states should 
appear in the spectrum around 2500 MeV. Searching for these is of obvious importance.

Szczepaniak also considers a molecular assignment for the $D_s(2317)$ as $D_s\pi$ bound state with 
a binding energy of approximately 40 MeV\cite{adamDS}. He examines the properties of such a state
with an effective lagrangian which incorporates the nonlinear realisation of chiral symmetry  and 
$U_A(1)$ symmetry breaking via coupling of pions to the field $Q = F\tilde F$. The coupling of $D_s$
mesons to pions is affected via a $Q^2D_s^2$ term in the lagrangian. Parameters and cutoffs are fit to
the mass of the $D_s(2317)$, and the formalism is used to predict other charged $D_s\pi^\pm$ modes, a
scalar $D\pi$  at 2150-2300 MeV, and a scalar $DK$ molecule at 2440-2550 MeV. All of these molecules are
predicted to have widths of order 10 MeV. Finally, Szczepaniak also suggests that $D^*\pi$ and $D^*K$ 
molecules may exist as well. Unfortunately, little evidence exists for this collection of predicted 
molecules. Generally, molecules containing pions are not considered viable because the small
mass of the pion leads to very small molecular reduced masses, which tends to destabilise the 
system.

{\bf Coupled Channels}

A less extreme version of the molecular picture postulates that the $D_s(2317)$ and $D_s(2460)$ are 
dominantly $c\bar s$ states which are heavily renormalised 
by mixing with the $DK$ and $DK^*$ continua. This approach has been followed by Hwang and Kim\cite{HwangKim}
who use the Cornell coupled channel model discussed in Sect. \ref{decaySec} to compute the mass shifts 
experienced by bare $c\bar s$ states upon coupling to the $DK$ and $DK^*$ continua. The authors find 
that
the bare scalar $D_s$ at 2480 MeV has its mass lowered to 2320 MeV, in agreement with the $D_s(2317)$. 
The same philosophy has been followed by van Beveren and Rupp\cite{vBR}  except that the
$^3P_0$ model is used to drive the transitions between $c\bar s$ and the $DK$ channel.
Note, however, that both papers do not consider coupling to other channels or renormalisation 
(see Sect. \ref{spectrumCritiqueSec} for further discussion on this issue).

{\bf Tetraquarks}

In general the $D_s$ has a Fock space expansion:

\be
D_s(2317) = Z_1^{1/2} | c \bar s\rangle + Z_2^{1/2} | c \bar s (u\bar u + d \bar d)\rangle + Z_3^{1/2} |D^0 K^+\rangle + 
Z_4^{1/2} |D^+ K^0\rangle + \ldots .
\ee
The models
presented here amount to different estimations of the relative strengths of the different
channels. Tetraquark models assume $Z_2 \gg Z_1, Z_3, Z_4$ and impose a specific structure on 
the internal configuration.
Several groups have pursued this scenario\cite{maianiA,Terasaki:2003qa,Cheng:2003kg,Nielsen:2005ia,Dmitrasinovic:2005gc}. 
Note that this is an 
old idea\cite{Zvi,Tuan} which builds on 
the proposed light tetraquarks of Jaffe\cite{jaffe,Maiani:2004uc}. 

\begin{figure}[h]
  \includegraphics[width=7 true cm, angle=0]{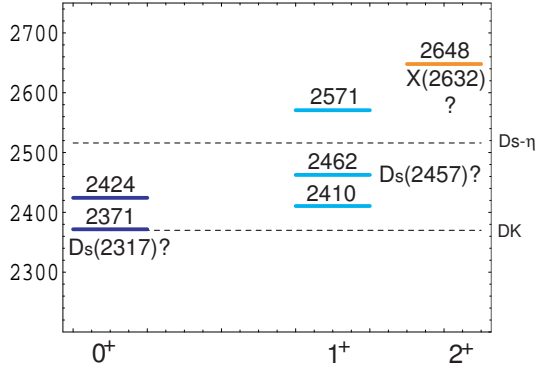}
  \caption{$c\bar s$ Tetraquark Spectrum\protect\cite{maianiA}.}
  \label{csqqFig}
\end{figure}

A large spectrum of $[cq][\bar s \bar q']$ tetraquark states exists and it is natural to attempt
identifying some of these with the anomalous $D_s$ states.  
The $[cq] \bar s \bar q']$ states may be obtained from Eq. \ref{tetraStates} upon making the replacement
$\bar c \to \bar s$. The resulting spectrum
of Maiani {\it et al.}\cite{maianiA} is shown in 
Fig. \ref{csqqFig}.
Possible assignments of the $D_s(2317)$, $D_s(2460)$, and $D_s(2632)$ (to be discussed in the following
section) are also indicated.

Along these lines, Chen and Li\cite{Chen:2004dy}
suggest modified four quark configurations of $c \bar s (u\bar u + d\bar d)$ for the $D_s(2317)$ and
$D_s(2460)$.
Nussinov\cite{Nussinov:2003uj} suggests that if the $D_s(2317)$ is a four quark state then 
a new isoscalar $D\bar D$ bound state with mass less than 3660 MeV must exist.

Browder {\it et al.}\cite{BPP} present a modified version of the tetraquark scenario, by assuming
that tetraquark states lie above 2400 MeV and mix with bare  $c\bar s$ states at a similar mass.
In this picture, the $D_s(2317)$ and
$D_s(2460)$ are predominantly $c\bar s$ states and narrow, while the four-quark states lie above
$DK$ threshold and are broad.

{\bf Lattice}

\begin{figure}[h]
  \includegraphics[width=7 true cm, angle=0]{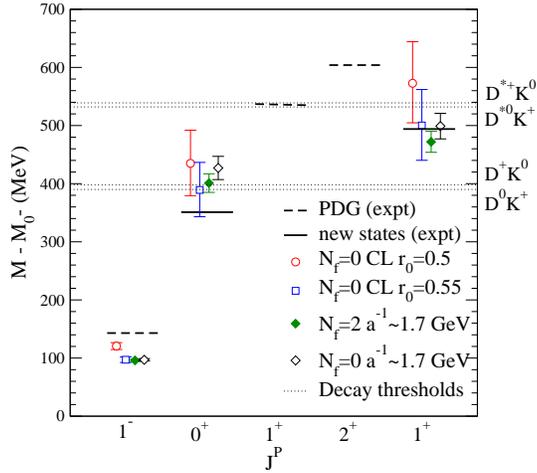}
  \caption{Lattice $D_s$ Spectrum \protect\cite{Dougall:2003hv}.}
  \label{DsLattFig}
\end{figure}

UKQCD\cite{Dougall:2003hv} have computed the $D_s$ spectrum in quenched and $N_f = 2$ unquenched
configurations. Their results are shown in Fig. \ref{DsLattFig}. The figure illustrates that
it  is difficult to come to definitive conclusions on masses in this sector due to statistical 
errors of order 100 MeV and systematic errors due to unquenching of similar size. 
Notice that the $D_{s0}$ mass is approximately 100 MeV higher than experiment -- in keeping with
naive quark model estimates (see the discussion in Sect. \ref{spectrumCritiqueSec}). We remark
that it is extremely unlikely that the lattice can differentiate the $D_{s1}$ and $D_{s1}'$ states.
Similar results are also reported by Bali\cite{Bali:2003jv}, who notes that significant finite
heavy quark mass corrections to the heavy quark mass limit are present at the charm mass scale,
non-degenerate sea quarks may be required, and quenched simulations of four-quark states
are desirable.

%
%
%
%
%

\subsubsection{Summary}

It appears likely that the $j_q = 1/2$ $(D_{s0},D_{s1})$ multiplet is difficult to 
interpret as simple $c\bar s$ states and novel dynamics are required. Of the continuum
of possibilities, tetraquarks are the most speculative and least constrained by theory
and experiment. Because of the unknown dynamics, tetraquark models tend to predict
extensive spectra. So far there is little evidence for tetraquark supernumerary
states and one must regard these models as unestablished.

Molecular interpretations are similar to tetraquark in the assumption that four quark
configurations dominate the $D_{s0}$ and $D_{s1}$. However the presumed diffuse spatial 
configurations are very different from those of tetraquark clusters (of course, the colour
and spin configurations can vary as well). Sufficiently assiduous experimental
study will be able to distinguish these scenarios.   Typical molecular models also suffer
from poor knowledge of the relevant dynamics -- in this case the nonperturbative gluodynamics
that drives meson coupling to the continuum. Thus, for example, if S-wave channels dominate
the putative $DK$ molecule one expects no charged partners.
However, these can exist if the $DK$ binding is driven by t-channel exchange.

The scenario which requires the weakest extension of the naive quark model assumes that mixing
of scalar (and axial) bare $c\bar s$ states to the $DK$ (and $DK^*$) continua substantially
modify the structure of these states. This mechanism is thought to be responsible for the
anomalous properties of the $f_0(975)$ and $a_0(980)$ and in general, should be important 
whenever bare quark states couple in S-wave to nearby continua. I regard this scenario as
the most likely explanation of the new $D_s$ states. 

Unfortunately, theoretical progress
will be difficult due to the previously mentioned unknown gluodynamics which drives Fock
sector mixing, and the difficulty in consistently renormalising the quark model. But it
is becoming clear that this effort will be required to understand a slowly expanding subset
of the hadronic spectrum.

\subsection{$D_s(2630)$}

\begin{figure}[h]
  \includegraphics[width=7 true cm, angle=0]{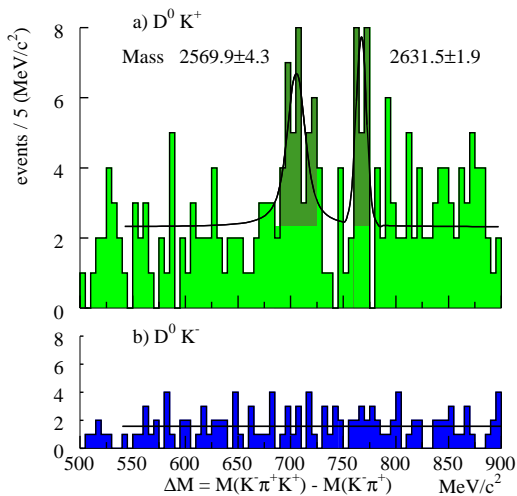}
  \caption{Observation of the $D_s(2630)$ in the $DK$ Mode\protect\cite{selex}.}
  \label{selexFig}
\end{figure}

The $D_s(2630)$ was discovered by the SELEX collaboration at FNAL\cite{selex} in the
final states
$D^0K^+$ and $D_s \eta$. The peak in the $DK$ mode can be seen in Fig. \ref{selexFig}. 
The combined measured mass is $2632.6 \pm 1.6$ MeV and the state is
surprisingly narrow with a width of
less than 17 MeV at the 90\% confidence level.

The ratio of the partial widths is measured to be 
\begin{equation}
{\Gamma(D_s \to D^0K^+) \over \Gamma(D_s \to D_s\eta)} = 0.16 \pm 0.06.
\end{equation}
As pointed out by the SELEX collaboration, this is an unusual result since the $DK$ mode is favoured
by phase space. 

 This observation has engendered speculation that the $D_s(2630)$ is an exotic 
state.
It is unlikely that this state is a $c\bar s$ hybrid since the mass of such a hybrid is expected to
be roughly 3170 MeV. Possible molecular states include a $D_s^*\eta$ system at 2660 MeV or 
$D_s^* \omega$ or $D^*K^*$ states at 2900 MeV. However the former is a P-wave which is not favoured
for binding, while the latter are too high in mass  to be plausible. 

The large number of possible tetraquarks makes it simple to fit the $D_s(2630)$ into tetraquark spectra.
For example, Chen and Li\cite{ChenLi} regard the new $D_s$ as a 
$[cs][\bar s \bar s]$ state -- an SU(3) partner
to the $D_s(2317)$. Claimed destructive interference in the decay amplitude leads to a narrow width for the state.
The authors predict an axial-vector tetraquark partner of the $D_s(2630)$ with a mass $M(D_{s1}) \approx 2770$ MeV. Note that the maximally attractive tetraquark channel, which is antisymmetric in colour
and in spin, is forbidden because of the identical anti-strange quarks\cite{gang}.

Alternatively, Maiani {\it et al.} suggest\cite{MaianiSelex}
that the $D_s(2630)$ is a $[cd][\bar d \bar s]$ tetraquark (see Fig. \ref{csqqFig}).
They speculate that this state does not mix with $[c u][\bar u \bar s]$ and hence breaks 
isospin symmetry. The decay to $D^0K^+$ would then be OZI suppressed. However the state should
decay to $D_s \pi^0$ and $D^+K^0$ with sizable branching fraction. Discovering the $D_s\pi$ mode
(along with the observed $D_s\eta$ mode) would indicate isospin breaking, as required in this 
scenario. Maiani {\it et al.} also predict the existence of a charge $+2$ state with mass near
2630 MeV, which decays to $D_s\pi^+$ and $D^+K^+$.

Finally, van Beveren and Rupp\cite{vanBeveren:2004ve} suggest that the $D_s(2630)$ is a heavily 
renormalised $c \bar s$ state with the aid of a K matrix coupled channel model. The model also predicts
${D_s^*}'$ states at 2720, 3030, and 3080 MeV with widths of 2, 50-60, and 8 MeV respectively.

\begin{figure}[h]
  \includegraphics[width= 6 true cm, angle=270]{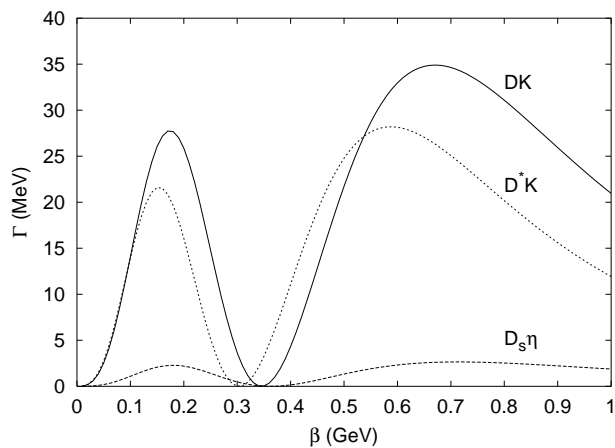}
  \caption{Theoretical Partial Widths for the $D_s(2630)$ as a Function of the SHO $D_s$ Scale
$\beta$ \protect\cite{gang}.}
  \label{betaSelexFig}
\end{figure}

If the predicted decay modes or partner states of the $D_s$ are not found, the
only remaining possibility is that the $D_s(2630)$ is a 
radially excited $c\bar s$ vector\cite{chao} (although it is some 100 MeV lighter than 
the quark model predictions of 2730 MeV). The peculiar decay ratio remains to be explained. 
Experience with the decay modes
of the $\psi(3S)$ to $DD$, $DD^*$ and $D^*D^*$ points to a possible resolution:
transition matrix elements for excited hadrons may have zeroes due to wavefunction nodes.
It is possible that such a node is suppressing the $DK$ decay mode.
Computation\cite{gang}, shown in Fig. \ref{betaSelexFig},  reveals that there is indeed a node 
but that it occurs at a 
wavefunction scale which
is 20\% lower than the scale expected in quark models.
Furthermore, the $DK$ mode is always larger than
the $D_s\eta$ mode so that explaining the data is difficult in this scenario.

\begin{figure}[h]
  \includegraphics[width=7 true cm, angle=0]{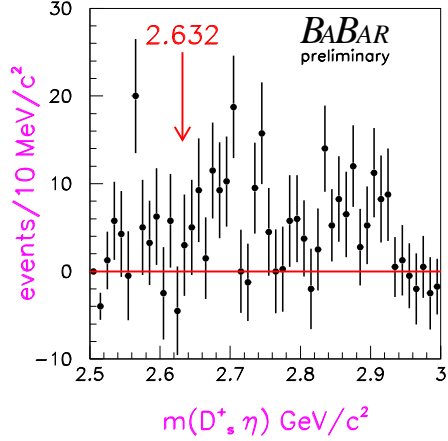}
  \caption{Non-observation of the $D_s(2630)$ in the $D_s\eta$ Mode by BaBar\protect\cite{babarA}.}
  \label{babar2630Fig}
\end{figure}

We have run out of options and must conclude that the $D_s(2630)$ is an experimental artefact.
This conclusion now appears likely because  searches by FOCUS\cite{focus},
BaBar\cite{babarA} (see Fig. \ref{babar2630Fig}), and CLEO\cite{cleoD} have found 
no evidence for the $D_s(2630)$.

\subsection{P-wave $B$ Mesons}
\label{PBSect}

\begin{figure}[ht]
\includegraphics[height=.25\textheight]{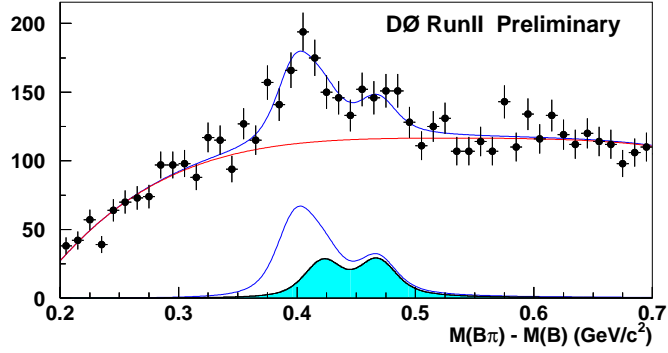}
\caption{Orbitally-excited P-wave B mesons, as reported by D{\O}. 
Fitted contributions from 
B$_1\to$B$\pi$,
B$_2^* \to$B$\pi$ and 
B$_2^* \to$B$^*\pi$ are shown \cite{D0_Bnote}.}
\label{B-p-waveFig}
\end{figure}

Preliminary results 
from CDF and D{\O} on orbitally excited B mesons were presented at Hadron 05\cite{Barnes05}.
The $B$ mesons were reconstructed in $J/\psi K \pi$ and sufficient events were collected
to discern the $B_2$ and $B_1$ contributions to the rate (see Fig. \ref{B-p-waveFig}).

The D{\O} mass results are
\begin{equation}
M(B_1) = 5724 \pm 4 \pm 7~{\rm MeV},
\label{B1_mass}
\end{equation}
\begin{equation}
M(B_2) - M(B_1) = 23.6 \pm 7.7 \pm 3.9~{\rm MeV}.
\label{B2_mass}
\end{equation}
Godfrey and Isgur\cite{GI} predict $M(B_2) = 5800$ MeV, $52 \pm 12 $ MeV too high. The 
$B_1$s are predicted to be ``within 40 MeV of the $B_2$''. Alternatively, the model of 
Matsuki and Morii\cite{MM}, which was so noticeably successful for the $D_s$ system, predicts
$M(B_2) = 5692$ MeV and $M(B_1) = 5740$ or $5679$ MeV.

The total widths, constrained in the fit to be equal (and predicted to be 
comparable theoretically) are
\begin{equation}
\Gamma = 23 \pm 12~{\rm MeV},
\label{B_width}
\end{equation}
which is consistent with theoretical expectations for the two narrow P-wave
B mesons. The experimental resolution is estimated to be about 10~MeV.

\subsection{$B_c$}

The $B_c$ pseudoscalar meson was originally seen in the semileptonic decay channels
$B_c \to J/\psi \ell \nu_\ell X$ at FNAL\cite{oldBC}. Unfortunately the low statistics and
unreconstructed final state did not permit an accurate determination of the mass:

\be
M(B_c) = 6400 \pm 390 \pm 130\ {\rm MeV}.
\ee

\begin{figure}[h]
  \includegraphics[width=7 true cm, angle=0]{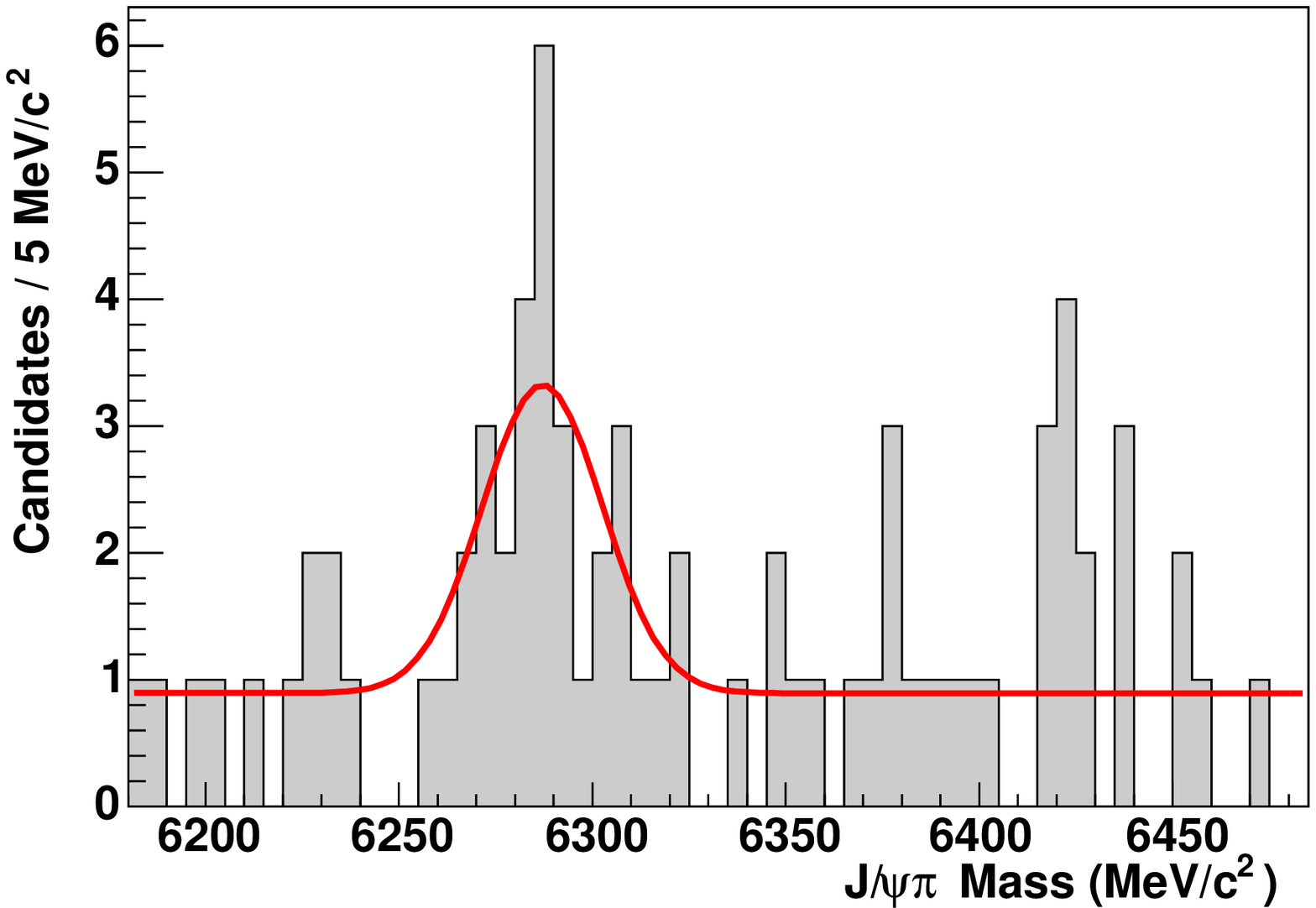}
  \caption{$B_c$ Signal\protect\cite{Acosta:2005us}.}
  \label{BcFig}
\end{figure}

In May 2005 the CDF collaboration\cite{Acosta:2005us} announced an accurate determination of the
$B_c$ mass. The state was
seen in $14.6 \pm 4.6$ events in the $J/\psi \pi^\pm$ mode (see Fig. \ref{BcFig}) with a 
reported mass of 

\be
M(B_c) = 6285.7 \pm 5.3 \pm 1.2\ {\rm MeV}.
\ee

Subsequently, the CDF collaboration has also reconstructed\cite{Spezziga:2005ec} 
 the $B_c$ in the decay mode $B_c \to J/\psi \ell$ with a significance of 5.2 $\sigma$ for $\ell= \mu$ (or 5.9 $\sigma$ for $\ell = e$). The following ratios of branching fractions have been 
measured:

\be
{\sigma(B_c) Br(B_c \to J/\psi \mu) \over\sigma(B) Br(B \to J/\psi K)} = 0.249 \pm 0.045 ^{+0.107}_{-0.076}
\ee
and
\be
{\sigma(B_c) Br(B_c \to J/\psi e) \over\sigma(B) Br(B \to J/\psi K)} = 0.282 \pm 0.038 \pm 0.074.
\ee
An updated fit to the $B_c \to J/\psi \pi$ decay mode yields\cite{Spezziga:2005ec}

\be
M(B_c) = 6287.0 \pm 4.8 \pm 1.1\ {\rm MeV}.
\ee

Lastly, the lifetime of the $B_c$ has been measured to be

\be
\tau(B_c) = 0.474 {}^{+0.073}_{-0.066} \pm 0.033 \ {\rm ps}.
\ee
This may be compared to theoretical expectations\cite{Kiselev:2003mp} of $0.55 \pm 0.15$ ps.
Remarkably, the inclusive computation of this lifetime considers 65 different decay modes.

The timing of the CDF announcement was arranged with the
HPQCD lattice group\cite{kronfeld} to permit them a prediction of its mass. The lattice prediction was
$6304 \pm 12 {}^{+18}_{-0}$ MeV.
Thus CDF may claim a state, the lattice 
may claim a victory, and the $B_c$ appears to carry no surprises. 
While there is justifiable pride in the present predictions and experiment, some
perspective may be gained when it is recalled that Godfrey and Isgur predicted a $B_c$ 
of 6270 MeV some twenty years ago\cite{GI}. As stated in Sect. \ref{spectrumCritiqueSec}, this 
should not 
astonish, the quark model and lattice gauge theory tend to agree on states with a simple 
Fock space expansion.

\section{Conclusions}

The moderately heavy charm quark makes charming spectroscopy especially interesting since it
lends confidence that the behaviour of soft QCD can be extracted from the data, yet it is not
so heavy that
one is simply exploring an analogue of the positronium spectrum.
Thus the new charming
mesons have generated much enthusiasm. The $X(3872)$, $D_s(2317)$, and $D_s(2460)$ remain enigmatic
and highlight a less than perfect understanding of QCD. Can we be seeing the expected breakdown of
the naive quark model? Where and how do gluonic degrees of freedom manifest themselves? Have we
entered a new era of `mesonic nuclear physics'?

If the $X(3872)$ is not canonical charmonium then the simplest explanation lies in expanding the
Fock space description of the state. With $1^{++}$ quantum numbers this is likely to mean 
$c\bar c q\bar q$. Whether a diffuse four-quark state dominates the $X$, as in the molecular
interpretation, or a compact state dominates as with tetraquarks, or the four-quark component
heavily renormalises an erstwhile $\chi_{c1}'$ is a matter of dynamics. This dynamics can be
subtle, but often leads to different and testable predictions.  Thus, for example,  an entire 
spectrum of 
tetraquarks is predicted while a modest number of additional molecular states are expected.
Alternatively, the idea that the $X$ is a purely kinematical cusp effect can be checked by 
measuring the rate for
$B^+ \to K^0 D^+ \bar D^{*0}$. If no enhancement is seen in this channel it implies that significant 
final state interactions are creating the $X$ signal.
Finally, pinning down the $DD\pi$ mode will be vital in establishing the magnitude of 
$c\bar c$--$c\bar c q\bar q$ mixing. And this will tell us something about the gluodynamics
which generates mixing.

The $X$, $Y$, $Z$ charmonium complex near 3940 MeV poses interesting problems. 
{\it Prima facie}, it has some
of the expected features of a 2P multiplet. However,
if the $Z$ is interpreted as the $\chi_{c2}'$ and the $X(3940)$ is the $\chi_{c1}'$ then
the multiplet displays an
inverted spin orbit splitting. This unusual situation would then highlight significant confusion
in our understanding of the Dirac structure of confinement.
Alternatively, the unusual $\omega J/\psi$ decay mode of the $Y(3940)$ has led to suggestions
that it is a charmonium hybrid. 
This suggestion does fit with (rather shaky) expectations for hybrid decays; however, the 
mass is much too low compared to (somewhat less shaky) quenched lattice computations and prudence
-- and more data -- are called for.
The evidence for the $Y(3940)$ is not especially robust and one must hope for 
additional experimental effort in determining  the properties of this state.
Lastly, if the $X(3940)$ is the $\chi_{c1}'$  then why is the $\chi_{c1}$ not apparent in the
same data set?

The $Y(4260)$ is an intriguing candidate for a vector charmonium hybrid. This possibility
may be pursued by searching for other $S+P$ decay modes such as $\chi_{c0}\omega$ and for
nearby pseudoscalar and $J^{PC} =1^{-+}$ hybrids. The resonance  interpretation of the $Y$
may be confounded by the proximity of the $D_1\bar D$ channel which will cause an enhancement
due to the threshold cusp effect and possible strong $D_1\bar D$ rescattering. This
possibility can be tested by measuring analogous strange systems such as $D_{s1}\bar D$ and 
$D_{s1}\bar D_s$. Finding manifestations of glue in the spectrum is a primary goal of 
nonperturbative QCD and one must hope that the $Y$ will be studied in greater detail. CLEO-c
may be able to assist substantially in this effort.

The masses of the $D_{s0}(2317)$ and $D_{s1}(2460)$ point to a significant lack of understanding
of QCD dynamics in the $c\bar s$ system. Possible interpretations
include tetraquark states or  molecular $DK$ or $DK^*$ states.
The scenario which requires the weakest extension of the naive quark model assumes that mixing
of scalar (and axial) bare $c\bar s$ states to the $DK$ (and $DK^*$) continua substantially
modify the structure of these states.  An alternative explanation is provided by the chiral 
doublet model, but it has been argued that this model is likely not relevant to low-lying
heavy-light states in Section \ref{DoubletModelSect}. Firmly establishing the analogous $D$
states will assist in understanding the novel dynamics which appears to be present in $P$-wave
heavy-light hadrons\cite{bianco}.

The advent of the $\Theta$ pentaquark states and the new mesons has led to increased interest
in diquark correlations as effective degrees of freedom. However, the subsequent collapse of the pentaquark
states raises serious concerns about the reliability of chiral soliton and other pentaquark models.
Similarly, an infinite collection of tetraquark states has been predicted. However, there is no
evidence for any of the {\it predicted} low lying states. Clearly a degree of circumspection is
called for.

Alternatively, many of the proffered explanations of the unusual states involve mixing with
higher Fock components. These explanations are economical in that new dynamics or degrees of
freedom need not be postulated. However, they must be carefully constructed because it is 
relatively easy to destroy previous agreement with experiment. A simple organising principle 
seems to be emerging: the presence of nearby S-wave meson-meson thresholds can strongly
affect meson structure. While this observation can serve as a useful rule of thumb, turning
it into a quantitative theory is a daunting task because it is driven by unknown nonperturbative
gluodynamics. Of course, understanding gluodynamics
is one of the great remaining challenges in the development of the Standard Model, and the effort
is worthwhile.

More than a decade ago Isgur stated\cite{Isgur:1989qa}, 
``Twenty five years after the birth of the quark model,
light quark ... spectroscopy is in a deplorable state".
While the situation is slowly improving, 
it is the formerly comfortable heavy quark sector which is now under siege. And new data seem 
to reveal new ignorance with each passing month.
It is clear that the simple constituent quark model must fail somewhere
 -- are we seeing this?
The new heavy meson spectroscopy
challenges our understanding of QCD. Can we meet the challenge?

\section*{Acknowledgments}

I am grateful to the organisers of the International Conference on QCD and Hadronic Physics and the
International Conference on Hadron Spectroscopy for invitations to speak about this fascinating
topic. These talks formed the nucleus of this review.
I would like to thank
Ted Barnes, Denis Bernard, Eric Braaten, Frank Close, Alexey Petrov, Walter Toki,
Steve Dytman, Steve Godfrey, Steve Olsen, 
and Steve Yzerman for inspiration and
many interesting discussions on heavy hadron phenomenology.
This work was supported by 
the US Department of Energy under contract DE-FG02-00ER41135, the National Science Foundation under grant 
NSF-PHY-244668, and PPARC grant PP/B500607 in the UK.


\appendix

\section{Bound State Decay Formalism}

While the use of the weak binding decay relations of Sect. \ref{XdecaysSect} are undoubtedly
accurate for the $X(3872)$, it is useful to examine molecular decays
in more  depth to assess the accuracy of the weak binding approximation. 
Furthermore, it was assumed
that the $\rho J/\psi$ decay to $\pi\pi J/\psi$ proceeded as if the $\rho J/\psi$ channel belonged
in the bound state instead of in the continuum. This simplifying assumption will also be justified
in the following.

%
%
%
%

\begin{figure}[h]
  \includegraphics[width=5 true cm, angle=0]{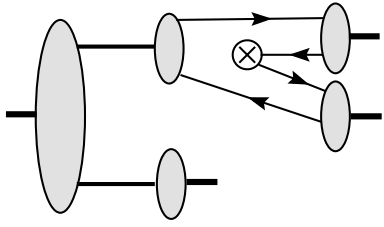}
  \caption{Generic $X \to (AB) \to \alpha\beta B$ Decay.}
  \label{GenXdecayFig}
\end{figure}

We first consider the generic strong decay of a molecule, $X$ to a final state $\alpha\beta B$ via a
bound state component, $AB$ illustrated in Fig. \ref{GenXdecayFig}.
In the $^3P_0$ model this amplitude is

\begin{equation}
{\cal A}(X \to AB; A\to \alpha\beta) = Z_{AB}^{1/2} \phi_{X(AB)}(q_\alpha+q_\beta) \int {d^3k\over (2\pi)^3} \phi_A(k)
\phi_\alpha^*(k + {q_\beta\over 2}) \phi_\beta^*(k - {q_\alpha\over 2}) \langle \sigma\rangle \cdot (\vec k - {\vec q_\alpha\over 2} + {\vec q_\beta \over 2}) 
\end{equation}
where $\langle \sigma \rangle$ is the matrix element of the Pauli matrices in the $A\alpha\beta$ spin
states and $q_\alpha$ and $q_\beta$ are the momenta of the mesons $\alpha$ and $\beta$ respectively.

The differential decay rate is given by 

\be
d\Gamma = |{\cal A}|^2 d\phi_3 = Z_{AB} |\phi_{X(AB)}(q_\alpha+q_\beta)|^2 {\cal A}(A \to \alpha\beta)|^2
\ee
where $d\phi_3$ is the three particle phase space. Notice that the use of the $^3P_0$ model is
no longer relevant. Integrating and using the recursion property of phase space (Eq. 37.12 of
Ref. \cite{pdg}) gives

\be
\Gamma(X \to \alpha\beta B) = Z_{AB} \int {d^3 q\over (2\pi)^3} |\phi_{X(AB)}(q)|^2 \Gamma(A \to \alpha\beta)
\ee
where the $A$ meson is taken to have energy $E_A = M(X) - E_B(q)$ and momentum $\vec q$. In the
weak binding limit the $X$ wavefunction  is strongly peaked at the origin and one obtains

\be
\Gamma(X \to \alpha\beta B) \approx Z_{AB} \Gamma(A\to \alpha\beta)
\ee
where now $\Gamma(A\to \alpha\beta)$ is the experimentally determined free space partial width of
the $A$. This identification requires $M(X) - M(B) \approx M(A)$ which is of course also true
in the weak binding limit. Thus the weak binding formulas are justified for the case where
the channel $AB$ is a bound state component of the molecule.

\begin{figure}[h]
  \includegraphics[width=7 true cm, angle=0]{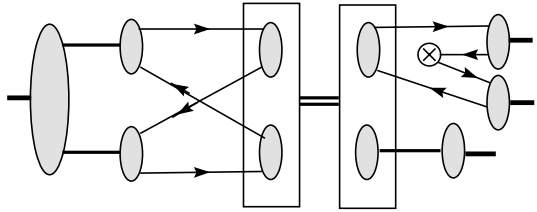}
  \caption{Second Order Transition $X(AB) \to CD \to \alpha\beta D$.}
  \label{GenXdecay2Fig}
\end{figure}

The use of the weak binding decay relation for $X \to \pi\pi J/\psi$ is more problematic 
because the $\rho J/\psi$ channel should be considered part of the continuum rather than part of the 
$X$ wavefunction as assumed in Sect. \ref{XdecaysSect}. Indeed, this process is second order
in perturbation theory, as illustrated in Fig. \ref{GenXdecay2Fig}, rather than first order
as above. In this figure, the boxes represent intermediate states which are summed.

The perturbative expression for the decay amplitude is

\begin{equation}
{\cal A}(X(AB) \to CD \to \alpha \beta D) = \tilde T(q_\alpha+q_\beta) { {\cal A}(B \to \alpha\beta))\over
M(X) - E_{CD} + i \epsilon}
\end{equation}
where

\begin{equation}
\tilde T(q) = \int {d^3p\over (2\pi)^3} \phi_{X(AB)}(p)T_{AB\to CD}(p\to q).
\end{equation}
and $T$ is the perturbative scattering matrix representing the subprocess $AB \to CD$ in Fig. \ref{GenXdecay2Fig}.

%
%

Integrating over three body phase space yields the following expression for the width

\begin{equation}
\Gamma(X \to \alpha\beta D) = \int {d^3Q\over (2\pi)^3}  \Gamma(C\to \alpha\beta) {|\tilde T(Q)|^2 \over 
(M(X) - E_{CD}(Q))^2 + \Gamma_C^2/4}
\label{XCD}
\end{equation}
where the total width of the $C$ meson has been restored to the propagator and it assumed that the $D$ meson is stable. The
width for the $C$ is to be evaluated at the kinematical point $E_C = M(X) - E_D(Q)$.

In the limit where the $C$ particle is narrow this approaches

\begin{equation}
\Gamma(X \to CD) = 2 \pi Q {E_C E_D\over M(X)} Br \int |\tilde T|^2 d\Omega
\end{equation}
as expected. Furthermore, if the secondary amplitude $C \to \alpha\beta$ is independent of momentum
Eq. \ref{XCD} simplifies to

\be
\Gamma(X \to CD \to \alpha\beta D) = \Gamma(C \to \alpha\beta)  \int {|\tilde T|^2\over (M(X)-E_{CD})^2} = Z_{CD} \Gamma(C \to \alpha\beta).
\ee
Thus the weak binding expression is recovered under the conditions that the production side of Fig. \ref{GenXdecay2Fig} and the decay side factorise. In general one expects no dependence of the $C$ width on the
momentum of the $C$ so the constraint is certainly plausible.

\section{Pion Exchange Potentials}

The standard quark-pion interaction (as used in nuclear physics) is given by

\be
L = -\frac{g}{\sqrt{2}f_\pi} \int d^3x \bar \psi \gamma_\mu \gamma_5 \vec \tau \psi \cdot \partial^\mu \vec \pi
\ee
where $f_\pi = 92$ MeV and $g$ is the pion-quark coupling constant\cite{NAT}.
Effective pion exchange potentials are derived from the nonrelativistic limit of the
appropriate $t$-channel scattering amplitude where the vertices are computed
from the pion-quark interaction given above.

Extraction of the $\pi HH$ vertex is simplified by using the Hamiltonian\cite{Ericson:1993wy}

\be 
H = i \sum_q \frac{g}{\sqrt{2}f_\pi} \vec \sigma \cdot \vec k \vec \tau \cdot \vec \pi
\ee
where the sum refers to a sum over quark flavours.
The computations may be simplified further by 
considering forward propagating neutral pions:

\be
H = i \sum_q \frac{g}{\sqrt{2}f_\pi} \sigma_z k \tau_z \pi^0
\ee

For example, the effective one-pion-exchange induced NN interaction is obtained from the
$\pi NN$ vertex which is given by

\be 
\langle p_\uparrow | H | p_\uparrow\rangle = {5\over 3} i \frac{g}{\sqrt{2}f_\pi}  k \pi^0.
\ee
This relation has been derived using the proton wavefunction

\be
|p_\uparrow\rangle = {1\over \sqrt{6}}\left( |u_\uparrow u_\downarrow d_\uparrow\rangle + | u_\downarrow u_\uparrow d_\uparrow \rangle - 2 |u_\uparrow u_\uparrow d_\downarrow \rangle\right).
\ee

The NN scattering amplitude is then given by

\be
{\cal M} = - \frac{25}{9} \left(\frac{g}{\sqrt{2}f_\pi}\right)^2 \vec \tau_1\cdot \vec \tau_2 {\sigma_1\cdot \vec k \, \sigma_2\cdot\vec k\over k^2 + m_\pi^2}
\ee
which yields the relationship $g_{NN\pi} = \frac{5}{3} \frac{g m_N \sqrt{2}}{f_\pi}$.

Fourier transforming gives the generic pion-induced interaction

\be
V_{ope} = \xi V_0 \vec \tau_1 \cdot\vec\tau_2 \left[ \sigma_1\cdot\sigma_2 C(r) + (3 \sigma_1\cdot\hat r \vec 
\sigma_2\cdot \hat r - \sigma_1\cdot \sigma_2) T(r) \right]
\ee
where $\xi$ is a numerical factor determined by the channel in question and

\be
V_0 = {g^2 m_\pi^3 \over 24 \pi f_\pi^2} \approx 1.3 \ {\rm MeV},
\ee

\be
C(r) =  {{\rm e}^{-m_\pi r}\over m_\pi r},
\ee
and

\be
T(r) = \left( {3\over m_\pi^2 r^2} + {3  \over m_\pi r} + 1\right) C(r).
\ee

Nucleon-nucleon interactions may now be evaluated using $\xi = \frac{25}{9}$ and $\langle \vec \tau_1 \cdot \vec \tau_2\rangle = 2I(I+1) - 3$. Thus all possible $NN$ interactions are as listed in Table \ref{NNTab}.

\begin{table}
\caption{One-pion-exchange $NN$ Interactions}
\begin{tabular}{cccl}
\hline
isospin & $J^P$ & channels & potential \\
\hline
\hline
1 & $0^+$ & $^1S_0$ &   $-{25\over 3} V_0 C(r)$\\
1 & $0^-$ & $^3P_0$ &  $ {25\over 9} V_0 (C(r) - 4 T(r))$\\
0 & $1^-$ & $^1P_1$ &  $ 25 V_0 C(r)$ \\ 
0 & $1^+$ & $^3S_1/{}^3D_1$ (the deuteron) & Eq. \ref{VdEqn}  \\
\hline
\hline
\end{tabular}
\label{NNTab}
\end{table}

\be
V = -{25\over 3} V_0 \left[ \pmatrix{1 & 0 \cr 0  & 1}C(r) + \pmatrix{0 & \sqrt{8} \cr \sqrt{8} & -2 } T(r) \right]
\label{VdEqn}
\ee
This corrects errors in Table 1 and Eq. 30 of Ref.\cite{NAT}. 

The interaction is often regulated in an attempt to mimic unknown short range dynamics.
It is traditional to  do so with a dipole form factor such as $(\Lambda^2 - m_\pi^2)/(\Lambda^2 + t)$. 
The resulting softening of the potential is crucial to reproducing deuteron properties. 
In principle, the form factor can be computed by convoluting the spatial pion and nucleon 
wavefunctions. However, attempts to do so yield unsatisfactory results and the dipole form factor
has been retained in $X$ the molecular computations reported here. Finally, we note that the 
form factor induces a very short range repulsive core (figures 1 and 2 of Ref. \cite{NAT} are
misleading) that, in part, explains the phenomenological usefulness of this method. The cutoff
has the further benefit of providing a factorisation scale between long range pion exchange
and short range quark exchange dynamics.

Computations  for meson-meson interactions proceed in a similar way. For example vertices for 
$K$ or $K^*$ scattering are given by 

\be
\langle K | H | K \rangle = \frac{g}{\sqrt{2}f_\pi} k \pi^0
\ee
and
\be
\langle \bar K | H | \bar K \rangle = -\frac{g}{\sqrt{2}f_\pi} k \pi^0.
\ee
Forming the $t$-channel scattering amplitude and Fourier transforming then yields the effective
one-pion-exchange meson-meson interactions shown in Table \ref{MMTab}.

\begin{table}
\caption{One-pion-exchange $MM$ Interactions}
\begin{tabular}{ccccl}
\hline
state & isospin & $J^{PC}$ & channels  & potential \\
\hline
$D\bar D^*$ & 0 & $0^{-+}$ & $^3P_0$ & $-3 V_0 (C(r) +2 T(r))$ \\
$D\bar D^*$ & 0 & $0^{--}$ & $^3P_0$ & $+3 V_0 (C(r) +2 T(r))$ \\
$D\bar D^*$ & 0 & $1^{-+}$ & $^3P_1$ & $-3 V_0 (C(r) - T(r))$ \\
$D\bar D^*$ & 0 & $1^{--}$ & $^3P_1$ & $+3 V_0 (C(r) - T(r))$ \\
$D\bar D^*$ & 0 & $1^{++}$ & $^3S_1$, $^3D_1\ $ & $+3\cdot$ (Eq \ref{MMEqn})\\
$D\bar D^*$ & 0 & $1^{+-}$ & $^3S_1$, $^3D_1\ $ & $-3\cdot$ (Eq \ref{MMEqn})\\
\hline
$D\bar D^*$ & 1 & $0^{-+}$ & $^3P_0$ & $+V_0 (C(r) +2 T(r))$ \\
$D\bar D^*$ & 1 & $0^{--}$ & $^3P_0$ & $-V_0 (C(r) +2 T(r))$ \\
$D\bar D^*$ & 1 & $1^{-+}$ & $^3P_1$ & $+V_0 (C(r) - T(r))$ \\
$D\bar D^*$ & 1 & $1^{--}$ & $^3P_1$ & $-V_0 (C(r) - T(r))$ \\
$D\bar D^*$ & 0 & $1^{++}$ & $^3S_1$, $^3D_1$ & $-$(Eq \ref{MMEqn})\\
$D\bar D^*$ & 0 & $1^{+-}$ & $^3S_1$, $^3D_1$ & +(Eq \ref{MMEqn})\\
\hline
$D^* \bar D^*$ & 0 & $0^{++}$ & $^1S_0$, $^5D_0$ & 6 $\cdot$ (Eq \ref{MM2Eqn})\\
$D^* \bar D^*$ & 0 & $0^{-+}$ & $^3P_0$ & $-3 V_0 (C(r) + 2 T(r))$ \\
$D^* \bar D^*$ & 0 & $1^{++}$ & $^5D_1$ & $+3 V_0 (C(r) - T(r))$ \\
$D^* \bar D^*$ & 0 & $1^{-+}$ & $^3P_1$ & $-3 V_0 (C(r) - T(r))$ \\
$D^* \bar D^*$ & 0 & $1^{+-}$ & $^3S_1$, $^3D_1$ & 3$\cdot$ (Eq \ref{MMEqn}) \\
$D^* \bar D^*$ & 0 & $1^{--}$ & $^1P_1$, $^5P_1$, $^5F_1$ & 6$\cdot$ (Eq \ref{MM3Eqn}) \\
$D^* \bar D^*$ & 0 & $2^{++}$ & $^1D_2$, $^5S_2$, $^5D_2$, $^5G_2$ $\ $  & 6$\cdot$ (Eq \ref{MM4Eqn}) \\
$D^* \bar D^*$ & 0 & $2^{+-}$ & $^3D_2$ & $-3 V_0 (C(r) - T(r))$ \\
\hline
$D^* \bar D^*$ & 1 & $0^{++}$ & $^1S_0$, $^5D_0$ & -2 $\cdot$ (Eq \ref{MM2Eqn})\\
$D^* \bar D^*$ & 1 & $0^{-+}$ & $^3P_0$ & $+V_0 (C(r) + 2 T(r))$ \\
$D^* \bar D^*$ & 1 & $1^{++}$ & $^5D_1$ & $-V_0 (C(r) - T(r))$ \\
$D^* \bar D^*$ & 1 & $1^{-+}$ & $^3P_1$ & $+V_0 (C(r) - T(r))$ \\
$D^* \bar D^*$ & 1 & $1^{+-}$ & $^3S_1$, $^3D_1$ & -(Eq \ref{MMEqn}) \\
$D^* \bar D^*$ & 1 & $1^{--}$ & $^1P_1$, $^5P_1$, $^5F_1$ & -2$\cdot$ (Eq \ref{MM3Eqn}) \\
$D^* \bar D^*$ & 1 & $2^{++}$ & $^1D_2$, $^5S_2$, $^5D_2$, $^5G_2$ $\ $  & -2$\cdot$ (Eq \ref{MM4Eqn}) \\
$D^* \bar D^*$ & 1 & $2^{+-}$ & $^3D_2$ & $V_0 (C(r) - T(r))$ \\
\hline
\hline
\end{tabular}
\label{MMTab}
\end{table}

\be
V = -V_0 \left[ \pmatrix{1 & 0 \cr 0  & 1}C(r) + \pmatrix{0 & -\sqrt{2} \cr -\sqrt{2} & 1 } T(r) \right].
\label{MMEqn}
\ee

\be
V = -V_0 \left[ \pmatrix{1 & 0 \cr 0  & -\frac{1}{2}}C(r) + \pmatrix{0 & \frac{1}{\sqrt{2}} \cr \frac{1}{\sqrt{2}} & 1 } T(r) \right].
\label{MM2Eqn}
\ee

\be
V = -V_0 \left[ \pmatrix{1 & 0 & 0 \cr 0  & -\frac{1}{2} & 0 \cr 0 & 0 & -\frac{1}{2} } C(r) + 
\pmatrix{0 & \sqrt{4\over 5} & -\sqrt{6\over 5} \cr \sqrt{4\over 5} & -\frac{7}{5} & \sqrt{6\over 25} 
\cr -\sqrt{6 \over 5} & \sqrt{6\over 25} & -\frac{8}{5} } T(r) \right].
\label{MM3Eqn}
\ee

\be
V = -V_0 \left[ \pmatrix{1 & 0 & 0 & 0 \cr 
                         0 & -\frac{1}{2} & 0 & 0  \cr 
                         0 & 0 & -\frac{1}{2} & 0 \cr 
                         0 & 0 & 0 & -\frac{1}{2} } C(r) + 
                 \pmatrix{0 & \sqrt{1\over 10} & -\sqrt{1\over 7} & \sqrt{9\over 35} \cr 
                        \sqrt{1\over 10} & 0 & -\sqrt{7\over 10} & 0 \cr
                        -\sqrt{1\over 7} & -\sqrt{7\over 10} & -\frac{3}{14} & -\frac{6\sqrt{5}}{35} \cr
                        \sqrt{9\over 35} & 0 & -\frac{6\sqrt{5}}{35} & \frac{5}{7} } T(r) \right].
\label{MM4Eqn}
\ee

\end{document}